\documentclass[12pt]{oxarticle}
\pdfoutput=1

\usepackage{graphicx, float, array, xspace, amscd, amsmath, amsthm, amssymb, latexsym, relsize, bbm}
\usepackage[letterpaper, left=1.8in, right=0.3in, top=0.8in, bottom=1.2in]{geometry}
\usepackage[bbgreekl]{mathbbol}
\usepackage{amsfonts, tikz-cd}
\usepackage[T1]{fontenc}
\PassOptionsToPackage{linecolor=blue,backgroundcolor=blue!25,bordercolor=blue,textsize=scriptsize}{todonotes}
\usepackage{tikz}
\usetikzlibrary{positioning, arrows, arrows.meta, shapes.multipart, fit}

\usetikzlibrary{arrows.meta, positioning,backgrounds}
\usepackage[framemethod=tikz]{mdframed} 
\AtBeginEnvironment{mdframed}{%
\tikzset{every picture/.style={}}%
}
\mdfsetup{roundcorner=.5ex}
{\begin{mdframed}[backgroundcolor=white]}%
{\end{mdframed}}

\DeclareSymbolFontAlphabet{\mathbb}{AMSb}
\DeclareSymbolFontAlphabet{\mathbbl}{bbold}

\usepackage[sort&compress, comma, square, numbers]{natbib}
\usepackage[all,cmtip]{xy}
\usepackage{color}
\definecolor{MyDarkBlue}{rgb}{0.15,0.25,0.45}
\usepackage[hypertexnames=false,linktocpage=true]{hyperref}
\hypersetup{colorlinks=true, citecolor=blue, linkcolor=blue, urlcolor=blue, pdfauthor={}, pdftitle={},breaklinks=true}

\let\SS=\S 

\renewcommand{\#}{^{\sharp}}

\newcommand{\Ric}{{\rm Ric}}

\newcommand{\ol}[1]{{\overline{#1}}}

\newcommand{\LC}{\text{\tiny LC}}
\newcommand{\CH}{\text{\tiny CH}}
\newcommand{\Hu}{{\text{\tiny H}}}
\newcommand{\Bi}{{\text{\tiny B}}}

\newcommand{\Ch}{\text{\tiny CH}}

\renewcommand{\sb}{{\overline{\sigma}}}

\newcommand{\rb}{{\overline{ r}}}

\newcommand{\cc}{\text{c.c.}}

\newcommand{\SO}{{\rm SO}}
\newcommand{\SU}{{\rm SU}}

\newcommand{\w}{{\,\wedge\,}}
\newcommand{\wt}{\widetilde}
\newcommand{\wh}{\widehat}

\newcommand{\fD}{{\mathfrak{D}}}

\newcommand{\delslash}{\ensuremath \raisebox{0.025cm}{\slash}\hspace{-0.23cm} \del}
\newcommand{\Hslash}{\hspace{0.1cm}\ensuremath \raisebox{0.03cm}{\slash}\hspace{-0.30cm} H}

\newcommand{\Pslash}{\hspace{0.1cm}\ensuremath \raisebox{0.03cm}{\slash}\hspace{-0.28cm} P}
\newcommand{\Qslash}{\hspace{0.1cm}\ensuremath \raisebox{0.03cm}{\slash}\hspace{-0.28cm} Q}

\newcommand{\Fslash}{\hspace{0.1cm}\ensuremath \raisebox{0.025cm}{\slash}\hspace{-0.28cm} F}

\newcommand{\half}{\frac{1}{2}}
\newcommand{\qrt}{\frac{1}{4}}

\def\CS{{\text{CS}}}

\newcommand{\ab}{{\overline\alpha}}
\newcommand{\bb}{{\overline\beta}}

\renewcommand{\a}{\alpha}
\renewcommand{\b}{\beta}
\newcommand{\g}{\gamma}\newcommand{\G}{\Gamma}
\renewcommand{\d}{\delta}\newcommand{\D}{\Delta}
\newcommand{\ve}{\varepsilon}

\newcommand{\Th}{\Theta}

\renewcommand{\k}{\kappa}
\renewcommand{\l}{\lambda}\renewcommand{\L}{\Lambda}
\newcommand{\m}{\mu}
\newcommand{\n}{\nu}

\renewcommand{\r}{\rho}
\newcommand{\s}{\sigma}\renewcommand{\S}{\Sigma}
\renewcommand{\t}{\tau}

\renewcommand{\o}{\omega}\renewcommand{\O}{\Omega}

\DeclareFontFamily{OT1}{pzc}{}
\DeclareFontShape{OT1}{pzc}{m}{it}{<-> s * [1.200] pzcmi7t}{}
\DeclareMathAlphabet{\mathpzc}{OT1}{pzc}{m}{it}

\newcommand{\ccD}{\mathpzc D}
\newcommand{\cE}{\mathcal{E}}
\newcommand{\cF}{\mathcal{F}}
\newcommand{\cG}{\mathcal{G}}

\newcommand{\cL}{\mathcal{L}}

\newcommand{\cO}{\mathcal{O}}

\newcommand{\cR}{\mathcal{R}}

\newcommand{\cT}{\mathcal{T}}\newcommand{\ccT}{\mathpzc T}

\newcommand{\ccZ}{\mathpzc Z}

\newcommand{\ccZb}{{\overline \ccZ}}

\DeclareFontFamily{U}{bbold}{}
\DeclareFontShape{U}{bbold}{m}{n}
 {  <-5.5> s*[1.05] bbold5
    <5.5-6.5> s*[1.05] bbold6
    <6.5-7.5> s*[1.05] bbold7
    <7.5-8.5> s*[1.05] bbold8
    <8.5-9.5> s*[1.05] bbold9
    <9.5-11.5> s*[1.05] bbold10
    <11.5-16> s*[1.05] bbold12
    <16-> s*[1.05] bbold17
 }{}

\newcommand{\IR}{\mathbbl{R}}

\newcommand{\ITheta}{\mathbbl{\Theta}}
\newcommand{\ITh}{\mathbbl{\Theta}}
\newcommand{\IThb}{\ol{\mathbbl{\Theta}}}

\font\eightrm=cmr8 at 8pt
\font\csc=cmcsc10

\newcommand{\beq}{\begin{equation}}
\newcommand{\eeq}{\end{equation}}
\newcommand{\beqnn}{\begin{equation*}}
\newcommand{\eeqnn}{\end{equation*}}
\newcommand{\bea}{\begin{eqnarray}}
\newcommand{\eea}{\end{eqnarray}}
\newcommand{\bean}{\begin{eqnarray*}}
\newcommand{\eean}{\end{eqnarray*}}

\newcommand{\sref}[1]{\SS\ref{#1}}

\newcommand{\nn}{\nonumber}

\newcommand{\norm}[1]{\left\| #1\right\|}
\newcommand{\ee}{\text{e}}
\newcommand{\ii}{\text{i}}

\newcommand{\place}[3]{\vbox to0pt{\kern-\parskip\kern-7pt
                             \kern-#2truein\hbox{\kern#1truein #3}
                             \vss}\nointerlineskip}

\DeclareFontFamily{U}{wncy}{}
\DeclareFontShape{U}{wncy}{m}{n}{<->wncyr10}{}
\DeclareSymbolFont{mcy}{U}{wncy}{m}{n}
\DeclareMathSymbol{\sha}{\mathord}{mcy}{"58}

\newcommand{\capt}[3]{\parbox{#1}{\renewcommand{\baselinestretch}{1.0}
                                                           \caption{\label{#2}\small\it #3}}}

\newcommand{\del}{{\partial}}
\newcommand{\delb}{{\overline{\partial}}}

\newcommand{\lb}{{\overline\lambda}}
\newcommand{\nb}{{\overline\n}}
\newcommand{\mb}{{\overline\m}}

\renewcommand{\aa}{\mathfrak{a}}

\newcommand{\dd}{{\text{d}}}

\newcommand{\K}{K\"ahler\xspace}

\renewcommand{\H}{\text{H}}

\newcommand{\vol}{\dd^6 x \sqrt{g}\, }
\newcommand{\tr}{\text{tr}\hskip2pt}
\newcommand{\Tr}{\text{Tr}\hskip2pt}

\newcommand{\tb}{{\overline{\tau}}}

\newcommand{\Sb}{{\overline{\mathcal{S}}\,}}
\renewcommand{\S}{{{\mathcal{S}}\,}}

\newcommand{\ap}{{\a^{\backprime}\,}}
\renewcommand{\sb}{{\overline{\sigma}}}

\renewcommand{\rb}{{\overline{\rho}}}
\renewcommand{\=}{\;=\;}

\makeatletter
\g@addto@macro\bfseries{\boldmath}
\makeatother

\newcommand{\citeM}{\cite{Candelas:2016usb}\xspace}

\newcommand{\citeSG}{\cite{McOrist:2019mxh}\xspace}

\newcommand{\citeMMS}{\cite{Melnikov:2014ywa}\xspace}

\newcommand{\citeBdR}{\cite{Bergshoeff:1989de}\xspace}
\newcommand{\citeH}{\cite{Hull:1986kz}\xspace}
\newcommand{\citeS}{\cite{Strominger:1986uh}\xspace}

\newcommand{\citeSPURs}{\cite{McOrist:2021dnd,Chisamanga:2024xbm,McOrist:2024ivz,McOrist:2024zdz,McOrist:2025sdy} \xspace}
\newtheorem{prop}{Proposition}

\renewcommand{\baselinestretch}{1.1}
\numberwithin{equation}{section}
\proofmodefalse
\begin{document}
\pagestyle{empty}      
\ifproofmode\underline{\underline{\Large Working notes. Not for circulation.}}\else{}\fi

\begin{center}
\null\vskip0.2in
{\Huge Stringy Corrections to Heterotic SU(3)-Geometry \\[0.5in]}

{\csc Jock McOrist$^{*\,1}$ and
Sebastien Picard$^{\dagger \,2}$\\[0.2in]}

{\it $^*$Department of Mathematics\hphantom{$^2$}\\
School of Science and Technology\\
University of New England\\
Armidale, 2351, Australia\\[3ex]

$^\dagger$ Department of Mathematics\\
University of British Columbia\\
 1984 Mathematics Road\\
  Vancouver, BC, Canada\\
}

\footnotetext[1]{{\tt  jmcorist@une.edu.au} \hfil
\hspace*{6.21cm}$^2\,${\tt spicard@math.ubc.ca}}
%
\vspace{0.4cm}

\vfill
{\bf Abstract\\[-12pt]}
\end{center}

We analyse the $\ap^2$ corrections to the supersymmetry transformations constructed by Bergshoeff--de Roo for heterotic compactifications on $\SU(3)$ manifolds. The internal geometry remains complex and conformally balanced.  The graviton equation of motion receives an explicit $\ap^2$ correction because the tangent-bundle connection appearing in the action is the composite Hull connection. We show that supersymmetry and the Bianchi identity imply all the equations of motion, including this correction. No instanton condition is required.

\vskip90pt
\newgeometry{left=1.9in, right=0.5in, top=0.75in, bottom=0.8in}
%
\newpage

{\renewcommand{\baselinestretch}{1.4}\tableofcontents}
\restoregeometry
\setcounter{page}{1}
\pagestyle{plain}
\renewcommand{\baselinestretch}{1.3}
\null\vskip-10pt

\section{Introduction}

Non-\K geometry naturally arises in the study of the heterotic string on $\IR^{3,1}\times X$. In modern notation, the action for this theory is~\citeBdR~\footnote{In \citeBdR there are two coupling constants for the $F$ and $R^\H$ curvatures, which from supergravity perspective can be different. In string theory the Green-Schwarz anomaly cancellation fixes them in terms of $\ap$.}
\beq\label{eq:BdRaction}
\begin{split}
 S \= \frac{1}{2} \int \dd^{10} x \sqrt{g} \, e^{-2\Phi} \left\{ R - \frac{1}{2} |H|^2 + 4 (\del_m\Phi)^2 + \frac{\ap}{4} \tr |F|^2 -\right. &\left.\frac{\ap}{4} \tr |R^\H|^2   + \cdots\right\} \\
 & + \cO(\ap^3) ~,
\end{split}
\eeq
The ellipsis denotes fermionic terms. Here $g$ is the ten-dimensional metric, $\Phi$ the dilaton, $F$ the curvature of a connection $A$ on an $E_8\times E_8$ or $SO(32)$ gauge bundle $E$, and $H$ a three-form satisfying the Bianchi identity
\beq\label{eq:Bianchi2}
\dd H \= \frac{\ap}{4} \tr F\w F - \frac{\ap}{4} \tr R^\H \w R^\H~.
\eeq
Here $R^\H$ is the curvature of the composite Hull connection $\Th^\H=\Th^\LC+\half H$, constructed from $(g,H)$, and $R$ is the ten-dimensional Ricci scalar. We suppress Newton's constant. In the BdR field basis, \eqref{eq:Bianchi2} is exact and receives no further explicit $\ap$ corrections; its form is fixed by the Green--Schwarz anomaly-cancellation mechanism \cite{Green:1984sg}. Although no separate bosonic $\cO(\ap^2)$ term appears explicitly in \eqref{eq:BdRaction}, the composite quantities $H$ and $\Th^\H$ generate definite $\cO(\ap^2)$ contributions when expanded and varied.

Green--Schwarz anomaly cancellation fixes the characteristic class entering the Bianchi identity, but not the connection used to evaluate it. Once the $E_8\times E_8$ and $SO(32)$ heterotic spectra were known \cite{Green:1984sg}, the remaining problem was to construct a local, supersymmetric and ghost-free action order by order in $\ap$ realising this spectrum and anomaly cancellation. Building on Chapline--Manton \cite{Chapline:1982ww}, Bergshoeff--de Roo showed that the Hull connection $\Th^\H$ and its fermionic partner, the supercovariant gravitino curvature, transform like a  Yang--Mills connection and gaugino respectively \cite{Bergshoeff:1988nn,Bergshoeff:1989de}. The pair must be composite: an independent, non-compact $SO(9,1)$ multiplet would contain wrong-sign kinetic term, making the energy unbounded below. The BdR construction determines the curvature-squared terms and their supersymmetric completion through $\cO(\ap^2)$. Thus Green--Schwarz fixes the topological condition, while BdR provides its local supersymmetric realisation.

The BdR construction has two independent checks. The action agrees with string amplitudes through $\cO(\ap)$ \cite{Cai:1986sa,Gross:1986mw}, with partial checks at $\cO(\ap^2)$ \cite{Metsaev:1986yb}. Up to field redefinitions, world-sheet conformal invariance gives the same target-space equations through $\cO(\ap)$ \cite{Hull:1986xn,Ross:1986ra,Hull:1987pc,Metsaev:1987zx}, with agreement extending to $\cO(\ap^2)$ \cite{Foakes:1988wy,Foakes:1987bn,Ellwanger:1988cc,Ellwanger:1987zx}. Together, these independent, non-trivial results support the BdR action \cite{Chemissany:2007he}.

Our starting point is the BdR supersymmetry transformations \citeBdR, which leave the full boson--fermion action invariant through order $\ap^2$; uncancelled variations begin at $\ap^3$. The theory is therefore defined order by order in $\ap$, rather than by a finite truncation. We apply these transformations through order $\ap^2$  to $\IR^{3,1}\times X$ and derive the resulting constraints on the compact six-manifold $X$. We restrict to backgrounds with a smooth, non-degenerate $\ap\to0$ limit, so that the usual perturbative sigma-model description applies.\footnote{This excludes the torsional $T^2\to K3$ examples whose metric degenerates as $\ap\to0$; they lie outside the perturbative regime considered here.} We show that the supersymmetry equations imply $H=\cO(\ap)$ and $\dd\Phi=\cO(\ap)$, so the zeroth-order geometry is Calabi--Yau. The corrected metric need not be \K: higher-order terms generate torsion, and its fundamental form is no longer closed. The first-order non-\K corrections were studied by Hull \citeH and Strominger \citeS.

The BdR supersymmetry equations require $X$ to admit a globally defined nowhere-vanishing spinor $\eta$ satisfying $\nabla^\Bi\eta=0$, where $\nabla^\Bi$ has torsion $T$. The holonomy of $\nabla^\Bi$ is therefore contained in $SU(3)$, and the spinor bilinears define a global three-form $\O$ and an almost complex structure $J$ satisfying $\nabla^\Bi J=0$. This covariant constancy does not by itself imply that $J$ is integrable. In \citeS, it is shown that the torsion satisfies
$$
T \= \ii (\del - \delb)\o + N~,
$$
with $N$ the Nijenhuis tensor for $J$. The analysis of \citeS holds to first order in $\ap$, where setting the gravitino variation to zero implies $T=H$, setting the dilatino variation to zero implies $H^{0,3}=0$, and combining these equations yields $N=0$. To second order in $\ap$, the calculation is more involved as the dilatino variation implies $H^{0,3} + 3\ap P^{0,3} = 0$ while the gravitino variation implies
\beq\label{eq:HPrelationa}
H\=T + 2\ap P~,\qquad{\rm where} \qquad P_{mab}=\frac14e^{2\Phi}(\nabla^-)^p(e^{-2\Phi}\dd H)_{pmab}  ~.
\eeq
Here $\nabla^-$ is the metric connection with torsion opposite to that of $\nabla^\H$; the connections are summarised at the end of Section~1. Nevertheless, we show that the Nijenhuis tensor still vanishes.  

With the manifold complex through order $\ap^2$, define
\beq\label{eq:hcE}
 h\=\tr|F|^2-\tr|R^\H|^2~,\qquad \cE^\H_{mn}\=\tr(F_{mp}F_n{}^p)-\tr(R^\H_{mp}R^{\H}{}_n{}^p)~.
\eeq
The corrected geometry is most naturally expressed in the simultaneously redefined metric and dilaton
\beq\label{eq:omega-field-redefinition}
 \widehat g_{\m\nb}=g_{\m\nb}+\frac{\ap^2}{4}\cE^\H_{\m\nb}~,\qquad \widehat\o=\o+\frac{\ii\ap^2}{4}\cE^\H_{\m\nb}\dd x^{\m\nb}~,\qquad \widehat\Phi=\Phi+\frac{3\ap^2}{32}h~.
\eeq
Here $\o$ is the fundamental form of the corrected $SU(3)$ structure determined by the Killing spinor, while $\widehat\o$ is the fundamental form of $\widehat g$ with respect to the same complex structure. 
Thus both forms contain $\ap$ corrections: the hat labels the field basis rather than distinguishing corrected from uncorrected geometry.

The complex manifold $X$ admits a closed  holomorphic volume form $\Omega$. We prove that in these variables the flux--torsion, holomorphic-volume and conformally balanced relations take the form
\beq\label{eq:introconf}
 H\=\ii(\del-\delb)\widehat\o+\cO(\ap^3)~,\qquad \log\norm{\Omega}_{\widehat g}=-2\widehat\Phi+\cO(\ap^3)~,\qquad \dd\left(e^{-2\widehat\Phi}\widehat\o^2\right)=\cO(\ap^3)~.
\eeq

The relations in \eqref{eq:introconf} have the same explicit form as the first-order Strominger relations only after the curvature-dependent field redefinition \eqref{eq:omega-field-redefinition}. This does not remove the $\ap^2$ corrections; it packages them into the hatted fields.

The  Green--Schwarz Bianchi identity remains \eqref{eq:Bianchi2}, with $R^\H$ the curvature of the  Hull connection. Combining that identity with the flux--torsion relation and the Chern--Hull transgression gives the resolved equation
\beq\label{surprising-bianchi}
 2\ii\delb\del\widehat\o
 =\frac{\ap}{4}\left(\tr F\w F-\tr R^\Ch\w R^\Ch\right)
 +\cO(\ap^3)~.
\eeq
Thus the appearance of $R^\Ch$ in \eqref{surprising-bianchi} should not be read as replacing $R^\H$ in either the action or the physical Bianchi identity. It is a derived reformulation valid through order $\ap^2$. Whether an analogous reformulation persists at higher orders remains open. A geometric interpretation of this equation with Chern connection was given in \cite{McOrist:2024zdz} with follow-up work by \cite{2409.04382}: the equation defines an elliptic complex on sections of an auxiliary vector bundle $Q \rightarrow X$ and this result is surveyed in Section \ref{sec:ddbar-ap2}.

Our central  observation is that the composite nature of the Hull connection produces a genuine $\ap^2$   correction to the graviton equation of motion, even when written in the hatted basis. Varying $g$ or $B$ also varies $\Th^\H$ inside $|H|^2$ and $\tr|R^\H|^2$ producing an additional chain-rule contribution. The BdR lemma \cite{Bergshoeff:1989de} shows that this contribution is proportional to $\ap$ times the zeroth-order equations of motion. It may therefore be omitted through order $\ap$,  but when evaluated on the first-order-corrected fields it gives the non-zero $\ap^2$ term $\Delta^{\rm conn}_{mn}$.

At first order in $\ap$, the Hull curvature satisfies the instanton condition:
\beq
 \ap R^\H_{mnpq}\Gamma^{mn}\eta=\cO(\ap^2)~.
\eeq
 At the next order, however, supersymmetry relates both its $(0,2)$ component and its hermitian trace to $\dd H$; see \eqref{eq:RhdH}. Thus $R^\H$ is not an instanton. In fact, this non-instanton property is crucial to the gravitino integrability calculation; it contributes a term that is central to reproducing the $\ap^2$--graviton equation of motion. 
 
Hence, imposing the instanton condition as an additional equation is neither required nor innocuous. On a compact background with a smooth $\ap\to0$ limit, \eqref{eq:RhdH}, the trace relation and the Lefschetz isomorphism would force
\beq
 \dd H=\cO(\ap^3)~,\qquad H=\cO(\ap^3)~.
\eeq
The auxiliary instanton condition therefore eliminates torsion through $\ap^2$. Even in the standard embedding, where $H=0$ and the leading-order background is Calabi--Yau, the four-loop beta function generates an $\ap^3$ correction for which the tangent-bundle connection no longer satisfies the instanton condition \cite{Grisaru:1986dk,Gross:1986iv}. The instanton condition should therefore be regarded as a leading-order statement, rather than a property of the full string theory.

In \sref{s:eomgeometry}, we reformulate our results in a manner more accessible to a mathematical audience. We begin by assuming that the supersymmetry conditions—expressed as tensorial equations—hold identically. Under this assumption, we demonstrate that the equations of motion can be recovered purely from these tensor conditions using standard results from Hermitian geometry. This provides an independent consistency check of our analysis, and also a foundation for further mathematical exploration of the system. 

The distinguished role of the Hull connection also appears in heterotic--type IIA duality, off-shell six-dimensional supergravity and generalised-geometry constructions \cite{Liu:2013dna,Novak:2017wqc,Coimbra:2014qaa}. These independent results further support the role of $\Th^\H$ in organising higher-derivative supergravity.

The paper is organised as follows. Section~2 reviews the BdR action and supersymmetry transformations through $\cO(\ap^2)$ and specialises them to an $SU(3)$ structure. Section~3 derives the corrected geometry of $X$. Section~4 obtains the equations of motion from supersymmetry integrability, while Section~5 gives an independent derivation using Hermitian geometry. Section~6 concludes with comments on heterotic geometric flows.


 \subsection{Equations of motion} \label{sec:eom}
In the remaining sections, to avoid notational clutter, we suppress $\cO(\ap^3)$ unless possible confusion may arise and understand equality signs are to this order in the $\ap$ expansion. For example, $\nabla^\Bi_m \ve + \cO(\ap^3) = 0$ is written as $\nabla_m^\Bi \ve = 0$.  

We state the $d=10$ equations of motion and supersymmetry variations that leave \eqref{eq:BdRaction} invariant. Let $B_{mn}$ be the local 2-form potential satisfying $H = \dd B + \frac{\ap}{4} (\CS(A)-\CS(\Th^\H))$. The equations of motion, listed in the following order, come from \eqref{eq:BdRaction} by varying the fields $\d g_{mn}, \d B_{mn}, \d A, \d \Phi$. The result is:
\begin{align}
&&\Ric_{mn}+2\nabla_m\nabla_n\Phi-\frac14H_{mpq}H_n{}^{pq} +\frac{\ap}{4}\cE^\H_{mn}+\Delta^{\rm conn}_{mn}\=0~,\label{eq:RicEOM}\\[6pt]
&&\nabla^\LC{}^m(\ee^{-2\Phi} H_{mnp}) \=0~, \label{eq:HEOM}\\[8pt]
&&\ap \ccD^{-\,m} (\ee^{-2\Phi} F_{mn}) \=0~,\label{eq:FEOM} \\[3pt]
&&  R - 4(\nabla \Phi)^2 + 4 \nabla^2 \Phi - \half |H|^2 + \frac{\ap}{4} \big(\tr |F|^2 - \tr |R^\H|^2\big)\= 0~.\label{eq:DilatonEOM}
\end{align}
where $\cE$ is defined in \eqref{eq:hcE} and we have simplified the graviton equation of motion using the dilaton and $H$-equations of motion. At order $\ap$, these equations of motion are standard. At order $\ap^2$ however, there is one new contribution to the graviton equation:
\beq\label{eq:Delta-connection_b}
\begin{split}
 \Delta^{\rm conn}_{mn}  &\= -\frac{\ap}{4}e^{2\Phi} \nabla^p \Bigl( e^{-2\Phi}\left(
 \nabla_m\cG_{np}  +\nabla_n\cG_{mp}  -2\nabla_p\cG_{mn}  \right) \Bigr) +\cO(\ap^3)~.\\
\end{split}
\eeq
This is the chain-rule contribution from the Hull connection to the equations of motion provided by a lemma in BdR, see equations (3.1)-(3.3) of ~\citeBdR.  Let us briefly explain its origin. The lemma decomposes a field variation, say a metric variation $\d g$, into an explicit variation with $\Theta^\H$ held fixed together with a term from $\Th^\H$ being composite. The former gives the first four terms in \eqref{eq:RicEOM}. The latter arises from the occurrence of $\Theta^\H$ in $|H|^2$ and in $\tr |R^\H|^2$:
\beq
 \left.\d S\right|_{\d\Theta^\H} \=\! -\frac{\ap}{4}\int\dd^{10}x\,\sqrt g\,e^{-2\Phi}\, \d\Theta^\H_{mab}
 \left( \nabla^a\cG^{mb}-\nabla^b\cG^{ma}+2P^{mab} \right) +\cO(\ap^3)~,
\eeq
where 
$$
\delta\Theta^\H_{mab}\=\nabla_{[b}\left(\delta g_{m|a]}+\d B_{m|a]}\right)+\cO(\ap)~, \quad \cG_{mn}\=\Ric_{mn}+2\nabla_m\nabla_n\Phi-\frac14H_{mpq}H_n{}^{pq}~.
$$
We remark that this contribution can be omitted when working to first order in $\ap$. It is relevant at $\ap^2$ for the graviton equation of motion, but not the B-field equation of motion. The first-order graviton equation of motion gives
\beq\label{eq:cGcE}
\cG_{mn}\=\!-\frac{\ap}{4}\cE^\H_{mn}+\cO(\ap^2)~.
\eeq
Putting \eqref{eq:cGcE} into \eqref{eq:Delta-connection_b} gives
\beq\label{eq:Delta-connection}
\begin{split}
 \Delta^{\rm conn}_{mn}  &\= \frac{\ap^2}{16}e^{2\Phi} \nabla^p \Bigl( e^{-2\Phi}\left(
 \nabla_m\cE_{np}  +\nabla_n\cE_{mp}  -2\nabla_p\cE_{mn}  \right) \Bigr) +\cO(\ap^3)~.\\
\end{split}
\eeq
To date this $\ap^2$--correction appears to have been missed and it will be crucial for our analysis.

We add a few more remarks on the conventions in \eqref{eq:RicEOM}. The indices $m,n$ are real coordinates in $d=10$, which will later be specialised to the compact manifold $X$. The minus connection $\ccD^-$ acts as $\Theta^- = \Theta^\LC - \half H$ on tangent bundle indices. The Laplacian $\nabla^2$, Ricci curvature $\Ric$ and scalar curvature $R$ are with respect to the Levi-Civita connection $\nabla^\LC$. This is the field basis of \citeBdR. The components of a $p$-form $Q$ are defined as
\beq\label{eq:formcomp}
 Q \= \frac{1}{p!} Q_{m_1\cdots m_p} \dd x^{m_1} \w \cdots \w \dd x^{m_p}~, \quad |Q|^2 = \frac{1}{p!} Q_{m_1\cdots m_p} Q^{m_1\cdots m_p}~,
\eeq
and we will now often omit the wedge `$\w$' for compactness. 

Next, let $S=S^+ \oplus S^-$ be a spinor bundle over the 10-dimensional spacetime so that we may add fermionic superpartners to the theory. To the bosonic action \eqref{eq:BdRaction}, we add the following fermionic fields: the gravitino $\psi \in \Gamma(S^+ \otimes T^*)$, the dilatino $\lambda \in \Gamma(S^- )$ and the gaugino $\chi \in \Gamma(S^+ \otimes {\rm End} \, E)$. We will omit the explicit form of the inclusion of the fermions $(\psi,\lambda,\chi)$ into the action \eqref{eq:BdRaction}, but this can be found in \citeBdR. The full action is such that it is invariant under the supersymmetry transformations, which upon setting fermions to zero take the form:
\beq
\begin{split} \label{eq:fermi}
 \d \psi_k &\= \nabla^{\LC}{}_k \ve - \frac{1}{8} H_k{}^{mn} \gamma_{mn} \ve + \frac{\ap}{4} P_k{}^{mn} \gamma_{mn} \ve  ~,\\
\d \lambda &\= \bigg( \delslash \Phi - \half  \Hslash + \ap \frac{3}{2} \Pslash \bigg) \ve ~,\\
\d \chi &\= -\half \Fslash \ve +\cO(\ap^2)~.
\end{split}
\eeq
There are also supersymmetry transformations for the bosonic fields $(\d g, \d \Phi, \d A, \d B)$ which we list to leading order below in \eqref{eq:susy1}. Here $\ve \in \Gamma(S^+)$ is a non-vanishing spinor generating the local supersymmetry algebra, and $P = - \frac{1}{4} e^{2 \Phi} \nabla^-{}^\dagger (e^{-2 \Phi} dH)$. We are using the slash convention $\Qslash = \frac{1}{p!} Q_{m_1\cdots m_p} \g^{m_1\cdots m_p}$, where $Q$ is a $p$-form and $\g^m$ are the gamma matrices.

A background is supersymmetric if $\delta (\psi,\lambda,\chi) = 0$. As usual, the background is bosonic and so to check supersymmetry we require that it remains bosonic. Hence, we just need to check the fermionic variations vanish. We give a more detailed summary of this result in the next section. 

\begin{table}[H]
\begin{center}
\setlength{\extrarowheight}{3pt}
\begin{tabular}{|m{4cm}|c|c|c|}
\hline
\hfil ~~Connection~~ &~~Symbol~~ & ~~Appearance~~& ~~Properties~~\\[3pt]
\hline\hline
~Levi-Civita  & \hfill $\Th^\LC$ \hfill ~ &  $g$-EOM & Torsion-free, $\nabla^\LC J \neq 0$ \\[3pt]
\hline
~Hull   &  $\Th^\H = \Th^\LC + \half H$ &   $g$-EOM, Bianchi & $\nabla^\H J \neq 0$ \\[3pt]
  \hline
 ~Minus ~ & $\Th^- =  \Th^\LC - \half H$& $F$-EOM, $P$ & $\nabla^- J \neq 0$ \\[3pt]
\hline
~Bismut   & $\Th^\Bi= \Th^\LC - \half T$ & Gravitino variation & $\nabla^\Bi J = 0$ \\[3pt]
\hline
~Chern  ~ & $\Th^\Ch$ & Complex geometry & $\nabla^\Ch_{\bar{\mu}} = \bar{\partial}_{\bar{\mu}}$, $\nabla^\Ch J =0$ \\[3pt]
\hline
\end{tabular}
\capt{6.0in}{tab:connections}{The connections that appear in heterotic compactifications on $SU(3)$ manifolds. }
\end{center}
\end{table}

 \section{Supersymmetry algebra and action for \texorpdfstring{$\SU(3)$}{SU3} } 
 \label{sec:susyalg}

 \subsection{Bergshoeff--de Roo in  \texorpdfstring{$d=10$}{10d}} \label{sec:BdR}
We give a brief summary of the supersymmetry algebra and action constructed in BdR \cite{Bergshoeff:1989de}. Notation is translated according to \eqref{eq:BdRdict} and we refer the reader to the reference \citeBdR for a more in-depth analysis. 

In \citeBdR, building on \cite{Bergshoeff:1988nn}, Bergshoeff and de Roo (BdR) use the correspondence between the Yang--Mills and gravitational sectors suggested by Green--Schwarz anomaly cancellation to construct the locally supersymmetric action containing $F^2$ and $R^2$ terms. At zeroth order in $\ap$, the supersymmetry commutator identifies the relevant Lorentz connection in their field basis as the Hull connection $\Theta^\H$. Its supersymmetry variation pairs it with a composite fermion $\psi^{ab}$, and the two transform respectively like an $\SO(9,1)$ Yang--Mills connection and gaugino. Neither is an independent field; both are built from the supergravity fields. At higher orders, varying $g$, $B$ and $A$ induces corrections to their transformation laws, so the Yang--Mills correspondence is not exact beyond zeroth order. BdR compute these corrections and construct the action and supersymmetry transformations iteratively in $\ap$, as we now briefly review.\footnote{\citeBdR also constructs the part of the quartic action obtained by supersymmetrising the Yang--Mills and Lorentz Chern--Simons forms. It does not include the separate $R^4$ invariant with no Yang--Mills counterpart.}

To zeroth order in $\ap$, the supersymmetry rules for coupled $d=10$ and YM fields are
\beq\label{eq:susy1}
\begin{split}
 \d_0 e_m^a \= \half \bar \ve \G^a \psi_m~, \qquad \qquad & \d_0 \psi_m \= \left(\del_m + \frac{1}{4}\Th^-_{mab}\G^{ab}\right)\ve~,\\
  \d_0 B_{mn} \=\bar\ve \G_{[m}\psi_{n]}~, \quad \qquad& \d_0 \l \= -\frac{1}{2\sqrt{2}}\left\{\delslash\Phi-\half\Hslash\right\}\ve~,\\
\d_0 \Phi \= -\frac{1}{\sqrt{2}} \bar\ve \l~,  \qquad \qquad& \d_0 A_m \= \bar\ve \G_m \chi~, \qquad \d_0 \chi \= - \frac{1}{4} F_{mn} \G^{mn}  \ve~,
\end{split}
\eeq
where we have not written terms cubic in fermions. Here $\ve$ is a Majorana--Weyl $d=10$ spinor and $e_m{}^a$ is the vielbein for the $d=10$ metric. Since the background is bosonic, supersymmetry requires the gravitino, dilatino and gaugino variations to vanish: $\d_0\psi_m=\d_0\l=\d_0\chi=0$.

The Chapline-Manton coupling of Yang-Mills to gravity \cite{Chapline:1982ww} requires an additional transformation for the $B$-field at first order in $\ap$---in the original paper the coupling is $g_{YM}^{-2}$ as it predates Green Schwarz anomaly cancellation. This transformation law reads
\beq\label{eq:susy2}
\d_\ap B_{mn} \= \frac{\ap}{2} \tr A_{[m} \d_0 A_{n]}  - \frac{\ap}{2} \tr \Th^\H_{[m} \d_0 \Th^\H_{n]}~,
\eeq
and follows the guiding principle that there should be a symmetry between the $A$ terms and the $\Theta$ terms. The total supersymmetry variation is the sum of these two contributions, so for example $\d B \= \d_0 B + \d_\ap B$. 
 
The transformations \eqref{eq:susy1}-\eqref{eq:susy2} leave the action $\cL(R) + \cL(F^2) + \cL(R^2)$ invariant through $\ap$, with a remainder beginning at $\ap^2$, where
\bea
\cL(R) &\=&  \half e^{-2\Phi}  \left\{ R - \frac{1}{12} H_{mnp} H^{mnp} + 4 (\del_m\Phi)^2 \right\} + {\rm quadratic~fermions} ~,\\
\cL(F^2) &\=& \frac{\ap}{4}   \tr |F|^2 + {\rm quadratic~fermions} ~,\\
\cL(R^2) &\=&\!- \frac{\ap}{4}   \tr |R^\H|^2 + {\rm quadratic~fermions} ~.
\eea
A key feature  is that  $\Theta^\H$ and the fermion $\psi^{ab}$ transform respectively like an $\SO(9,1)$ Yang--Mills connection and gaugino to zeroth order in $\ap$: \footnote{Other field bases are possible \cite{Bergshoeff:1988nn}, but obscure this structure. A basis adapted to the Chern connection, for example, requires simultaneous redefinitions of $g$ and $B$, and the resulting metric variable no longer has the usual tensorial transformation law \citeMMS.} 
\beq\label{eq:susy3} 
\d_0\Theta^\H_m{}^{ab}
=-\half\bar\ve\Gamma_m\psi^{ab}~,\qquad
\d_0\psi^{ab}
=\frac14\Gamma^{mn}\ve R_{mn}{}^{ab}(\Theta^\H)~.
\eeq
Here $\psi^{ab}$ is a composite fermion built out of the fundamental bosons and fermions  and plays the role of the gaugino. Since $H$ contains the Lorentz Chern--Simons form built from $\Theta^\H$, this identification is recursive in $\ap$ and determines higher-order corrections iteratively.
  
The terms $\cL(R)+\cL(F^2)+\cL(R^2)$ formally resemble $G\times\SO(9,1)$ Yang--Mills coupled to supergravity. The Lorentz sector is not independent, however: $\cL(R^2)$ is gravitational, and $\Theta^\H$ is determined by $(g,H)$, with $H$ depending recursively on $B$, $A$ and $\Theta^\H$. This dependence induces corrections to the transformations of $\Theta^\H$ and $\psi^{ab}$ at order $\ap$:
\beq\label{eq:susy4}
\d_\ap \Th^\H_m{}^{ab} \= \frac{3}{2} \G_{[m} X_{ab]}~, \qquad \qquad \d_\ap \psi^{ab} \= \frac{3}{4} \G_{cd} \ve (\dd H)^{abcd}
\eeq
where $X_{ab} =\frac{\ap}{4} \tr (F_{ab} \chi) - \frac{\ap}{4} \tr ( R_{ab}(\Th^\H) \psi)$. These are the first corrections to the zeroth-order Yang--Mills--Lorentz correspondence.

On an $SU(3)$ manifold, the background being supersymmetric means $(\d_0 + \d_\ap )\psi^{ab} =0$ and this gives 
\beq\label{eq:RhdH}
\left(R^\H_{\m\nb}{}^{\a\bb}- \frac{1}{2} (\dd H)_{\m\nb}{}^{\a\bb}\right)g^{\m\nb} + \cO(\ap^2) \=0~, \qquad R^\H_{\mb\nb}- \frac{1}{2} (\dd H)_{\mb\nb} + \cO(\ap^2) \= 0~.
\eeq
Using conformally balanced complex geometry, \eqref{eq:RH-inst} independently verifies \eqref{eq:RhdH}: the condition $(\d_0+\d_\ap)\psi^{ab}=0$ follows from the gravitino, dilatino and gaugino variations and is not independent. Since $\Theta^\H(g,H)$ is composite, it has no separate tangent-bundle equation of motion or independent degrees of freedom \citeSPURs; its metric variation instead produces $\Delta^{\rm conn}_{mn}$ in \eqref{eq:Delta-connection}.

We see from \eqref{eq:RhdH} that the Hermitian-Yang-Mills condition does not hold for the curvature on $TM$. Let us compare with what is known for the standard embedding; this is the case when $A$ is taken to be $\Theta$, which implies $\dd H = 0$ and, when compactified to six dimensions, implies that $H=0$, $\Phi=const$ and the compact space is Calabi-Yau. It is known in string theory that explicit corrections at the standard embedding through $\ap^3$ lead to violations of the Hermitian-Yang–Mills (HYM) condition for the connection on $TM $ \cite{Grisaru:1986dk,Gross:1986iv,Melnikov:2014ywa}. Away from the standard embedding, \eqref{eq:RhdH} already indicates this phenomenon occurs earlier on in the $\ap$-expansion. 

The supersymmetry transformations retained through first order in $\ap$ leave the action invariant through the corresponding perturbative order, but generate a remainder through $\ap^2$. There is therefore no exact truncation of the action and supersymmetry transformations at first order in $\ap$.

The  lemma in \cite{Bergshoeff:1989de} concerns the variation of the action with respect to the
composite pair $\Th^\H$ and $\psi^{ab}$. On a bosonic background, the variation induced
by $\delta\Theta^\H$ is proportional to the zeroth order in $\ap$ supergravity equations
of motion and to $P\sim(\nabla^-)^\dag \dd H$. The lemma carries an explicit factor of $\ap$. Consequently its contribution may be omitted when deriving the field equations through first order in $\ap$. At the next order, the contribution is non-trivial. Using  the  chain-rule together with  \eqref{eq:cGcE}   produces the nonzero term labelled  $\Delta^{\rm conn}_{mn}$ through $\ap^2$. In the supersymmetry construction, the lemma also determines the corrections to the transformation rules needed to cancel the order-$\ap^2$ remainder.

When applied to the variation \eqref{eq:susy4}, the resulting $\ap^2$ terms fall into two categories: those proportional to the equations of motion and those involving the Bianchi identity. The former can be cancelled by introducing appropriate $\ap^2$ corrections to the supersymmetry transformations:
\beq\label{eq:susy5}
\begin{split}
 \d_{\ap^2} \Psi_m &\=  \frac{\ap}{4} P_{mab}\G^{ab} \ve~,\quad P_{mab}\= \frac{1}{4} e^{2\Phi} \nabla^{-\,p} (e^{-2\Phi} (\dd H)_{pmab})~,\\
 \d_{\ap^2} \l &\= -\frac{3\ap}{4\sqrt{2}} \Pslash ~, \\
 \d_{\ap^2} \chi &\=\! 0~.
 \end{split}
\eeq
together with $\d_{\ap^2} e_m{}^a, \d_{\ap^2} B_{mn}, \d_{\ap^2} \Phi$ which we have not written here but are in \citeBdR. There are also additional $\ap^2$ terms added to the action which are quadratic in fermions; we do not write these as we only ever need the bosonic terms. 

If we evaluate the supersymmetry transformations \eqref{eq:susy1}, \eqref{eq:susy2}, and \eqref{eq:susy5} the action is invariant through $\ap^2$ with a  $\ap^3$ remainder. BdR consider how to modify the action and supersymmetry transformations at $\ap^3$, but we do not discuss this here except to reiterate that the process does not truncate and one needs an order-by-order analysis.

Just as \eqref{eq:susy1}-\eqref{eq:susy2} induces the transformations \eqref{eq:susy3}, \eqref{eq:susy4}, the $\ap^2$ corrections \eqref{eq:susy5} induce an $\ap^2$ correction to $\d \Th^\H, \d \psi^{ab}$ which will further modify the extent to which  the connection $\Th^\H$ departs from  being  an instanton.

\subsection{Bergshoeff--de Roo on  \texorpdfstring{$SU(3)$}{SU3}}

We now compactify from the $d=10$ spacetime $\IR^{3,1}\times X$ to the 6-dimensional compact manifold $X$. For this, we assume the metric splits as $g_{10} = ds^2(\IR^{3,1}) + g_X$ where the metric on $\IR^{3,1}$ is the Minkowski metric, and write the generator of supersymmetry as $\ve = \zeta \otimes \eta + c.c.$, where $\eta$ is a normalized positive chirality spinor on $X$ and $\zeta$ is a spinor on Minkowski space $\IR^{3,1}$. We assume that $(H,A,\Phi)$ have no components along $\IR^{3,1}$. In this setup, we may treat the fields $(g,H,A,\Phi)$ as defined over the compact manifold $X$. Setting the fermionic variations \eqref{eq:fermi} to zero gives the constraints on $X$ required for a supersymmetric background.
\beq\label{eq:fermionicvariations}
\begin{split}
 \nabla^{\LC}{}_k \eta - \frac{1}{8}  H_k{}^{ab} \gamma_{ab} \eta + \frac{\ap}{4}  P_k{}^{ab} \gamma_{ab} \eta &\= 0~,\\
 \bigg( \delslash \Phi -\half  \Hslash + \ap \frac{3}{2}  \Pslash \bigg) \eta &\= 0~,\\
\frac{\ap}{16}F_{mn}\gamma^{mn}\eta+\cO(\ap^3)=0~,
\end{split}
\eeq
where $m,n,k,a,b$ are real coordinates on the compact 6-manifold $X$ and
\beq\label{eq:Pdef}
P_{mab} \= \frac{1}{4} e^{2 \Phi} (\nabla^-)^q (e^{-2 \Phi} \dd H)_{qmab}~,
\eeq
and
\beq\notag
\nabla^\pm{}_k V^q \= \nabla^{\LC}{}_k V^q \pm \half H_k{}^q{}_p V^p.
\eeq
This is to be supplemented by a Bianchi identity for $H$ given in \eqref{eq:Bianchi2}. Our strategy is to follow the approach of \cite{Candelas:1985en} and later \cite{Hull:1986kz, Strominger:1986uh} in rewriting these equations in terms of conditions on the geometry of the manifold and vector bundle.

\subsubsection{Gravitino variation}
Let $(X,g)$ be a compact six-dimensional Riemannian manifold equipped with a spinor bundle $S \rightarrow X$. Let $H \in \Omega^3(X)$ and $\Phi \in C^\infty(X)$. Suppose $\eta \in \Gamma(S)$ is a nowhere vanishing positive chirality spinor satisfying $\eta^\dagger \eta = 1$. In this section, we study how the gravitino equation affects the geometry of $X$. Setting the gravitino variation to zero gives the following gravitino equation:
\beq \label{parallel-spinor}
\nabla^{\LC}{}_k \eta - \frac{1}{8} H_k{}^{ab} \gamma_{ab} \eta + \frac{\ap}{4}   P_k{}^{ab} \gamma_{ab} \eta \= 0~.
\eeq

As noticed by Candelas-Horowitz-Strominger-Witten \cite{Candelas:1985en} and Strominger \cite{Strominger:1986uh}, the existence of a normalized non-vanishing positive chirality spinor $\eta$ defines an almost-complex structure on $(X,g)$ given by
\beq\notag
J^k{}_\ell \= \ii \eta^\dagger \gamma^k{}_\ell \eta~,
\eeq
and satisfying $g(V,W) = g(JV,JW)$. In \cite{Candelas:1985en} and \cite{Strominger:1986uh}, this $J$ is integrable and so $X$ is a complex manifold. Our first goal is to understand whether $J$ is integrable once the $\ap^2$-correction term in the supersymmetry equation is included. Equation \eqref{parallel-spinor} implies
\beq\notag
\nabla^{\Bi} J \= 0~,
\eeq
where the connection $\nabla^{\Bi}$ acts on vector fields by
\beq \label{corrected-connection}
\nabla^{\Bi}{}_k V^q \= \nabla^{\LC}{}_k V^q - \half  H_k{}^q{}_p V^p + \ap P_k{}^q{}_p V^p, \quad V \in \Gamma(TX)~.
\eeq
The constraint $\nabla^{\Bi} J = 0$ gives a relation between $(X,g,J)$ and the field $H$. The following general mathematical statement is well-known in the literature: see for example \cite{gauduchon1997hermitian,FriedIvanov,Garcia-Fernandez:2015hja}. We give here the full proof to establish conventions.

\begin{prop} \label{uniqueness-bismut}
  Let $(X,g,J)$ be an almost-complex manifold, and let $\omega(V,W) = g(JV,W)$. Suppose $\nabla^{\Bi} J = 0$, where $\nabla^{\Bi}$ is a connection defined by a 3-form $T \in \Omega^3(X)$ in the following way:
\beq \label{bismut-conn}
\nabla^{\Bi}{}_k V^i \= \nabla^{\LC}{}_k V^i - \half  {T}_k{}^i{}_p V^p, \quad V \in \Gamma(TX)~.
\eeq
Then we have the relations
\begin{align}
 {T}_{ijk} &\=   4N_{kij} + (d \o)_{Ji,Jj,Jk} \label{locked-in-1}\\
T &\= \ii (\partial- \bar{\partial})\omega + N \label{locked-in-2}
\end{align}
where $\partial \omega = (d \omega)^{2,1}$ and $\bar{\partial} \omega = (d \omega)^{1,2}$, and $N \in \Omega^2(T^*X)$ is the Nijenhuis tensor.
\end{prop}

\begin{proof}
In components, our conventions are
\[
\omega \= \half  \omega_{ij} \dd x^{ij}, \quad J^i{}_p J^p{}_j \=\!- \delta^i{}_j, \quad \omega_{ij} \= J^p{}_i g_{p j}, \quad \omega_{ij} \= g_{Ji,j}~.
\]
Here, and throughout the proof, we use the following short-hand notation:
\beq \notag
J^k{}_i \beta_k := \beta_{Ji}, \quad \beta \in \Omega^1(X)
\eeq
so that for example $g_{Ji, j} := J^k{}_i g_{kj}$ and $g(J \cdot, J \cdot) = g(\cdot, \cdot)$ reads $g_{Ji,Jj}=g_{ij}$. The Nijenhuis tensor is defined by
\beq\notag
N \= \half  N_{pij} \, \dd x^p \otimes \dd x^{ij}~, \quad N_{pij} \= g_{pk} N^k{}_{ij}~,
\eeq
where
\beq \label{Nijenhuis}
N^p{}_{ij} \= \frac{1}{4}  \bigg( J^k{}_i \nabla^{\LC}{}_k J^p{}_j + J^p{}_k \nabla^{\LC}{}_j J^k{}_i - (i \leftrightarrow j) \bigg)~.
\eeq
\par $\bullet$ The first step is to use $\nabla \omega = 0$ to compute $d \omega$ in terms of $T$. We start from 
\[
\dd \omega \= \half  \nabla^{\LC}{}_k \omega_{ij} \dd x^{kij}, \quad (d \omega) = \frac{1}{3!} (d \omega)_{kij} \, \dd x^{kij}~.
\]
From $\nabla^{\Bi} \omega = 0$, we use the expression \eqref{bismut-conn} to convert $\nabla^{\LC}$ to $\nabla^{\Bi}$ and obtain
\[
\dd \omega \= \half  \bigg( -\half  T_k{}^\ell{}_i \omega_{\ell j} - \half  T_k{}^\ell{}_j \omega_{i \ell} \bigg) \dd x^{kij}~.
\]
We skew-symmetrize this expression and derive
\beq\notag
(\dd \omega)_{kij} \=\!-T_k{}^\ell{}_i \omega_{\ell j} -T_j{}^\ell{}_k \omega_{\ell i} -T_i{}^\ell{}_j \omega_{\ell k}~.
\eeq
Since $\omega_{ij} = - g_{i,Jj}$, this is
\beq\notag
(\dd \omega)_{kij} \=T_k{}^\ell{}_i g_{\ell, Jj} + T_j{}^\ell{}_k g_{\ell, Ji} + T_i{}^\ell{}_j g_{\ell, Jk}~.
\eeq
and so
\beq \label{domega-H}
(\dd \omega)_{kij} \=T_{k, Jj,i} + T_{j,Ji,k} + T_{i,Jk,j}~.
\eeq
\par $\bullet$ The second step is to use $\nabla^{\Bi} J = 0$ to compute $N$ in terms of $T$. We can use the expression \eqref{bismut-conn} to derive the relation
\beq\label{eq:NabLCJ}
\nabla^{\LC}{}_k J^p{}_j  \= \half  T_k{}^p{}_r J^r{}_j - \half  T_k{}^r{}_j J^p{}_r~.
\eeq
Substituting this into \eqref{Nijenhuis} gives
\beq\notag
N^p{}_{ij} \= \frac{1}{8} \bigg( J^k{}_i (T_k{}^p{}_r J^r{}_j - T_k{}^r{}_j J^p{}_r) +  J^p{}_k (T_j{}^k{}_r J^r{}_i - T_j{}^r{}_i J^k{}_r) -  (i \leftrightarrow j) \bigg)~,
\eeq
where $T_j{}^{Jp}{}_j \cong J^p{}_k T_j{}^k{}_j$ and Nijenhuis becomes
\beq\label{eq:Nijreal}
N^p{}_{ij} \= \frac{1}{4}  (T_{Ji}{}^p{}_{Jj} - T_{Ji}{}^{Jp}{}_j + T_j{}^{Jp}{}_{Ji} + T_j{}^p{}_i )~.
\eeq

\par $\bullet$ We now derive \eqref{locked-in-1}. Returning to \eqref{domega-H}, we have
\beq \label{domega-H2}
(\dd \omega)_{Ji,Jj,Jk} \= T_{ Ji, j, Jk} + T_{i,Jj,Jk} + T_{ Ji, Jj, k}~.
\eeq
Lower an index on $N^p{}_{ij}$ in \eqref{eq:Nijreal} and use $g_{kp} T_{Ji}{}^{Jp}{}_j = - T_{Ji,Jk,j}$
\beq\notag
\begin{split}
 N_{kij} &\= \frac{1}{4}   (T_{Ji,k,Jj} + 2 T_{Ji,Jk,j}   + T_{kij} )\\
 &\=\!- \frac{1}{4}   (T_{Ji,Jj,k} +T_{Ji,j,Jk} +  T_{i,Jj,Jk}   - T_{ijk} )\\
 &\=\!- \frac{1}{4}   \big( (\dd \o)_{Ji,Jj,Jk}   - T_{ijk} \big)~.
\end{split}
\eeq
We have used that $N^p{}_{ij} = - N^p{}_{ji}$. The identity \eqref{locked-in-1} is then established.

\par $\bullet$ We now prove \eqref{locked-in-2}. Indeed, let $\{ e_\alpha \}_{\alpha=1}^3$ be a local frame for $\ccT_X^{(1,0)}$, where $\ccT_X^{(1,0)}$ is the $+i$ eigenspace of $J$ so that e.g. $A_{J\alpha} = i A_\alpha$. Evaluating \eqref{domega-H} in this frame gives
\[
(\dd \omega)_{\alpha \beta \gamma} \=\! -3\ii T_{\alpha \beta \gamma}, \quad (\dd \omega)_{\alpha \bar{\beta} \gamma} \=\!-\ii T_{\alpha \bar{\beta} \gamma}~.
\]
Therefore $T_{\alpha \bar{\beta} \gamma} \= (\ii \partial \omega)_{\alpha \bar{\beta} \gamma}$. Evaluating \eqref{eq:Nijreal} in this frame gives 
\begin{equation} \label{Nijenhuis-H}
N_{\alpha \beta \gamma} \= T_{\alpha \beta \gamma}~, \quad N_{\bar{\alpha} \beta \gamma} \= N_{\alpha \bar{\beta} \gamma}\= 0~.
\end{equation}
Combining these identities proves \eqref{locked-in-2}. Lastly, we note that \eqref{locked-in-1} and \eqref{locked-in-2} are consistent since the above equations imply $(d \omega)_{\alpha \beta \gamma} = -3i N_{\alpha \beta \gamma}$.

\end{proof}

\subsubsection*{Supersymmetry at second order in  \texorpdfstring{$\ap$}{ap2}}
We will apply Proposition \ref{uniqueness-bismut} to the connection \eqref{corrected-connection} that appears in the gravitino equation. This amounts to replacing $T$ with $H_k{}^i{}_p - 2 \ap P_k{}^i{}_p$. We conclude this subsection with the following statement: suppose $\nabla^\Bi J = 0$ where $\nabla^\Bi$ is defined in \eqref{corrected-connection}. Then 
\beq \label{eq:H-locked}
H \= \ii (\partial - \bar{\partial}) \omega + N + 2 \ap P~.
\eeq
We remark that the complex structure on the compact manifold $X$ is not automatic from $\nabla^\Bi J = 0$, and more constraints are required to force $N=0$. To determine whether the Nijenhuis tensor vanishes, we include the dilatino equation.

\subsubsection{Dilatino variation}
The dilatino equation sets the following constraint
\beq \label{dilatino}
\bigg( \delslash \Phi - \half  \Hslash + \ap \frac{3}{2}  \Pslash \bigg) \eta \= 0~.
\eeq
To extract information from it, we use the following identity noticed by Strominger.
\begin{prop} \label{prop-dilatino} \cite{Strominger:1986uh}
  Let $(X,g,J)$ be an almost-complex 6-manifold with complex structure defined by a normalized non-vanishing positive chirality spinor $\eta$, so that $J^k{}_\ell = i \eta^\dagger \gamma^k{}_\ell \eta$. Suppose
  \begin{equation} \label{gen-dilatino-eq}
(\Qslash +  \delslash \Phi) \eta \= 0~,
\end{equation}
for $Q \in \Omega^3(X)$ and $\Phi \in C^\infty(X)$ and e.g. $\Qslash \eta = \frac{1}{3!} \gamma^{ijk} Q_{ijk} \eta$. Then
  \bea 
Q_{\mu \nu \lambda} &=& 0 \label{dilatino-cor1}~,\\
g^{\mu \bar{\nu}} Q_{\mu \bar{\nu} \lambda} &=& \partial_\lambda \Phi ~,\label{dilatino-cor2}
\eea
where Greek indices refer to a local frame $\{ e_\mu \}$ for $\ccT_X^{(1,0)}$. 
\end{prop}

\begin{proof}
  The fact that $J$ comes from $\eta$ means that $\gamma^{\bar{\alpha}} \eta = 0$ and $\gamma_\alpha \eta = 0$. Here we use the following convention for indices: Greek indices (e.g. $\mu$, $\mu$) denote directions along $\ccT_X^{(1,0)}$ while Roman indices (e.g. $i$, $j$) denote directions spanning all of $T_{\mathbb{C}} X = \ccT_X^{(1,0)} \oplus \ccT_X^{(0,1)}$.

  We start by applying $\gamma_\mu$ to \eqref{gen-dilatino-eq}.  This gives
  \begin{equation} \label{gen-dilatino-eq2}
\frac{1}{6}  Q_{\alpha \beta \delta} \{ \gamma_\mu, \gamma^{\alpha \beta \delta} \} \eta + \half  \bigg( Q_{\bar{\alpha} \beta \delta} \{ \gamma_\mu, \gamma^{\bar{\alpha} \beta \delta} \} + Q_{\bar{\alpha} \bar{\beta} \delta} \{ \gamma_\mu, \gamma^{\bar{\alpha} \bar{\beta} \delta} \} \bigg) \eta \=\! - \Phi_\alpha \gamma_\mu \gamma^\alpha \eta~,
  \end{equation}
  since $Q_{\bar{\alpha} \bar{\beta} \bar{\gamma}} \gamma^{\bar{\alpha} \bar{\beta} \bar{\gamma}} \eta = 0$. We can apply the commutator identity \eqref{eq:gamma_rel1} to verify the following identities:
  \bea
 \frac{1}{6}  Q_{\alpha \beta \delta} \{ \gamma_\mu, \gamma^{\alpha \beta \delta} \} \eta &\=& Q_{\mu \alpha \beta} \gamma^{\alpha \beta} \eta~,\\\label{eq:PropDuoUno}
  \half  Q_{\bar{\alpha} \beta \delta} \{ \gamma_\mu, \gamma^{\bar{\alpha} \beta \delta} \} \eta &\=& 2 Q_{\bar{\alpha} \beta \mu} \gamma^{\bar{\alpha} \beta} \eta~,\\
  Q_{\bar{\alpha} \bar{\beta} \delta} \{ \gamma_\mu, \gamma^{\bar{\alpha} \bar{\beta} \delta} \} \eta &\=& 0~.
   \eea
   Substituting these into \eqref{gen-dilatino-eq2}, we obtain
   \begin{equation}
Q_{\mu \alpha \beta} \gamma^{\alpha \beta} \eta + 2 Q_{\bar{\alpha} \beta \mu} g^{\bar{\alpha} \beta} \eta \=\! -2 \Phi_\mu \eta~.
   \end{equation}
   The first term is in fact zero. Indeed, multiplying this equation through by $\gamma_\nu$ gives
   \[
Q_{\mu \alpha \beta} [\gamma_\nu, \gamma^{\alpha \beta}] \eta \= 0~,
   \]
   and from the commutator identity \eqref{eq:gamma_rel1} we derive $Q_{\mu \alpha \beta} = 0$. Substituting this back into \eqref{gen-dilatino-eq2} completes the proof of the Proposition.
\end{proof}

We now apply Proposition \ref{prop-dilatino} to the dilatino equation \eqref{dilatino}. The result is
\bea 
H_{\mu \nu \lambda} - 3 \ap P_{\mu \nu \lambda} &\=& 0~, \label{eq:dilatino1}\\
g^{\mu \bar{\nu}} ( H_{ \bar{\nu} \mu  \lambda} - 3 \ap P_{ \bar{\nu}\mu  \lambda}) &\=& 2\partial_\lambda \Phi\label{eq:dilatino2}
\eea
where indices $\mu,\nu,\lambda$ denote a local frame $\{ e_\mu \}_{\mu=1}^3$ generating $\ccT_X^{(1,0)}$ with respect to the almost-complex structure $J$.

Next, we recall that $\nabla^{\Bi} J = 0$ implies \eqref{Nijenhuis-H}, which combined with \eqref{eq:H-locked} reads
    \beq\notag
N_{\lambda \mu \nu} \= H_{\lambda \mu \nu} - 2 \ap P_{\mu \nu \lambda}~.
\eeq
So we deduce that $\nabla^{\Bi} J = 0$ together with the dilatino equation \eqref{dilatino} imply
\beq\label{eq:N30}
N \= N^{3,0} + N^{0,3}~, \qquad N_{\lambda \mu \nu} \= \ap P_{\mu \nu \lambda}~.
\eeq
From here, we conclude that $N = \cO(\ap^2)$, recovering the result of Strominger \cite{Strominger:1986uh} that the manifold is complex at linear order in $\ap$. In particular, $X$ admits an integrable complex structure obtained by setting $\ap=0$. However, the almost-complex structure denoted $J^k{}_\ell = i \eta^\dagger \gamma^k{}_\ell \eta$ comes from a spinor $\eta$ satisfying the gravitino and dilatino equations with $\ap^2$-terms included, and in the following section we verify that the Nijenhuis tensor of this perturbed $J$ vanishes at second order in $\ap$.

\section{Complex geometry at second order}
In the previous section, we started from the equations obtained by setting the fermionic variations to zero \eqref{eq:fermionicvariations}, and derived various constraints on the fields $g$, $H$, $\Phi$, and $J$. In the current section, we interpret the constraints as equations of complex geometry. We will find that $X$ is a non-K\"ahler Calabi-Yau manifold with a conformally balanced metric satisfying a nonlinear constraint on $i \partial \bar{\partial} \omega$. This extends Strominger's analysis \cite{Strominger:1986uh} from linear order in $\ap$ to second order $\ap^2$.

\subsection{Zeroth order analysis} \label{sec:orderzero}
Before delving into the equations of complex geometry to order $\ap^2$, we begin by rederiving the well-known result that at zeroth order in supersymmetry the manifold is complex with K\"ahler metric \cite{Candelas:1985en}. We take a family of our fields $(g,H,\Phi,A,\eta)$ flowing in $\ap$, e.g.
$$
g \= g_\ap \= g^{(0)} + \ap g^{(1)} + \ap^2 g^{(2)} + \cdots, \quad \eta \= \eta_{\ap} \= \eta^{(0)} + \ap \eta^{(1)} + \cdots~,
$$
and consequently the bilinear $J$ also admits an $\ap$ expansion and so does its Nijenhuis tensor $N$. We assume $g^{(0)}$ is a metric tensor as well as $g_\ap$ at each value of the parameter $\ap$.

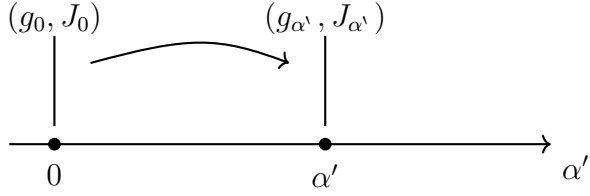
\begin{figure}[ht]
\centering
\begin{tikzpicture}[scale=1.2]

\draw[->, thick] (-0.5, 0) -- (5.5, 0) node[below right] {\(\alpha'\)};

\fill (0,0) circle (2pt);
\node[below] at (0,-0.1) {\(0\)};

\fill (3,0) circle (2pt);
\node[below] at (3,-0.1) {\(\alpha'\)};

\node at (0,1.4) {\((g_0, J_0)\)};
\draw[thick] (0,0.2) -- (0,1.2);

\draw[->, thick] (0.4,0.9) .. controls (1.5,1.2) and (1.8,1.2) .. (2.6,0.9);

\node at (3,1.4) {\((g_\ap, J_\ap)\)};
\draw[thick] (3,0.2) -- (3,1.2);


\end{tikzpicture}
\caption{A family of tensors \((g_\ap, J_\ap)\) over a real parameter $\ap$. The background $(g_0,J_0)$ is K\"ahler, while $\ap$ terms add non-K\"ahler corrections.}
\end{figure}

In this section, we prove that if the fields $(g,H,\Phi,\eta)$ satisfy \eqref{eq:fermionicvariations}, then
\beq \label{eq:orderzero}
H^{(0)} \= 0~, \quad \Phi^{(0)} \= {\rm const}, \quad N^{(0)} \= 0~,
\eeq
and $(X,J^{(0)}, g^{(0)})$ is a complex K\"ahler manifold. We will often use in the following sections a consequence of this result, which is that
\beq\notag
H \= \cO(\ap)~, \quad \nabla \Phi \= \cO(\ap)~.
\eeq
To see \eqref{eq:orderzero}, send $\ap \rightarrow 0$ in \eqref{eq:H-locked}, \eqref{eq:dilatino2}, \eqref{eq:dilatino1}, \eqref{eq:Bianchi2}. We obtain
\begin{align}
  H^{(0)} -i (\partial-\bar{\partial})_{J^{(0)}} \omega^{(0)} -N_{J^{(0)}} &\= 0~, \\
 i \Lambda_{\omega^{(0)}} H^{(0)} + 2 (\partial - \bar{\partial})_{J^{(0)}} \Phi^{(0)} &\= 0~, \\
  (H^{(0)})^{(3,0)} &\= 0~, \\
  d H^{(0)} &\= 0~.
\end{align}
It follows that
\beq\notag
(N^{(0)})^{(3,0)} \= (H^{(0)})^{(3,0)} \= 0~,
\eeq
and so $N_{J^{(0)}}=0$ since the Nijenhuis tensor is purely $(3,0)$-type. We conclude that $J^{(0)}$ defines an analytic complex structure by the Newlander-Nirenberg theorem. Therefore
\beq \label{eq:orderzero0}
H^{(0)} \= \ii (\partial - \bar{\partial})\omega^{(0)}~, \quad  (H^{(0)})_\mu{}^\mu{}_\lambda \=\! - 2 \partial_\lambda \Phi^{(0)} ~, \quad i \partial \bar{\partial} \omega^{(0)}\=0~.
\eeq
It is well-known \cite{Ivanov:2000fg, Papadopoulos:2002gy} that \eqref{eq:orderzero0} implies that $\omega^{(0)}$ is K\"ahler. To see this explicitly, we can take the trace of the exchange relation on Chern curvatures \eqref{exchange-relation} to derive
\beq \notag
(\ii \partial \bar{\partial} \omega)_\mu{}^{\mu \alpha}{}_{\alpha} \= g^{\alpha \bar{\beta}} \partial_\alpha (d^c \omega)_\mu{}^\mu{}_{\bar{\beta}} - g^{\alpha \bar{\beta}} \partial_{\bar{\beta}} (d^c \omega)_\mu{}^\mu{}_\alpha + (d^c \omega)_{\alpha \mu \bar{\nu}} (d^c \omega)^{\mu \bar{\nu} \alpha}~, \quad d^c \omega \= \ii (\partial - \bar{\partial}) \omega~,
\eeq
which is a general identity that holds on any hermitian manifold $(X,\omega)$. Substituting \eqref{eq:orderzero0} into the above identity, we obtain
\beq\notag
4 g^{\alpha \bar{\beta}} \partial_\alpha \partial_{\bar{\beta}} \Phi^{(0)} \= \frac{1}{4}  (H^{(0)})_{\alpha \mu \bar{\nu}} (H^{(0)})^{\mu \bar{\nu} \alpha} ~\geq~ 0~.
\eeq
The maximum principle for elliptic equations on a compact manifold implies that $\Phi^{(0)}$ is a constant and $H^{(0)}=0$. Therefore, at zeroth order the geometry is K\"ahler Calabi-Yau, and we will see that once $\ap$-corrections are included the geometry is perturbed to a non-K\"ahler Calabi-Yau structure.

\subsection{The manifold is complex at  \texorpdfstring{$\ap^2$}{ap2}}
The first order analysis in \cite{Strominger:1986uh} proves that $J^{(0)} + \ap J^{(1)}$ is integrable. We now show the Nijenhuis tensor vanishes to $\ap^2$. The only component   to check is $N^{3,0}$ and its complex conjugate.  From \eqref{eq:N30}, 
\beq\label{eq:N30dH}
 N_{\m\n\l} \=  \frac{\ap}{4}  \Big( (\nabla^-)^\r (\dd H)_{\r\m\nu \lambda} + (\nabla^-)^\rb (\dd H)_{\rb\m\nu \lambda}\Big)~.
\eeq
where $\{ e_\m \}_{\m=1}^3$ is a local frame for $\ccT_X^{(1,0)}$ over an almost-complex manifold $X$.

From \citeS, it is shown that  $F^{0,2} = \cO(\ap^2)$. The curvature of the Hull connection satisfies $R^{\H\,0,2} = \cO(\ap)$ using the  complex geometry valid at first order in $\ap$. We do this explicitly \eqref{eq:ChernHullRelation1}-\eqref{eq:RH02} below.  Consequently, 
\beq\notag
\frac{\ap}{4} \tr F\w F \= \frac{\ap}{4}( \tr F \w F )^{2,2} + \cO(\ap^3)~, \quad \frac{\ap}{4} \tr R^\H \w R^\H \= \frac{\ap}{4}( \tr R^\H \w R^\H )^{2,2} + \cO(\ap^2)~.
\eeq
The Bianchi identity \eqref{eq:Bianchi2} then implies $(\dd H)^{3,1} = \cO(\ap^2)$, and $(\dd H)^{4,0} = 0$ by dimensionality. Hence, \eqref{eq:N30dH} gives
\beq\notag
N_{\m\n\l} \= \cO(\ap^3)~.
\eeq
The manifold is complex at $\ap^2$ with $J = J^{(0)} + \ap J^{(1)} + \ap^2 J^{(2)}$ integrable.

Hence, \eqref{eq:H-locked} becomes 
\beq\notag
T \= \ii (\del-\delb)\o \= H - 2\ap P~.
\eeq

\subsection{The constraint for  \texorpdfstring{$i \partial \bar{\partial} \omega$}{deldelbom} at \texorpdfstring{$\ap^2$}{ap2} } \label{sec:ddbar-ap2}

The Bianchi identity \eqref{eq:Bianchi2} repeated here is determined by anomaly cancellation and so unexpectedly not modified by $\ap$ corrections:
\beq\notag
\dd H\= \frac{\ap}{4}\Big(\tr F \w F - \tr R^\H \w R^\H \Big) ~.
\eeq
What is corrected is the relation between $\del\delb\o$ and the gauge bundle. We calculate this here. For this reason it is important that we refer to \eqref{eq:Bianchi2} as the Bianchi identity and not its expression in terms of $\o$. 
We will derive
\beq\label{eq:delbdel-o}
2\ii\delb\del\widehat\o-\frac{\ap}{4}\left[\tr F\w F-\tr \widehat{R}^\Ch\w \widehat{R}^\Ch\right]+\cO(\ap^3)\=0~,
\eeq
where $\widehat\o$ is defined in \eqref{eq:omega-field-redefinition}.
To start, we first improve on the previous subsection by showing $[\dd H]^{3,1} =0 + \cO(\ap^3)$.  
Write  $\Th^\H = \Th^\Ch + S$ where $S$ is some tensor
 \beq\label{eq:ChernHullRelation1}
\begin{split}
R^\H &\= R^\CH + \dd_\Th S + S^2~, \\
 \tr R^\H \w R^\H &\=  \tr R^\CH \w R^\CH + \dd\left[ \tr \left(2S\w R^\CH +S\w \dd_{\Th^\Ch} S+ \frac{2}{3} S^3 \right)\right]~.
\end{split}
 \eeq
In  complex coordinates  
\beq\label{eq:RH02}
R^{\H\,0,2}{\,}^{\bar\sigma}{}_\rho
=\dd x^{\bar\mu\bar\nu}
 \nabla^\Ch_{\bar\mu}T_{\bar\nu}{}^{\bar\sigma}{}_\rho
 +\cO(\ap^2)~.
\eeq
 $R^\H $ fails to be a holomorphic instanton at first order in $\ap$, consistent with   \eqref{eq:RhdH}.

Split $S$ into holomorphic and antiholomorphic components $S = \S + \Sb$, where from \sref{app:ConnectionSymbols} the only non-vanishing components are
\beq\label{eq:cScSb}
\S^\n{}_\rb \=  H_\m{}^\n{}_\rb~ \dd x^\m~, \qquad \Sb^\nb{}_\r \=  H_\mb{}^\nb{}_\r~ \dd x^\mb ~.
\eeq
Using $R^{\Ch\,0,2} = 0$ we find (to save space, we write $\Th$  in this subsection to mean $\Th^\Ch$):
\beq\notag
\begin{split}
 \tr S\w R^\Ch &\= 0~,\qquad  \tr S^3 \=  0~,\\
 \tr S\w \dd_\Th S      &\= 2  \S^\n{}_\rb\w \big( \del_\Th  \Sb^\rb{}_\n + \delb_\Th  \Sb^\rb{}_\n\big) +  \dd\left(\, \tr    \S \w \Sb \right) ~,
\end{split}
\eeq 
we find
\beq\label{eq:trRHtrRCh}
\begin{split}
 \tr R^\H \w R^\H - \tr R^\Ch \w R^\Ch 
 &\=  2( (\del_\Th + \delb_\Th) \S^\n{}_\rb)\w \big( \del_\Th  \Sb^\rb{}_\n + \delb_\Th  \Sb^\rb{}_\n\big) \\[3pt]
 &\qquad +  2 \S^\n{}_\rb\w \big( R^\Ch{}^\rb{}_\sb \Sb{}^\sb{}_\n - R^\Ch{}^\s{}_\n \Sb{}^\rb{}_\s \Big)~.
\end{split}
\eeq
 We can  isolate the components that are not $(2,2)$ in the Bianchi:
\beq\label{eq:trRR31}\notag
[\tr R^\H \w R^\H ]^{(1,3)} \= 2\delb \left(\S{}^\n{}_\rb \w \delb_\Th \Sb{}^\rb{}_\n  \right) ~, \qquad\qquad [\tr R^\H \w R^\H ]^{(0,4)}  \= 0~.\\[3pt]
\eeq
and so 
\beq\notag
[\dd H]^{1,3}  \=\!- \frac{\ap}{2} \delb \left(\S{}^\n{}_\rb \w \delb_\Th \Sb{}^\rb{}_\n  \right) \= \cO(\ap^3) ~,\qquad \qquad [\dd H]^{0,4} \= 0 ~.\label{eq:dH13}\\[3pt]
 \eeq

Now we find the $\ap^2$-corrected relation between $\o$ and $F$ using the Bianchi identity and the gravitino equation. Substitute \eqref{eq:H-locked} into \eqref{eq:Bianchi2}, with terms manifestly $(2,2)$  on the left hand side,
\beq\label{eq:Balp3}
\begin{split}
  \dd \dd^c\o - \frac{\ap}{4} \tr (F \w F)  & \=\! -\frac{\ap}{4}\Big(  \tr R^\H \w R^\H \Big)  - 2\ap \dd P~.
\end{split}
\eeq

Using $H=\cO(\ap)$:
$$
\ap \tr R^\H \w R^\H - \ap \tr R^\Ch \w R^\Ch \= \cO(\ap^3)~,
$$
from \eqref{eq:trRHtrRCh}. 

Using \eqref{eq:apP}, we get 
\beq
\ap\dd P\=\!-\frac{\ap^2}{4}\delb\del\left(\cE^\H_{\m\nb}\dd x^{\m\nb}\right)+\cO(\ap^3)~.
\eeq
Substitution into \eqref{eq:Balp3}, together with the Chern--Hull comparison above, gives \eqref{eq:delbdel-o}.

{\bf Remark:} The nonlinear constraint on $i\partial\bar{\partial}\widehat\omega$, taken as the stand-alone mathematical equation
\beq\label{eq:delbdel-o1}
2\ii\delb\del\widehat\o-\frac{\ap}{4}\left[\tr F\w F-\tr \widehat{R}^\Ch\w \widehat{R}^\Ch\right]\=0~,
\eeq
is given a differential geometric interpretation in \cite{McOrist:2024zdz}, building on \cite{delaOssa:2014cia, McOrist:2021dnd}. Let $E \rightarrow X$ be a holomorphic bundle over a complex manifold $X$ with $F$ the Chern curvature of a hermitian metric on the bundle $E$. There is a differential operator
\beq\notag
\bar{D} : \Omega^{(0,p)}(Q) \rightarrow \Omega^{(0,p+1)}(Q) , \quad Q= \ccT_X^{*(1,0)} \oplus {\rm End} \, E \oplus \ccT_X^{(1,0)}
\eeq
of the form
\beq\notag
\bar{D}= \begin{bmatrix}
\delb & - \frac{\ap}{4} \cF^* & \cT - \frac{\ap}{4} \cR \nabla \\
0 & \delb_A & \cF \\
0 & 0 & \delb
\end{bmatrix}~,
\eeq
which satisfies $\bar{D}^2=0$ if and only if \eqref{eq:delbdel-o1} holds. The definitions are:
\begin{align}
  \cF \colon \Omega^{(0,p)}(\ccT_X^{(1,0)}) 
  &\longrightarrow \Omega^{(0,p+1)}({\rm End} \, E)~, \nonumber \\
  \D 
  &\longmapsto F_{\mu \bar{\nu}} \, \dd x^{\bar{\nu}} \wedge \D^\mu~, \label{eq:F_def} \\[1ex]
  \cF^* \colon \Omega^{(0,p)}({\rm End} \, E) 
  &\longrightarrow \Omega^{(0,p+1)}( \ccT_X^{*(1,0)}), \nonumber \\
  \aa 
  &\longmapsto {\rm Tr}\bigl(F_{\mu \bar{\nu}} \, \dd x^\mu \otimes \dd x^{\bar{\nu}} \wedge \aa\bigr), \label{eq:Fstar_def} \\[1ex]
  \cT \colon \Omega^{(0,p)}(\ccT_X^{(1,0)}) 
  &\longrightarrow \Omega^{(0,p+1)}(\ccT_X^{*(1,0)})~, \nonumber \\
  \D 
  &\longmapsto (d^c \widehat\omega)_{\rho \bar{\nu} \mu} \, \dd x^\rho \otimes (\dd x^{\bar{\nu}} \wedge \D^\mu), \label{eq:H_def} \\[1ex]
  (\cR \nabla) \colon \Omega^{(0,p)}(\ccT_X^{(1,0)}) 
  &\longrightarrow \Omega^{(0,p+1)}(\ccT_X^{*(1,0)})~, \nonumber \\
  \D 
  &\longmapsto - \frac{1}{p!} \widehat{R}^\Ch{}_{\rho \bar{\mu}}{}^\sigma{}_\lambda \, 
    \widehat\nabla^\Bi_\sigma \D^\lambda{}_{\bar{\kappa}_1 \dots \bar{\kappa}_p} \, 
    \dd x^\rho \otimes \dd x^{\bar{\mu} \bar{\kappa}_1 \cdots \bar{\kappa}_p}~.\label{eq:R_def}
\end{align}
The operator $\bar{D}$ does not quite define a holomorphic structure on the vector bundle $Q$ due to the $\ap \cR \nabla$ term. Given \eqref{eq:delbdel-o1}, then the differential operator $\bar{D}$ defines an elliptic complex
\beq\notag
0 \longrightarrow \Gamma(Q) \overset{\bar{D}}{\longrightarrow} \Omega^{(0,1)}(Q) \overset{\bar{D}}{\longrightarrow} \Omega^{(0,2)}(Q) \overset{\bar{D}}{\longrightarrow} \cdots~.
\eeq
We refer to \cite{Chisamanga:2024xbm, 2409.04382} for cohomological calculations and physical interpretations of this complex. There is also a mathematical interpretation given in the language of generalized geometry \cite{Garcia-Fernandez:2015hja, Garcia-Fernandez:2018ypt} of related equations of the form $i \partial \bar{\partial} \omega = c \langle F \wedge F \rangle$ where $c$ is a bi-invariant symmetric pairing on the Lie algebra of the structure group of the bundle. For more on interactions between generalized geometry and heterotic string theory, see \cite{Ashmore:2019rkx, Garcia-Fernandez:2020awc}.

\subsection{Holomorphic volume form}
We have shown that the supersymmetry equations \eqref{eq:fermionicvariations} lead to an integrable complex structure $J$ through $\ap^2$. Next, we investigate holomorphicity of a nowhere vanishing 3-form through $\ap^2$. Consider the 3-form $\Psi \in \Omega^3(X)$ determined by the positive chirality spinor $\eta$ given by
\beq \label{eq:3form}
\Psi_{ijk} \=\bar{\eta}^T \g_{ijk} \bar{\eta}~.
\eeq
One can check using $\gamma_7 \eta = \eta$ and $\gamma_\alpha \eta = 0$ in holomorphic directions that $\Psi \in \Omega^{3,0}(X,\mathbb{C})$. The gravitino equation \eqref{parallel-spinor} implies
\beq \label{parallel-3form}
 \quad \nabla^{\Bi} \Psi \= 0~,
\eeq
where the connection $\nabla^{\Bi}$ is defined in \eqref{corrected-connection}. Since the connection $\nabla^{\Bi}$ also satisfies $\nabla^{\Bi} g = 0$ and $\nabla^{\Bi} J =0$, its holonomy is contained in the group $SU(3)$.
\beq\notag
{\rm Hol}(\nabla^{\Bi}) \subseteq SU(3)~.
\eeq
The pair $(\omega,\Psi)$ forms an $SU(3)$-structure, namely
\beq\notag
\omega \wedge \Psi \= 0~, \quad i \Psi \wedge \bar{\Psi} \= \frac{\omega^3}{3!}~, \quad |\Psi|_g \= 1~.
\eeq
We note that neither $\omega$ nor $\Psi$ are closed. The constant norm 3-form $\Psi$ will not be a holomorphic volume form through $\ap^2$, but we will arrange that $\Omega = e^f \Psi$ solves $\bar{\partial} \Omega = 0$ for some conformal factor $e^f$.

Let us rewrite the gravitino equation for $\nabla^{\LC} \eta$ in a different way for later use. Using $\g_\alpha \eta = 0$ in holomorphic directions and $H^{3,0}= 3\ap P^{3,0} = \cO(\ap^3)$, from the gravitino equation \eqref{parallel-spinor} we derive
\beq\notag
\nabla^{\LC}{}_\alpha \eta = \frac{1}{4} H_\alpha{}^{\mu \bar{\nu}} g_{\mu \bar{\nu}} \eta - \frac{\ap}{2} P_\alpha{}^{\mu \bar{\nu}} g_{\mu \bar{\nu}} \eta~,
\eeq
in holomorphic indices. We next substitute the dilatino equation \eqref{eq:dilatino2}.
\beq \label{eq:dilatino+grav0}
\nabla^{\LC}{}_\alpha \eta \= \frac{1}{2} (\partial_\alpha \Phi) \eta + \frac{\ap}{4} P_\alpha{}^{\mu \bar{\nu}} g_{\mu \bar{\nu}} \eta~.
\eeq
We can also combine the dilatino equation \eqref{eq:dilatino2} with the locking relation for $H$ given in \eqref{eq:H-locked}, knowing now that the Nijenhuis tensor is zero, to see
\beq \label{eq:dilatino+grav}
(i \partial \omega)_{\mu}{}^\mu{}_{\lambda} - \ap P_{\mu}{}^\mu{}_{\lambda} \=\! -2\partial_\lambda \Phi~.
\eeq
Combining \eqref{eq:dilatino+grav0} and \eqref{eq:dilatino+grav} yields the identity
\beq \label{eq:dilatino+grav1}
\nabla^{\LC}{}_\alpha \eta \=\!-\frac{1}{4}  (i \partial \omega)_{\mu}{}^\mu{}_{\alpha} \eta~.
\eeq

With \eqref{eq:dilatino+grav1}, we can now compute the covariant derivative of the 3-form $\Psi$ defined in \eqref{eq:3form}.
\beq\notag
\nabla^{\LC}{}_{\bar{\alpha}} \Psi_{ijk} \= 2 \bar{\eta}^T \g_{ijk} \nabla^{\rm LC}{}_{\bar{\alpha}} \bar{\eta} \=\! -\half   (i \bar{\partial} \omega)_{\mu}{}^\mu{}_{\bar{\alpha}} \Psi_{ijk}~.
\eeq
The Levi-Civita connection induced on top forms $\Psi \in \Omega^{3,0}(X)$ is
\beq\notag
\nabla^{\LC}{}_{\bar{\alpha}} \Psi_{ijk} \= \partial_{\bar{\alpha}} \Psi_{ijk} - \Gamma^{\LC}{}_{\bar{\alpha}}{}^\mu{}_\mu\Psi_{ijk} \= \partial_{\bar{\alpha}} \Psi_{ijk}- \frac{1}{2} (i \bar{\partial} \omega)_{\bar{\alpha}}{}^\mu{}_\mu \Psi_{ijk}.
\eeq
Therefore
\beq\notag
 \partial_{\bar{\alpha}} \Psi_{ijk} \=\! - (\ii \bar{\partial} \omega)_{\mu}{}^\mu{}_{\bar{\alpha}} \Psi_{ijk}~.
 \eeq
 Substituting \eqref{eq:dilatino+grav} gives
\beq\notag
 \partial_{\bar{\alpha}} \Psi_{ijk} \= (2 \partial_{\bar{\alpha}} \Phi+ \ap P_\mu{}^\mu{}_{\bar{\alpha}}) \Psi_{ijk}~.
 \eeq
 As expected, $\Psi$ is holomorphic to zeroth order in $\ap$, but to higher order it must be corrected by setting $\Omega = e^f \Psi$ and solving $\bar{\partial} \Omega = 0$ for a suitable conformal factor. 
 
 Using \eqref{eq:trP}, we have
\beq\label{eq:trP2}
 \ap P_\m{}^\m{}_\l=-\frac{\ap^2}{16}\partial_\l h+\cO(\ap^3)~.
\eeq
Consequently,
\beq
 \Phi_\omega=\Phi+\frac{\ap^2}{32}h~,\qquad \Omega=e^{-2\Phi_\omega}\Psi~,\qquad \log\norm{\Omega}_g=-2\Phi_\omega+\cO(\ap^3)~,
\eeq
It is convenient to pass to the hatted variables \eqref{eq:omega-field-redefinition}:
\beq\label{eq:norm-Omega}
 \log\norm{\Omega}_{\widehat g}=-2\widehat\Phi+\cO(\ap^3)~.
\eeq
 
Hence, $(X,\omega)$ is a complex hermitian manifold with holomorphic volume form $\Omega$ but because $d \omega \neq 0$, it is a non-\K Calabi-Yau threefold.

\subsection{Conformally balanced equation}
The supersymmetry equations on spinors have produced a pair $(\omega, \Omega)$, where $\omega$ is a hermitian metric on a complex manifold and $\Omega$ is a holomorphic volume form. We now derive the conformally balanced relation $\dd ( \norm{\Omega}_{\wh g} \wh\omega^2)=0$. Substituting \eqref{eq:trP2}  into \eqref{eq:dilatino+grav} gives 
\beq \notag
\frac{1}{2} (\ii \partial \omega)_{\mu}{}^\mu{}_{\lambda}  + \frac{\ap^2}{32}\nabla_\l h \=\! -\partial_\lambda \Phi~.
\eeq
In the hatted  basis \eqref{eq:omega-field-redefinition} this becomes
\beq\notag
\frac{1}{2} (\ii \partial\widehat \omega)_{\mu}{}^\mu{}_{\lambda}  \=\! -\partial_\lambda \widehat\Phi~.
\eeq
and thus by \eqref{eq:dconfbal} we obtain
\beq \label{eq:confbal}
\dd \Bigl(  e^{-2\widehat\Phi} \,\widehat \omega^2 \Bigr) \= 0~.
\eeq
We can write this conformal factor $e^{-2 \widehat\Phi}$ in terms of the holomorphic volume form $\Omega$ by substituting \eqref{eq:norm-Omega}, and the result is 
\beq
\dd ( \norm{\Omega}_{\widehat g} \,\widehat \omega^2 )=0~.
\eeq

The mathematical significance of the conformally balanced equation becomes apparent when the gaugino equation \eqref{eq:fermionicvariations} is included, which states $\ap g^{\mu \bar{\nu}} F_{\mu \bar{\nu}}=0$; we can pass to $\widehat g$ at the expense of $\ap^3$ corrections. The triple $(\widehat \omega,\Omega,F)$ then solves
\beq\notag
\dd ( \norm{\Omega}_{\widehat g}  \widehat \omega^2 )\=0, \quad F \wedge \widehat \omega^2\= 0~.
\eeq
The (conformal) closedness of $\widehat\omega^2$ is what allows the numerical pairing $[\norm{\Omega}_{\widehat g} \widehat \omega^2] \cdot c_1(S)$ to be defined on subsheaves $S \subseteq E$, and this is what allows a well-defined notion of slope stability as a criterion for the non-K\"ahler Donaldson-Uhlenbeck-Yau theorem \cite{Uhlenbeck1986,0529.53018,0664.53011}: if $[\norm{\Omega}_{\widehat g} \,\widehat \omega^2] \cdot c_1(S) < 0$ for all strict torsion-free coherent subsheaves $S \subseteq E$, then there exists a solution to $F \wedge\widehat \omega^2= 0$.

 \section{Integrability and the equations of motion } \label{sec:integrability}
 A supersymmetric background admits a globally defined spinor satisfying \eqref{eq:fermionicvariations}. In particular, $\nabla^\Bi_m\eta=0$ implies the integrability equation
$$
[\nabla^\Bi_m,\nabla^\Bi_n]\eta\=\qrt R^\Bi_{mnpq}\gamma^{pq}\eta\=0~,
$$
which is an identity, not an additional condition. Together with the dilatino and gaugino equations and the Bianchi identity, we show this calculation reproduces the bosonic equations of motion, including the explicit $\ap^2$ term $\Delta^{\rm conn}_{mn}$. Section~5 confirms this result using complex geometry.

 \subsection{Preliminaries}
Let $\ve$ be the ten-dimensional Majorana Weyl spinor. Recall, there is the torsion of the Bismut connection $T$, its relation to $H$ and a new three-form $\hat H$ that appears in the dilatino variation:
\beq\label{eq:wtHhatH}
T_{mab} \= H_{mab} - 2\ap P_{mab} \= \wt H_{mab} + \ap P_{mab}~, \qquad \wt H_{mab} \= H_{mab} - 3 \ap P_{mab}~.
\eeq
The gravitino variation implies a spinor covariantly constant with respect to the $T$ connection. Take a second covariant derivative of the gravitino variation with respect to Levi-Civita:
\beq
\nabla_n\nabla_m \ve - \frac{1}{8} \nabla_n T_{mab} \gamma^{ab} \ve - \frac{1}{64}T_{mab} T_{ncd} \gamma^{ab} \gamma^{cd} \ve \= 0~,
\eeq
where in this section unless otherwise noted $\nabla = \nabla^\LC$. 
Antisymmetrise using
$
[\nabla_m,\nabla_n] \= \frac{1}{4} R_{mnab} \gamma^{ab}~,
$
where $\g^a$ are the gamma matrices and use the first line of \eqref{eq:gamma_rel2} to give
\beq\label{eq:integ0a}
\dd x^m \w \dd x^n \left( R_{mnab} - \nabla_{[m} T_{n]ab} + \half T_{[m|ac} T_{n]}{}^c{}_b \right) \gamma^{ab} \ve \= 0~.
\eeq
This result can be derived in a second way using Cartan's formulation. Write the gravitino variation as
\beq
 \nabla^\Bi_m \ve \= \del_m \ve + \frac{1}{4} \Th^\Bi_{m\,ab} \gamma^{ab}~, \qquad  \Th^\Bi_m{}^a{}_b \= \Th^\LC_m{}^a{}_b - \half T_m{}^a{}_b~.
\eeq
The gravitino--integrability identity comes from skew-symmetrising the derivatives
\beq\label{eq:integ0}
[ \nabla^\Bi_m, \nabla^\Bi_n] \ve \= \frac{1}{4}  R^\Bi_{mnab} \gamma^{ab}\ve\=0~, \qquad  R^\Bi \= \dd \Th^\Bi +  \Th^\Bi \w  \Th^\Bi~.
\eeq
Now, as $\Th^\Bi = \Th^\LC - \half T$ we have
$$
 R^\Bi \= R^\LC - \half \dd_{\nabla^\LC} T +\qrt  T \w T~, 
$$
where $\nabla^\LC$ acts only on the tangent space indices labelled $a,b$. Evaluating this gives 
\beq
 R^\Bi_{ab}  \= \half \dd x^m \w \dd x^n \left( R_{mnab}^\LC - \nabla^\LC_{[m} T_{n]ab} + \half T_{[m|a}{}^c T_{n]cb} \right)~,
\eeq
which when put in \eqref{eq:integ0} gives \eqref{eq:integ0a} as promised.

As tempting as it is, we cannot  identify the terms in parenthesis with zero in \eqref{eq:integ0a}. Instead, multiply \eqref{eq:integ0a} by $\g^n$ and use $\g^n \g^{ab} =   g^{na} \g^b - g^{nb} \g^a + \g^{nab}$ via the middle line of \eqref{eq:gamma_rel1} to give
\beq\label{eq:integrab2}
\begin{split}
 \half R^\Bi_{mnab} \g^n\g^{ab} \ve \=&\left(\Ric_{mn}  - \qrt T_{mab} T_{n}{}^{ab}  +\half  \nabla^{\LC\,p} T_{pmn}\right) \g^n\ve \\
&\qquad \quad +\frac{1}{4} \left(\nabla^\LC_m T_{nab} - \nabla^\LC_n T_{mab}-  T_{ma}{}^c T_{bnc}   \right) \gamma^{nab} \ve\= 0~.
\end{split}
\eeq
Note we have used $H^p{}_{pq} = 0$, $\Ric_{mn} = R_{mpn}{}^p$ and the Bianchi identity for the Riemann curvature $R_{m[nab]}=0$ has been used. Our goal is to massage the term proportional to $\g^n \ve$ to be as close to the equations of motion as possible.

We need the covariant derivative of the dilatino equation, which is the second line of \eqref{eq:fermionicvariations}. Taking a derivative and using $\nabla^\Bi \ve = 0$, we find
$$
 \left( 2  \nabla^\Bi_m \nabla_n \Phi\, \g^n-  \frac{1}{6}\nabla^\Bi_m \wt H_{nab}\, \g^{nab}\right) \ve \= 0~.
$$
In components, this becomes
\beq\label{eq:covderivdil}
\begin{split}
  \left( 2 \nabla_m \nabla_n \Phi + H_m{}^p{}_n \nabla_p \Phi \right) \g^n\ve \= \frac{1}{6}\left( \nabla_m \wt H_{nab} - \frac{3}{2} H_{mn}{}^p H_{abp} \right)\g^{nab}\ve ~,
\end{split}
\eeq
where we use $\hat H$ in \eqref{eq:wtHhatH}.

Finally, multiply the gaugino equation $\Fslash \ve = 0$ by $F_{pn}\g^n$ to obtain
\beq\label{eq:gaugeinstantongamma}
F_{pn} F_{ab} \g^{nab} \ve + 2 F_p{}^a F_{an} \g^n \ve \= 0~.
\eeq
Hence, 
\beq\label{eq:gaugino3}
\begin{split}
 &\frac{1}{12} \tr (F \w F)_{mnab} \g^{nab} \ve \= F_m{}^a F_{na}\g^n \ve~, \qquad {\rm where}~~\\[5pt]
&\qquad\qquad \half \tr (F\w F)_{mnab} = F_{mn}  F_{ab} - F_{ma} F_{nb} - F_{mb} F_{an} ~.
\end{split}
\eeq
Note that
\beq\label{eq:Hconfdiv}
- e^{2\Phi} \nabla^p (e^{-2\Phi} H_{pmn} ) \g^n \ve \= - \nabla^p H_{pmn} \g^n \ve + 2(\nabla^p \Phi) H_{pmn} \g^n \ve~.
\eeq

With this collection of results in hand we return to the gravitino--integrability identity \eqref{eq:integrab2}.  
Using \eqref{eq:wtHhatH}, \eqref{eq:gaugino3} and
\begin{align}
 & (\dd  H)_{mnab} \= \nabla_m H_{nab} - \nabla_n H_{abm} + \nabla_a H_{bmn} - \nabla_b H_{mna}~, \notag\\
& (\dd H)_{mnab} \g^{nab}  \= (\nabla_m H_{nab} - 3 \nabla_n H_{abm} )\g^{nab}~,\notag
\end{align}
gives
\beq\notag
\begin{split}
& \left(\Ric_{mn} - \qrt  H_{mab}  H_{n}{}^{ab}\right) \g^n\ve    -\qrt \left(  H_{ma}{}^c  H_{bnc}   \right) \gamma^{nab} \ve\\
&\quad\=\!-  \frac{1}{12} \left(\nabla^\LC_m T_{nab} - 3\nabla^\LC_n T_{abm}  + 2\nabla^\LC_m \hat H_{nab}  +2\ap \nabla^\LC_m  P_{nab} \right) \g^{nab}\ve\\
&\qquad\qquad +\ap \nabla^p P_{pmn} \g^n \ve -  H_{pmn} ( \nabla^p \Phi)  \g^n \ve - \half e^{2\Phi} \nabla^p (e^{-2\Phi}  H_{pmn} ) \g^n \ve\\[5pt]
\end{split}
\raisetag{3.5cm}
\eeq
Using \eqref{eq:covderivdil}, together with \eqref{eq:Pdef2} that $\ap P \propto \ap \dd^\dag (\dd H) + \cO(\ap^3)$, so that  $\dd^\dagger P=\cO(\ap^2)$, and hence $\ap\dd^\dagger P=\cO(\ap^3)$, this becomes 
\beq\notag
\begin{split}
& \left(\Ric_{mn} + 2 \nabla_m \nabla_n \Phi - \qrt  H_{mab}  H_{n}{}^{ab}\right) \g^n\ve  +  \half e^{2\Phi} \nabla^p (e^{-2\Phi}  H_{pmn} ) \g^n\ve   \\
&\quad\=\!- \frac{1}{12} (\dd T)_{mnab} \g^{nab} \ve- \frac{\ap}{6} \nabla_m P_{nab} \g^{nab} \ve~. 
\end{split}
\raisetag{3.5cm}
\eeq
Using $T = H - 2\ap P$ and \eqref{eq:gaugeinstantongamma}-\eqref{eq:gaugino3} 
\beq\label{eq:integrab4}
\begin{split}
& \left(\Ric_{mn} +   2\nabla_m \nabla_n \Phi - \qrt  H_{mab}  H_{n}{}^{ab} + \frac{\ap}{4}\tr  F_m{}^a F_{na} \right) \g^n\ve +   \half e^{2\Phi} \nabla^p (e^{-2\Phi}  H_{pmn} ) \g^n\ve\\
&\quad\=\!- \frac{1}{12}\Big(   (\dd  H)_{mnab} - \frac{\ap}{4} (\tr F^2)_{mnab} \Big) \g^{nab} \ve   + \frac{\ap}{6} \Big(  
 (\dd P)_{mnab} -\nabla^\LC_m  P_{nab}   \Big) \g^{nab} \ve~.   
\end{split}
\raisetag{2cm}
\eeq
The first line contains the first order graviton and H equation of motion; the second contains the Bianchi identity and explicit $\ap^2$ corrections.

 We now add and subtract terms to get the graviton equation of motion and evaluate the heterotic Bianchi identity \eqref{eq:Bianchi2}:
\beq\label{eq:integrab5}
\begin{split}
& \left(\Ric_{mn} +   2\nabla_m \nabla_n \Phi - \qrt  H_{mab}  H_{n}{}^{ab} + \frac{\ap}{4}\tr  F_m{}^a F_{na} -  \frac{\ap}{4}\tr  R^\H{}_m{}^a R^\H{}_{na} \right) \g^n\ve \\[5pt]
&+   \half e^{2\Phi} \nabla^p (e^{-2\Phi}  H_{pmn} ) \g^n \ve  \=\! -   \frac{\ap}{4}\left(\tr   ( R^\H{}_m{}^a R^\H{}_{na} )\g^n\ve -  \frac{1}{12} \tr (R^\H \w R^\H)_{mnab}  \g^{nab} \ve \right) \\[5pt]
&~~~ + \frac{\ap}{6} \Big(  
 (\dd P)_{mnab} -\nabla^\LC_m  P_{nab}   \Big) \g^{nab} \ve ~.\\[5pt]  
\end{split}\raisetag{1.0cm}
\eeq
The left hand side contains the graviton and $H$ equations of motion to first order in $\ap$. The right hand side contains the $\ap^2$ corrections. 

\subsection{\texorpdfstring{$\SU(3)$}{SU3}  manifold}
Thus far we have not specified the dimension of the compactification. To make progress we  specialise \eqref{eq:integrab5} to $SU(3)$. This means we can introduce a complex structure, as we proved earlier that the manifold is complex. The ten-dimensional spinor is written as $\ve = \zeta\otimes \eta + \cc$, where $\eta$ is a Weyl spinor of $SO(6)\cong SU(4)$. As usual we focus just on the internal manifold and discard the $d=4$ spacetime component. 

\subsubsection{\texorpdfstring{$H$}{H} and \texorpdfstring{$F$}{F}  equations of motion} \label{sec:HFeom}

We comment on the $H$ and $F$ equations of motion. First, we derive from the gaugino variation $\Fslash \ve = 0$ the $F$ equation of motion. For an $SU(3)$ manifold, the gaugino equation reads
\beq
F_{\m\nb} \g^{\m\nb} \eta \= 0~,\qquad \o^{\m\nb} F_{\m\nb} \= 0~,
\eeq
where we use \eqref{eq:spinor_rel}. The gaugino is understood to come multiplied by an $\ap$ and so we keep terms up to order $\ap$. This means we can differentiate with respect to the Bismut connection and it will kill $\o$ leaving
\beq\label{eq:Feom}
\begin{split}
\o^{\m\nb} \nabla^\Bi_\r F_{\m\nb} &\= \o^{\m\nb}\left( \del^A_\r F_{\m\nb} -\G^\LC_\r{}^\s{}_\m F_{\s\nb} + \half H_\r{}^\s{}_\m F_{\s\nb} + H_\r{}^\sb{}_\nb F_{\m\sb}  \right)
\= \ii e^{2\Phi} \, \nabla^{\Bi\, \nb} (e^{-2\Phi} F_{\nb\r}) \= 0~,
\end{split}
\eeq
where $\nabla^A_\m F_{\m\nb}$ is the gauge covariant derivative and we use the Bianchi identity $\del_A F = 0$. We have replaced the torsion $T$ by $H$ as the difference is $\ap^2$, keeping in mind the gaugino comes with an implicit $\ap$, this difference is really $\ap^3$.  We also use $2\nabla^\sb \Phi = H^\r{}_\r{}^\sb + \cO(\ap^2)$. This means supersymmetry implies the Yang--Mills equation of motion and we can use it in the following.

Second, we show that the $H$ equations of motion follow directly from the supersymmetry conditions. Using that $\star \beta^{2,1} = \ii  \beta^{2,1} - \beta_\m{}^\m\w\o$, we find
\beq\label{eq:Tcalib}
\star T \= - e^{2\hat\Phi} \dd (e^{-2\hat \Phi} \o)~, \qquad \hat \Phi= \Phi + \frac{3\ap^2}{32} (\tr |F|^2 - \tr |R^\H|^2)~,
\eeq
where we use $T = \dd^c\o = H-2\ap P$, the dilatino equation \eqref{eq:dilatino2} applied to $T$, and the trace of $P$ \eqref{eq:trP} calculated below. Here we introduced the scalar function $\hat{\Phi}$ which is defined by \eqref{eq:Tcalib}. Hence,   
\beq
\begin{split}
 e^{-2\hat \Phi} \star H &\= 2\ap e^{-2\hat \Phi} \star  P -  \dd (e^{-2\hat \Phi} \o). \\
\end{split}
\eeq
From \eqref{eq:Pdef2} below we see that 
$$
\ap  \dd (e^{-2\hat \Phi} \star P) \=  \ap^2 (\dd^\dag)^2 (\tr F\w F - \tr R^\H \w R^\H)  -2 \dd \hat\Phi \w \star \ap P = \cO(\ap^3)
$$
 and so finally noting that $e^{-2\hat\Phi} H = e^{-2\Phi} H + \cO(\ap^3)$ we end up with
\beq\label{eq:Hcalib}
\dd \left( e^{-2 \Phi} \star H \right) \=0~,
\eeq
which is the $H$ equation of motion up to and including $\ap^2$. It follows directly from supersymmetry.

\subsubsection{Preliminary results on \texorpdfstring{$P$}{P} and \texorpdfstring{$H$}{H} }

We derive some results to be used in the following sections. In the following we assume supersymmetry holds. We have already shown to zeroth order \sref{sec:orderzero} that the $SU(3)$ manifold is complex, \K with Ricci-flat metric. In fact, the Bergshoeff--de Roo algebra has \eqref{eq:RhdH} which to zeroth order in $\ap$ shows $R^\H$ satisfies a Yang-Mills equation and that $R^{\H\,0,2} = \cO(\ap)$ (shown directly using \eqref{eq:susy3} for an $SU(3)$ manifold in the same way as $F$; or using \eqref{eq:RH02} with $H=\cO(\ap)$). Together with Yau's theorem, this implies that the metric is Ricci flat to zeroth order: $\Ric_{mn} = \cO(\ap)$. 

For any $p$-form $Q$, we write $Q_m = \frac{1}{(p-1)!} Q_{m n_2\cdots n_p} \dd x^{n_2\cdots n_p}$ and its divergence satisfies:
 \beq\label{eq:divBism}
\begin{split}
  \nabla^{\Bi\, m} Q_m &\= g^{mn} \left( \nabla^\LC_n Q_m + \half H_n{}^p{}_m Q_p  \right)  \= \nabla^{\LC\,m} Q_m + \cO(\ap) \\[3pt]
 &\= \left(\nabla^\Ch_\m Q_\nb + \nabla^\Ch_\nb Q_\m \right)g^{\m\nb} + \cO(\ap)~,
\end{split}
 \eeq
 where we use the dilatino equation \eqref{eq:dilatino2} with \eqref{eq:Pdef} and that $\dd H = \cO(\ap)$. 
 
The tensor  $P$ in \eqref{eq:Pdef} simplifies using \eqref{eq:divBism}
\beq\label{eq:Pdef2}
\begin{split}
 \ap P &\= \frac{\ap}{4} e^{2\Phi} \nabla^{\Bi \,m}\left( e^{-2\Phi} (\dd H)_m \right) \\
 &\=  \frac{\ap^2}{8} \nabla^m \left( \tr F_m \w F - \tr R^\H{}_m \w R^\H \right) + \cO(\ap^3) \=\!- \frac{\ap}{4} \dd^\dag \dd H ~.
\end{split}
\eeq
Due to the $\ap^2$, $\nabla^m$ can be any connection and the dilatino equation $\nabla_\r \Phi = \half H_\r{}^\t{}_\t + \cdots  =  \cO(\ap)$ allowed us to eliminate $e^{\Phi}$ from $P$.  Furthermore, $\ap \dd^\dag P = \cO(\ap^3)$. 

We can further simplify the divergence of the Chern class
\beq\label{eq:divTrF2}
\begin{split}
 \frac{\ap^2}{2}  &  \nabla^{\Bi\,m}\Big(\tr (F\w F)_m\Big) \=  \ap^2 \tr \left( \nabla_A^{m} F\w F_m + F\w \nabla_A^{m} F_m\right)\\[5pt]
 & \=\ap^2 \tr ( \dd_A F_m\w F^m)  +\cO(\ap^3)  \=\frac{\ap^2}{2} ( \delb - \del) \tr (F^\r \w F_\r) ~,
\end{split}
\eeq
 where we use $\dd_A F =  0$, the $F$ equation of motion (imposed via the gaugino variation). Repeat for the corresponding term in $R^\H$, which is straightforward due to $\nabla^m R^\H_m = \cO(\ap)$. This follows both directly from the BdR supersymmetry algebra \eqref{eq:RhdH}; and also hermitian geometry with the conformally balanced equation: \eqref{eq:R-inst} with \eqref{eq:RHRCh}, \eqref{eq:RH02}.  
 
 Consequently, 
 \beq\label{eq:apP}
 \ap P \= \frac{\ap^2}{8} (\delb -\del)\Big( \tr (F^\r \w F_\r) - \tr (R^{\H\,\r} \w R^\H_\r) \Big) ~.
 \eeq

We need some results for $\dd H$ and its trace. The trace of $(\dd H)^{3,1}$: 
\beq\label{eq:dH31}
\begin{split}
( \dd H)^{3,1}&\= \frac{1}{3!} (\dd H)_{\m\n\r\sb} \dd x^{\m\n\r\sb} \= \delb H^{3,0} + \del H^{2,1} \\
&\=  \frac{1}{3!} \left(- \nabla_\sb H_{\m\n\r} +  \nabla_\m H_{\n\r\sb} +  \nabla_\n H_{\r\m\sb} +  \nabla_\r H_{\m\n\sb}\right) \dd x^{\m\n\r\sb} ~.
\end{split}
\eeq 
Hence, 
\beq\label{eq:dHtr}
\begin{split}
(\dd H)_{\m\n\r}{}^\r  
&\=  \left(- \nabla^\r H_{\m\n\r} +  \nabla^\sb H_{\m\n\sb} -  \nabla_\m H{}_\n{}^\r{}_{\r}{} +  \nabla_\n H{}_\m{}^\r{}_{\r} \right)~.
\end{split}
\eeq
The dilatino equation \eqref{eq:dilatino2} with  \eqref{eq:wtHhatH} implies $\nabla_\n H_\m{}^\r{}_\r   = 2\nabla_\n \nabla_\m \Phi + 3\ap \nabla_\n P_\m{}^\r{}_\r$. 
Hence, with \eqref{eq:trP}, 
\beq\notag
\nabla_\n H_\m{}^\r{}_\r  - \nabla_\m H_\n{}^\r{}_\r   \= 0~.
\eeq

The dilatino equation \eqref{eq:dilatino2} implies $\hat H_{\m\n\r}=H_{\m\n\r}-3\ap P_{\m\n\r}=0$. Since $\ap P^{3,0}=\cO(\ap^3)$, this gives $H_{\m\n\r}=\cO(\ap^3)$. The Levi-Civita connection acts by
 $$
\nabla_{\bar{\sigma}} H_{\mu \nu \rho} \= \partial_{\bar{\sigma}} H_{\mu \nu \rho} - \Gamma_{\bar{\sigma}}{}^i{}_\mu H_{i \nu \rho} - \Gamma_{\bar{\sigma}}{}^i{}_\nu H_{\mu i \rho}  - \Gamma_{\bar{\sigma}}{}^i{}_\rho H_{\mu \nu i} 
 $$
and so $H_{\m\n\r} = \cO(\ap^3)$ does not imply $\nabla_\sb H_{\m\n\r}=\cO(\ap^3)$. Using \eqref{eq:LCsymbols}, we see that 
$$
-\nabla_\sb H_{\m\n\r} \dd x^{\m\n\r} \= \frac{3}{2} H_\sb{}^\lb{}_\m H_{\lb \n\r} \dd x^{\m\n\r}~, \qquad -\nabla^\r H_{\m\n\r} \= H^{\r\lb}{}_{[\m} H_{\n]\r\lb} - \half H^\r{}_{\r\l} H_{\m\n}{}^\l~.
$$
Together with the $H$ equation of motion and an application of dilatino equation, we find  
\beq\label{eq:divHdH}
\begin{split} 
(\dd H)_{\m\n\r}{}^\r  &\=  -2\nabla^\r H_{\m\n\r} +  \nabla^p H_{\m\n p} \=   -2\nabla^\r H_{\m\n\r} +  2(\nabla^p \Phi) H_{\m\n p}\\
&\= 2H^{\r\lb}{}_{[\m} H_{\n] \r\lb} ~.
\end{split}
\eeq

Now, we compute the trace of $(\dd H)^{2,2} = \delb H^{2,1} + \del H^{1,2}$. It is useful to record the form in components:
$$
(\dd H)_{\m\n\sb\rb}\= \left( \nabla_\sb H_{\m\n\rb} -  \nabla_\rb H_{\m\n\sb} \right)  + \left( \nabla_\m H_{\n\sb\rb} - \nabla_\n H_{\m\sb\rb} \right) ~.
$$
The trace is
\beq\label{eq:dHtrmixed}
\begin{split}
 (\dd H)_{\n}{}^\n{}_{\m\sb} &\=  \nabla_\sb H_\m{}^\n{}_{\n}    -\nabla_\m H_\sb{}^\n{}_{\n} -  \nabla^\n H_{\n\m\sb} + \nabla^\nb H_{\nb\m\sb}  ~.
\end{split}
\eeq 
The dilatino equation \eqref{eq:dilatino2} and \eqref{eq:trP} implies
\beq\notag
\begin{split}
 \nabla_\sb H_\m{}^\r{}_\r   &\= 2\nabla_\sb \nabla_\m \Phi + 3\ap \nabla_\sb P_\m{}^\r{}_\r \= \nabla_\sb \nabla_\m\left(2  \Phi + \frac{3\ap^2}{16}  \left( \tr|F|^2 - \tr |R^\H|^2 \right)\right)~,\\
\end{split}
\eeq
with $\nabla_\m H_\sb{}^\r{}_\r$ following by complex conjugation.

Putting this into \eqref{eq:dHtrmixed} gives
\beq\label{eq:dHtrmixed2}
(\dd H)_{\n}{}^\n{}_{\m\sb}\=4\nabla_\m\nabla_\sb\Phi-\nabla^\n H_{\n\m\sb}+\nabla^\nb H_{\nb\m\sb}+\cO(\ap^2)~.
\eeq

It is also useful to trace the Bianchi identity \eqref{eq:Bianchi2}. Using $F^{0,2}=F_\m{}^\m= \cO(\ap^2)$ with $R^\H{}^{0,2} = R^\H{}_\m{}^\m=\cO(\ap)$ and \eqref{eq:gaugino3}, 
\beq\label{eq:trdHtrF}
\begin{split}
  (\dd H)_{\n}{}^\n{}\=  (\dd H)_{\n}{}^\n{}_{\m\sb} \dd x^{\m\sb}  &\= \frac{\ap}{2} (\tr F^\n\w F_{\n} - \tr R^\H{}^\n \w R^\H_\n  )  + \cO(\ap^2)~,
\end{split}
\eeq

Substituting the traced Bianchi identity \eqref{eq:trdHtrF} into \eqref{eq:apP} gives
\beq\label{eq:apP2}
\begin{split}
\ap P&\=\frac{\ap}{4}(\delb-\del)(\dd H)_\n{}^\n\\
&\=\frac{\ap^2}{8}(\delb-\del)\left(\cE^\H_{\m\nb}\dd x^{\m\nb}\right)+\cO(\ap^3)~.
\end{split}
\eeq
Thus  $\ap P$ does not vanish through $\ap^2$; instead, its  contribution is absorbed by modifying the hermitian form, \eqref{eq:omega-field-redefinition}, giving
\beq\label{eq:Hreincarnated}
H\=\ii(\del-\delb)\widehat\o+\cO(\ap^3)~.
\eeq
Recall the scalar quantity $h=\tr|F|^2-\tr|R^\H|^2$.The double trace of the Bianchi identity gives
\beq
 -\frac{\ap}{2}h=\L^2\dd H+\cO(\ap^2)=\cO(\ap)~, 
\eeq
and so $h=\cO(1)$. The trace and divergence of $P$ therefore satisfy
\beq\label{eq:trP}
 \ap P^\b{}_{\b m}\dd x^m=\frac{\ap^2}{16}(\del-\delb)h+\cO(\ap^3)~,\qquad \ap\nabla^pP_{pab}=\cO(\ap^3)~.
\eeq
We have proved
\beq
H\=\ii(\del-\delb)\widehat\o+\cO(\ap^3)~. 
\eeq

\subsubsection{Double holomorphic direction}
The graviton equations of motion are
\beq\notag
  \Ric_{mn}+2\nabla_m\nabla_n\Phi-\qrt H_{mpq}H_n{}^{pq} + \frac{\ap}{4} \Bigl(\tr(F_{mp}F_n{}^p)-\tr(R^\H_{mp}R^\H_n{}^p)\Bigr)+\Delta^{\rm conn}_{mn}=\cO(\ap^3)~.
\eeq
Using 
\beq
\label{eq:E-trace-divergence}
 \half g^{mn}\cE^\H_{mn}\=h+\cO(\ap)~,\qquad \nabla^p\cE^\H_{mp}=\frac12\nabla_mh+\cO(\ap)~,
\eeq
we can evaluate \eqref{eq:Delta-connection} using $\Ric_{mn} = \cO(\ap)$ and
\beq
 \Delta^{\rm conn}_{mn}=\frac{\ap^2}{16}\nabla_m\nabla_nh-\frac{\ap^2}{8}\left(\nabla^p\nabla_p\cE^\H_{mn}+R_{mpnq}\cE^{\H\,pq}\right)+\cO(\ap^3)~.
\eeq
In complex coordinates, 
\beq
\label{eq:Delta-connection-unfixed-dilaton}
\begin{split}
 \Delta^{\rm conn}_{\m\n}
 &=\frac{\ap^2}{16}\nabla_\m\nabla_\n h+\cO(\ap^3)~,\\
 \Delta^{\rm conn}_{\n\mb}
 &=-\frac{\ap^2}{4}\nabla^\r\nabla_\r\cE^\H_{\n\mb}
 -\frac{\ap^2}{8}R_{\n\mb}{}^{\r\sb}\cE^\H_{\r\sb}
 +\frac{\ap^2}{16}\nabla_\n\nabla_\mb h
 +\cO(\ap^3)~.
\end{split}
\eeq

We put $m=\m$ in \eqref{eq:integrab5}:
\beq\label{eq:integrab5b}
\begin{split}
& \left(\Ric_{\m\n} +   2\nabla_\m \nabla_\n \Phi - \qrt  H_{\m ab}  H_{\n}{}^{ab} + \frac{\ap}{4}\tr  F_\m{}^a F_{\n a} -  \frac{\ap}{4}\tr  R^\H{}_\m{}^a R^\H{}_{\n a} \right) \g^\n\eta  \\[6pt]
& +   \half e^{2\Phi} \nabla^p (e^{-2\Phi}  H_{p\m\n} ) \g^\n \eta \=   \frac{\ap}{4} \tr\left(       \frac{1}{4} (R^\H \w R^\H)_{\m\n \a\bb}  \g^{\n \a\bb} \eta - \left(R^\H{}_\m{}^a R^\H{}_{\n a} \right) \g^\n\eta  \right)\\
&\qquad\qquad +\frac{\ap}{6}
\left((\dd P)_{\m nab}-\nabla^\LC_\m P_{nab}\right)
\g^{nab}\eta~.
\end{split}\raisetag{2.5cm}
\eeq
The left hand side is the  graviton equations of motion to first order in $\ap$ and the $H$ equation of motion. The right hand side are the $\ap^2$-corrections. The first is higher order
\beq\notag
\begin{split}
\ap  \tr  \left(R^\H{}_\m{}^a R^\H{}_{\n a} \right) \g^\n\eta  -  \frac{\ap}{4}\Big(\tr (R^\H \w R^\H)_{\m\n \a\bb} \Big) \g^{\n \a\bb} \eta  \=  \ap \tr (R^\H_{\m\n} R^\H{}_\a{}^\a )\g^\n \eta \= \cO(\ap^3) ~.
\end{split}
\eeq
We have used that $\tr R^\H \w R^\H$ has no $(4,0)$ component, that 
$$
\half \tr (R^\H\w R^\H)_{\m\n\a}{}^\a = \tr( R^\H{}_{\m\n}  R^\H{}_{\a}{}^\a -  R^\H{}_{\m}{}^{\ab} R^\H{}_{\n}{}_\ab -  R^\H{}_{\m}{}^\a R^\H{}_{\a \n}) ~,
$$
as well as $R^\H{}^{0,2} = \cO(\ap)$ from \eqref{eq:RH02} and $R^\H_\a{}^\a = \cO(\ap)$ from \eqref{eq:RhdH}. The terms involving $P$  give an $\ap^2$ correction:
\beq
\frac{\ap}{6}\Bigl((\dd P)_{\m n ab}-\nabla^\LC_\m P_{nab}\Bigr)\g^{nab}\eta\=\!-\frac{\ap^2}{16}\nabla_\m\nabla_\n h\,\g^\n\eta+\cO(\ap^3)
 \=\!-\Delta^{\rm conn}_{\m\n}\g^\n\eta+\cO(\ap^3)~,
\eeq

Hence, \eqref{eq:integrab5b} simplifies to
\beq\label{eq:integrab6}
\begin{split}
& \left(\Ric_{\m\n} +   2\nabla_\m \nabla_\n  \Phi - \qrt  H_{\m ab}  H_{\n}{}^{ab} + \frac{\ap}{4}\tr  F_\m{}^a F_{\n a} -  \frac{\ap}{4}\tr  R^\H{}_\m{}^a R^\H{}_{\n a}   +\D_{\m\n}^{\rm conn} \right) \g^\n\eta\= 0~. \\[5pt]
 \end{split}
\eeq

\subsubsection{Mixed component}
For $m=\mb$, using
$$
\left(R^\H_\a{}^\a\right)_{\r\sb} \= \half g^{\m\nb}(\dd H)_{\m\nb\r\sb} \= \frac{\ap}{4}\cE^\H_{\r\sb}+\cO(\ap^2),
$$
we find
\beq\label{eq:trRmixed}
\begin{split}
-\frac{\ap}{4}\tr\left( R^\H_{\mb\n}R^\H_\a{}^\a \right)\g^\n\eta
 &\=\! -\frac{\ap^2}{8}R_{\n\mb}{}^{\r\sb}\cE^\H_{\r\sb}\g^\n\eta+\cO(\ap^3)~.
\end{split}
\eeq
The $P$ term is
\beq
\label{eq:P-graviton-mixed-unfixed}
\begin{split}
 &\frac{\ap}{6}\left((\dd P)_{\mb n ab}-\nabla^{\LC}_{\mb}P_{nab}\right)\g^{nab}\eta\\
 &\qquad=\left[
 -\frac{\ap^2}{4}g^{\r\sb}
 \left\{\delb\del\left(\cE^\H_{\l\tb}\,\dd x^{\l\tb}\right)\right\}_{\r\sb\n\mb}
 -\frac{\ap^2}{16}\nabla_\mb\nabla_\n h
 \right]\g^\n\eta+\cO(\ap^3)\\
 &\qquad=\left[
 -\Delta^{\rm conn}_{\n\mb}
 +\frac{\ap^2}{8}R_{\n\mb}{}^{\r\sb}\cE^\H_{\r\sb}
 \right]\g^\n\eta+\cO(\ap^3)~.
\end{split}
\eeq
In passing to the third line we have used the trace and divergence identities in \eqref{eq:E-trace-divergence} and
\beq
\notag
\begin{split}
 &g^{\r\sb}\left\{\delb\del\left(\cE^\H_{\l\tb}\,\dd x^{\l\tb}\right)\right\}_{\r\sb\n\mb}
\=\!-\nabla^\r\nabla_\r\cE^\H_{\n\mb}
 -R_{\n\mb}{}^{\r\sb}\cE^\H_{\r\sb}
 +\cO(\ap)~.
\end{split}
\eeq
Adding \eqref{eq:trRmixed} and \eqref{eq:P-graviton-mixed-unfixed}, we find $ -\Delta^{\rm conn}_{\n\mb}\g^\n\eta$ and so \eqref{eq:integrab5} with $m=\mb$ is
\beq\label{eq:integrab6b}
\begin{split}
& \left(\Ric_{\mb\n} +   2\nabla_\mb \nabla_\n  \Phi - \qrt  H_{\mb ab}  H_{\n}{}^{ab} + \frac{\ap}{4}\tr  F_\mb{}^a F_{\n a} -  \frac{\ap}{4}\tr  R^\H{}_\mb{}^a R^\H{}_{\n a}   +\D_{\mb\n}^{\rm conn} \right) \g^\n\eta\= 0~. 
 \end{split}
\eeq

\subsection{Summary}
Equations  \eqref{eq:integrab6} and \eqref{eq:integrab6b} give the graviton equation of motion \eqref{eq:RicEOM}. Thus, assuming a smooth $\ap\to0$ limit and the Bianchi identity, supersymmetry implies the $H$, $F$ and graviton equations of motion through $\ap^2$. In~\sref{s:eomgeometry} we check this using complex geometry and derive the dilaton equation of motion.
%
%

The supersymmetry equations \eqref{eq:introconf} and resolved Bianchi identity \eqref{surprising-bianchi}  were conveniently expressed in the $\widehat g, \widehat\omega,\widehat\Phi$ field basis. We now express the equations of motion in this basis. Using the Palatini formula and the trace and divergence identities for
$\cE^\H$, the metric redefinition gives
\beq
\begin{split}
 \Ric(\widehat g)_{\m\n}-\Ric(g)_{\m\n}
 &=-\frac{\ap^2}{8}\nabla_\m\nabla_\n h+\cO(\ap^3)~,\\
 \Ric(\widehat g)_{\n\mb}-\Ric(g)_{\n\mb}
 &=-\frac{\ap^2}{4}\nabla^\r\nabla_\r\cE^\H_{\n\mb}
 -\frac{\ap^2}{4}R_{\n\mb}{}^{\r\sb}\cE^\H_{\r\sb}
 -\frac{\ap^2}{8}\nabla_\n\nabla_\mb h
 +\cO(\ap^3)~.
\end{split}
\eeq
Since
\beq
 2\nabla_m\nabla_n(\widehat\Phi-\Phi)
 =\frac{3\ap^2}{16}\nabla_m\nabla_n h+\cO(\ap^3)~,
\eeq
comparison with \eqref{eq:Delta-connection-unfixed-dilaton} gives
\beq
\begin{split}
 \Delta^{\rm conn}_{\m\n}
 &=\Ric(\widehat g)_{\m\n}-\Ric(g)_{\m\n}
 +2\nabla_\m\nabla_\n(\widehat\Phi-\Phi)
 +\cO(\ap^3)~,\\
 \Delta^{\rm conn}_{\n\mb}
 &=\Ric(\widehat g)_{\n\mb}-\Ric(g)_{\n\mb}
 +2\nabla_\n\nabla_\mb(\widehat\Phi-\Phi)
 +\frac{\ap^2}{8}
 R_{\n\mb}{}^{\r\sb}\cE^\H_{\r\sb}
 +\cO(\ap^3)~.
\end{split}
\eeq
Hence, the graviton equations of motion become
\beq
\label{eq:graviton-hatted-unfixed-dilaton}
\begin{split}
 0={}&\Ric(\widehat g)_{\m\n}
 +2\nabla_\m\nabla_\n\widehat\Phi
 -\frac14H_{\m ab}H_\n{}^{ab}
 +\frac{\ap}{4}\cE^\H_{\m\n}
 +\cO(\ap^3)~,\\
 0={}&\Ric(\widehat g)_{\n\mb}
 +2\nabla_\n\nabla_\mb\widehat\Phi
 -\frac14H_{\n ab}H_\mb{}^{ab}
 +\frac{\ap}{4}\cE^\H_{\n\mb}
 +\frac{\ap^2}{8}\widehat R^\H_{\n\mb}{}^{\r\sb}\cE^\H_{\r\sb}
 +\cO(\ap^3)~.
\end{split}
\eeq
As the difference between $g$ and $\widehat g$ is of order $\ap^2$, the covariant derivatives in the Hessian may be taken with either $\nabla$ or $\widehat\nabla$ at the expense of $\ap^3$ corrections.
The simultaneous metric and dilaton redefinition absorbs most of $\Delta^{\rm conn}$ but leaves an explicit $\ap^2$ correction. In real coordinates, the hatted variables obey the graviton equation of motion:
\beq\label{eq:graviton_hatted}
\Ric_{mn}(\widehat g) +2 \nabla_m \nabla_n\widehat\Phi -\frac14H_{mpq}H_n{}^{pq} +\frac{\ap}{4}{\cE}^{\H}_{mn} +\frac{\ap^2}{8} R^\H_{mpnq}{\cE}^{\H\,pq} +\cO(\ap^3) \= 0 ~.
\eeq
This change of variables is one-to-one, so we have proved that supersymmetry  in terms of the hatted variables \eqref{eq:introconf} implies the equations of motion in the hatted variables.

{\bf Remark}: Even in the hatted variables, there is still an explicit $\ap^2$ correction to the graviton equation of motion in this basis. This is despite \eqref{eq:introconf} not containing any explicit $\ap^2$ terms. In the next section we demonstrate that results from hermitian geometry, supersymmetry and the graviton equation of motion imply the dilaton equation. 

In contrast to e.g. \cite{delaOssa:2014msa, Martelli:2010jx, garcia2023, Ivanov:2009rh, LopesCardoso:2003dvb}, no instanton condition is required on $R$. Supersymmetry implies all the equations of motion without any auxiliary conditions.  Suppose one nevertheless imposes the tangent-bundle instanton condition.  The trace relation \eqref{eq:RhdH} first gives
$$
\Lambda\dd H\=\cO(\ap^2)~.
$$
through $\ap$, $\dd H$ has type $(2,2)$, and in complex dimension three
$\Lambda:\Omega^{2,2}\longrightarrow\Omega^{1,1}$ is an isomorphism. Hence
$\dd H=\cO(\ap^2)$. Compactness, together with the first-order flux--torsion and conformally balanced equations, then gives $H=\cO(\ap^2)$. Repeating the argument at the next order yields
$$
\dd H\=\cO(\ap^3),\qquad H\=\cO(\ap^3).
$$
Thus the instanton condition eliminates torsion up to and including order $\ap^2$ rather than supplying a missing equation of motion.
 
 \section{Equations of motion from hermitian geometry}
\label{s:eomgeometry}

\subsection{Statement of results}

In Section \ref{sec:susyalg}, we started from constraints on a spinor $\eta$ on a compact 6-manifold $X$ obtained by setting fermionic supersymmetry variations to zero, and from these equations on spinors we found a complex structure, a holomorphic volume form, and constraints for the metric tensor intertwining these structures at $\ap^2$.

This section tells a self-contained mathematical story based on our findings in earlier sections. The setup is as follows:

 \begin{enumerate}
 \item Let $X$ be a compact manifold of dimension 6. Let $(\widehat{g}_\ap, J_\ap, \Omega_\ap, F_\ap)$ be a family of tensors on $X$ smoothly varying with a parameter $\ap \in [0,\epsilon)$ for $\epsilon>0$ fixed, where $\widehat{g}_\ap$ is a metric tensor, $J_\ap$ is a complex structure, $\Omega_\ap$ is a holomorphic volume form, and $F_\ap$ is the curvature of a connection on a vector bundle $E \rightarrow X$. For example,
  \beq \notag
\widehat{g}_\ap \= \widehat{g}^{(0)} + \ap \widehat{g}^{(1)} + \ap^2 \widehat{g}^{(2)} + \dots~,
  \eeq
  and $\widehat{g}^{(0)}$ is a metric tensor.

\item Suppose $(\widehat{g}_\ap, J_\ap, \Omega_\ap, F_\ap)$ solves the following equations of complex geometry:
\begin{align}
  d  ( \norm{\Omega}_{\widehat g} \, \widehat\omega^2) &\= 0 \label{eq:holconfbal}\\
  F^{0,2} =F^{2,0} &\= 0~, \\
  F \wedge \widehat\omega^2 &\=  0 \label{eq:Fo2}~,\\
2 \ii \partial \bar{\partial} \widehat\omega + \frac{\ap}{4} [ \tr F \wedge F - \tr \widehat R^{\Ch} \wedge \widehat R^{\Ch} ] &\= 0 \label{eq:iddbo}~,
\end{align}
where $\widehat{\omega} = \widehat{g} (J \cdot, \cdot)$. Here and in what follows, we will often drop the subscript $\ap$ on tensors and simply write $\widehat{g} = \widehat{g}_{\ap}$, $\Omega=\Omega_\ap$, etc.

\item We will also use the notation
\begin{align}
  H &\= \ii (\partial-\bar{\partial}) \widehat\omega ~, \label{eq:truncatedH}  \\
  \widehat\Phi &\=\! - \half  \log \norm{\Omega}_{\widehat g}   ~. \label{eq:truncatedPhi}
\end{align}

For this section, we regard these equations as an exact mathematical system. Their identification with the BdR system is valid through $\cO(\ap)$. Earlier in the paper, we discussed how the physical fields $g$ and $\Phi$, which receive $\ap^2$-corrections and beyond, are determined from the hatted $\widehat{g}$ and $\widehat{\Phi}$ up to $\cO(\ap^3)$ corrections, and how the equation for $\ii \partial \bar{\partial} \widehat{\omega}$ given in \eqref{eq:iddbo} is consistent with the Bianchi identity $dH = \frac{\ap}{4} ( \tr FF - \tr R^\H R^\H)$ up to $\mathcal{O}(\ap^3)$ corrections; see Section \ref{sec:ddbar-ap2}.
\end{enumerate}
 
From a quadruple $(\widehat{g},\widehat{\Phi},H,F)$ solving the complex geometric system, we define
\beq \label{eq:unhatting}
\begin{split}
  g_{\mu \bar{\nu}} &\= \widehat{g}_{\mu \bar{\nu}} - \frac{\ap^2}{4} \bigg( \tr F_{\mu p} F_{\bar{\nu}}{}^p - \tr \widehat{R}^\H{}_{\mu p} \widehat{R}^\H{}_{\bar{\nu}}{}^p \bigg) ~,\\
  \Phi &\= \widehat{\Phi} - \frac{3 \ap^2}{32} \left( \tr |F|^2 - \tr | \widehat{R}^\H|^2 \right)~.
  \end{split}
\eeq
We will verify that $(g,\Phi,H,F)$ solves the equations of motion at order $\ap^2$.
\begin{equation}
\label{EOM-ap3}
\begin{aligned}
&\Ric_{mn}
 + 2 \nabla_m \nabla_n \Phi
 - \frac{1}{4} H_{mab} H_n{}^{ab} + \frac{\ap}{4}
   \Bigl(
      \tr F_{mr}F_n{}^r
      - \tr R^\H{}_{mr} R^\H{}_n{}^r
   \Bigr)
\\[-2pt]
&\qquad
-\frac{\ap}{4}e^{2\Phi} \nabla^p \Bigl( e^{-2\Phi}\left(
  \nabla_m\cG_{np}  +\nabla_n\cG_{mp}  -2\nabla_p\cG_{mn}  \right) \Bigr)
 \= \cO(\ap^3) ~,
\\[6pt]
&\nabla^m\bigl(\ee^{-2\Phi}H_{mnp}\bigr)
 \= \cO(\ap^3) ~,
\\[6pt]
&\ap\,\ccD^{-\,m}\bigl(\ee^{-2\Phi}F_{mn}\bigr)
 \= \cO(\ap^3) ~,
\\[6pt]
&R
 - 4\lvert\nabla\Phi\rvert^2
 + 4\nabla^2\Phi
 - \half\lvert H\rvert^2 + \frac{\ap}{4}
   \Bigl(
      \tr\lvert F\rvert^2
      - \tr\lvert R^\H\rvert^2
   \Bigr)
 \= \cO(\ap^3) ~.
\end{aligned}
\end{equation}
These equations were discussed in Section \ref{sec:eom}. The second line of the graviton equation, which is an $\ap^2$-level correction, uses the notation $\cG_{mn} = \Ric_{mn} + 2 \nabla_m \nabla_n \Phi  - \frac{1}{4} H_{mab} H_n{}^{ab}$. Recall that $R^\H$ is defined by $\Gamma_m^\H = \Gamma^\LC_m + \half H_m$ and $\ccD^-$ acts as $\Gamma_m^- = \Gamma^\LC_m - \half H_m$ on tangent bundle indices. In the current section, we simply write $\nabla$ instead of $\nabla^{\LC}$ for the Levi-Civita connection of $g$.

\begin{prop} \label{prop:herm2eom}
  Under the setup assumptions denoted above by 1, 2, and 3, the quadruple $(g,\Phi,H,F)$ solves the equations of motion \eqref{EOM-ap3}.
\end{prop}

We make three remarks on this Proposition:

\begin{itemize}
\item Section \ref{sec:integrability} analyzed the relation between the supersymmetry constraints on spinors and the equations of motion. The current section gives a second perspective by working only at the level of complex geometry and making no mention of spinors. 

\item Some other studies of heterotic supersymmetry and the equations of motion add the extra hypothesis that $R$ is an instanton e.g. \cite{delaOssa:2014msa, Martelli:2010jx, garcia2023, Ivanov:2009rh, LopesCardoso:2003dvb}, but this condition does not play a role in the proof of Proposition \ref{prop:herm2eom}. We remark in passing that this instanton condition is not consistent with supersymmetry at $\ap^2$ as explained by Bergshoeff--de Roo \cite{Bergshoeff:1989de}. This is because the condition
\beq\notag
g^{\mu \bar{\nu}} R_{\m\nb}+ \cO(\ap) \=0~, \qquad R_{\mb\nb} + \cO(\ap) \= 0~,
\eeq
which automatically holds for $R=R^\H$ or $R=R^{\Ch}$ under setup assumptions 1, 2 and 3, is refined in the BdR algebra at next order in $\ap$ to
\beq\notag
\left(R^\H_{\m\nb}{}^{\a\bb}- \frac{1}{2} (\dd H)_{\m\nb}{}^{\a\bb}\right)g^{\m\nb} + \cO(\ap^2) \=0~, \qquad R^\H_{\mb\nb}- \frac{1}{2} (\dd H)_{\mb\nb} + \cO(\ap^2) \= 0~.
\eeq
See Section \ref{sec:BdR} for more on this. This refinement is automatically implemented by setup assumptions 1, 2 and 3; see \eqref{eq:RH-inst}. Thus, once $\ap^2$-corrections are included in the supersymmetry constraints, the curvature $R$ of the tangent bundle fails to be an instanton. 

  \item Sequences solving assumptions 1 and 2 have been constructed by various methods; see \cite{liyau2005, andreas2012, Melnikov:2014ywa, collins2022}.

  \end{itemize}

In the proof of Proposition \ref{prop:herm2eom},  we will frequently use 
   \beq\notag
H \= \cO(\ap), \quad \dd \widehat\omega \= \cO(\ap), \quad \dd \widehat\Phi = \cO(\ap)~.
\eeq
These conditions are implied by our setup assumptions 1, 2, and 3, as proved in Section \ref{sec:orderzero}. Lastly, a comment about notation. In this section, we will distinguish equality up to error terms of order $\cO(\ap^3)$ and exact equality. We use the notation
  \beq\notag
A \overset{\ \ap^2}{=} B \quad \Leftrightarrow \quad A \= B + \cO(\ap^3)~.
\eeq
This is to separate identities in complex geometry from statements on the $\ap$-expansion.

  \subsection{Yang-Mills equation}
 It is well-known (see e.g. \cite{Gillard:2003jh}) that the equations
  \beq\notag
  d  (e^{-2
    \widehat \Phi} \widehat{\omega}^2) \=0,~ \quad \widehat{g}^{\mu \bar{\nu}} F_{\mu \bar{\nu}}\= 0, \quad F^{0,2} \= F^{2,0} \= 0,~
  \eeq
  imply
  \beq \label{eq:nonKahHYM}
  \widehat\ccD^{-\,m} (\ee^{-2 \widehat\Phi} F_{mn}) \= 0~,
  \eeq
  where
  $\widehat\Gamma^- = \widehat\Gamma^\LC - \half i (\partial-\bar{\partial}) \widehat\omega$.  We outlined the derivation earlier in Section \ref{sec:HFeom}.  As $g$, $\Phi$ differ from $\widehat g$, $\widehat \Phi$ up to terms of order $\ap^2$, we obtain
  \beq \notag
 \ap \ccD^{-\,m} (\ee^{-2 \Phi} F_{mn}) \overset{\ \ap^2}{=} 0~.
  \eeq
  
\subsection{Divergence of the 3-form}
\label{s:HEOM}
Next, we note that $d (e^{-2 \widehat \Phi} \widehat \omega^2) = 0$ implies
\beq \label{eq:div-H}
\widehat\nabla^m(\ee^{-2 \widehat\Phi} H_{mij}) \= 0~.
\eeq
This is well-known, and we give the proof from \cite{Martelli:2010jx, Gauntlett:2003cy}. The general identity goes as follows: let $\omega$ be a conformally balanced metric on a complex manifold $X$ of dimension 3 satisfying $d (e^{-2 \Phi} \omega^2) = 0$ for some scalar function $\Phi$. Then
\beq \label{eq:starT}
\star d^c \omega \= -  e^{2 \Phi} \dd (e^{-2 \Phi} \omega)~, \quad \dd \star (e^{-2 \Phi} d^c \omega) \= 0~, \quad d^c \= i (\partial -\bar{\partial}) ~.
\eeq
To show \eqref{eq:starT}, we can first use the primitive decomposition to derive the general formula
\beq \label{eq:genstar}
\star \chi \= J \chi - J(\Lambda_\omega \chi) \wedge \omega~, \quad \chi \in \Omega^3(X)~.
\eeq
Here $J$ acts on differential forms by dual pairings, so that for example in holomorphic coordinates $J \dd x = \ii \dd x$ and $J \dd \bar{x} = - \ii \dd \bar{x}$. Equation \eqref{eq:starT} then follows from letting $\chi = d^c \omega$ in \eqref{eq:genstar} and using $\Lambda_\omega d \omega = 2 d \Phi$ \eqref{def:confbal}. This proves \eqref{eq:div-H}. We now remove hats. 
 \beq \notag
 \nabla^m(\ee^{-2 \Phi} H_{mij}) \overset{\ \ap^2}{=} 0~.
  \eeq
This is because $H= \cO(\ap)$ and hat removal happens at order $\ap^2$.

\subsection{Graviton equation}
In this subsection, we show 
\beq \label{eq:einstein}
\widehat\Ric_{mn}+ 2 \widehat\nabla_m \widehat\nabla_n \widehat\Phi - \frac{1}{4} H_{mab} H_n{}^{ab}
+ \frac{\ap}{4} \cE_{mn} 
+ \frac{\ap^2}{8} R^\H{}_{m}{}^p{}_n{}^q \cE_{pq}
 \overset{\ \ap^2}{=} 0~,
\eeq
where
\beq
\cE_{mn} \= \tr F_{mp} F_n{}^p - \tr R^\H{}_{mp} R^\H{}_n{}^p ~.
\eeq
We explained previously in the paper how \eqref{eq:einstein} implies the graviton equation of motion, which is the first equation listed in \eqref{EOM-ap3}. We will recall this interpretation of \eqref{eq:einstein} later in Section \ref{s:riccihat}, but first, we assume setup assumptions 1, 2, and 3 and derive \eqref{eq:einstein}.

\subsubsection{Preliminaries}
Before starting the calculation of $\widehat\Ric$, we note a few identities. First, the conformally balanced equation \eqref{eq:holconfbal} gives via \eqref{def:confbal} the identity
\beq \label{eq:trT-improved}
H_\mu{}^\mu{}_\lambda \=\! - 2 \partial_\lambda \widehat\Phi~, \quad H_\mu{}^\mu{}_{\bar{\lambda}} \=  2 \partial_{\bar{\lambda}} \widehat\Phi~.
\eeq
Second, in the presence of a holomorphic volume form, the Chern curvature satisfies
\beq \notag
\widehat{R}^{\rm Ch}{}_{\alpha \bar{\beta}}{}^\mu{}_\mu \= \partial_\alpha \partial_{\bar{\beta}} \log \norm{\Omega}^2_{\widehat\omega} ~,
\eeq
and hence
\beq \label{eq:trChern}
\widehat{R}^{\rm Ch}{}_{\alpha \bar{\beta}}{}^\mu{}_\mu \= - 4 \partial_\alpha \partial_{\bar{\beta}} \widehat\Phi ~.
\eeq
Third, the Hermitian--Yang--Mills and conformally balanced equations imply the identity
\beq \label{eq:trFFRR1}
  \begin{split}
    & \ \frac{\ap}{8} (\tr F \wedge F - \tr \widehat{R}^\Ch \wedge \widehat{R}^\Ch)_\mu{}^\mu{}_{\bar{\beta} \alpha} \\
    &\= -  \frac{\ap}{4} \tr \bigg[ F_{\alpha \bar{\mu}} F_{\bar{\beta}}{}^{\bar{\mu}} - \widehat{R}^\Ch{}_{\alpha \bar{\mu}} \widehat{R}^\Ch{}_{\bar{\beta}}{}^{\bar{\mu}} \bigg] - \frac{\ap}{4} (\ii \Lambda_\omega \ii \partial \bar{\partial} \widehat{\omega})^m{}_n \widehat{R}^\Ch{}_{\alpha \bar{\beta}}{}^n{}_m \\
    &\ + \frac{\ap}{4} H^{\mu \bar{\nu} n} H_{m \mu \bar{\nu}} \widehat{R}^\Ch{}_{\alpha \bar{\beta}}{}^m{}_n ~.
  \end{split}
  \eeq
 To show \eqref{eq:trFFRR1}, we start with
\beq\notag
    (\tr F \wedge F)_{\mu \bar{\nu} \alpha \bar{\beta}} \= 2 \tr F_{\mu \bar{\nu}} F_{\alpha \bar{\beta}} - 2 \tr F_{\mu \bar{\beta}} F_{\alpha \bar{\nu}}~,
 \eeq
  which combined with $\widehat{g}^{\mu \bar{\nu}} F_{\mu \bar{\nu}}=0$ gives
   \beq \notag
(\tr F \wedge F)_{\mu}{}^\mu{}_{\bar{\beta} \alpha} \=  2 \tr F_{\mu \bar{\beta}} F_{\alpha}{}^\mu~.
\eeq
Similarly,
  \beq\notag
- (\tr \widehat{R}^\Ch \wedge \widehat{R}^\Ch)_{\mu}{}^\mu{}_{\bar{\beta} \alpha} \= 2  \widehat{g}^{\mu \bar{\nu}} \tr \widehat{R}^\Ch{}_{\mu \bar{\nu}} \widehat{R}^\Ch{}_{\alpha \bar{\beta}}- 2  \widehat{g}^{\mu \bar{\nu}} \tr \widehat{R}^\Ch{}_{\mu \bar{\beta}} \widehat{R}^\Ch{}_{\alpha \bar{\nu}}~.
\eeq
Thus
\beq \label{eq:trFFRR0}
 \begin{split}
    & \ \frac{\ap}{8} (\tr F \wedge F - \tr \widehat{R}^\Ch \wedge \widehat{R}^\Ch)_\mu{}^\mu{}_{\bar{\beta} \alpha} \nonumber\\
    &\= \!-  \frac{\ap}{4} \tr \bigg[ F_{\alpha \bar{\mu}} F_{\bar{\beta}}{}^{\bar{\mu}} - \widehat{R}^\Ch{}_{\alpha \bar{\mu}} \widehat{R}^\Ch{}_{\bar{\beta}}{}^{\bar{\mu}} \bigg]  + \frac{\ap}{4} \widehat{g}^{\mu \bar{\nu}} \tr \widehat{R}^\Ch{}_{\mu \bar{\nu}} \widehat{R}^\Ch{}_{\alpha \bar{\beta}} ~. 
 \end{split}
 \eeq
 The term $\widehat{g}^{\mu \bar{\nu}} R^\Ch_{\mu \bar{\nu}}$ was computed in the appendix \eqref{eq:R-inst}.
\beq \label{eq:badguy}
  \widehat{g}^{\mu \bar{\nu}} \widehat{R}^\Ch{}_{\mu \bar{\nu}}{}^\alpha{}_\beta \= -  (i \partial \bar{\partial} \widehat\omega)_\mu{}^{\mu \alpha}{}_\beta + H^{\mu \bar{\nu} \alpha} H_{\beta \mu \bar{\nu}} ~.
 \eeq
 Substitution of this identity proves \eqref{eq:trFFRR1}. Up to errors of order $\cO(\ap^3)$, we can convert Chern to Hull. Indeed, the explicit formulas in \eqref{defn:RH} show that $\widehat{R}^\H{}_{\alpha \bar{\beta}}{}^{\bar{\gamma}}{}_{\lambda} = \nabla^\Ch_\alpha H_{\bar{\beta}}{}^{\bar{\gamma}}{}_\lambda$ and 
 \beq\label{eq:RHRCh}
\widehat{R}^{\H}{}_{\alpha \bar{\beta}}{}^\lambda{}_\gamma ~= \widehat{R}^{\rm Ch}{}_{\alpha \bar{\beta}}{}^\lambda{}_\gamma + H_\alpha{}^\lambda{}_{\bar{\sigma}} H_{\bar{\beta}}{}^{\bar{\sigma}}{}_\gamma~.
\eeq
Thus \eqref{eq:trFFRR1} implies
\beq \label{eq:trFFRR}
\begin{split}
    & \ \frac{\ap}{8} (\tr F \wedge F - \tr \widehat{R}^\Ch \wedge \widehat{R}^\Ch)_\mu{}^\mu{}_{\bar{\beta} \alpha} \\
    &\overset{\ \ap^2}{=} -  \frac{\ap}{4} \tr \bigg[ F_{\alpha \bar{\mu}} F_{\bar{\beta}}{}^{\bar{\mu}} - R^\H{}_{\alpha \bar{\mu}} R^\H{}_{\bar{\beta}}{}^{\bar{\mu}} \bigg] - \frac{\ap}{4} (\ii \Lambda_\omega \ii \partial \bar{\partial} \widehat{\omega})^m{}_n R^\Ch{}_{\alpha \bar{\beta}}{}^n{}_m~.
\end{split}
\eeq
We note that $\widehat{R}$ can be converted to $R$ up to errors of order $\cO(\ap^2)$.

\subsubsection{Double holomorphic directions}
We start by examining the graviton equation \eqref{eq:einstein} when $m,n$ are both holomorphic indices. The first step is to note the general formula for the Levi-Civita Ricci curvature of a hermitian metric $(X,\omega)$ along two holomorphic indices:
\beq\notag
\Ric_{\alpha \beta} \= \half  \bigg( \nabla_\alpha^{\Ch} (d^c \omega)_\mu{}^\mu{}_\beta + \nabla_\beta^{\Ch} (d^c \omega)_\mu{}^\mu{}_\alpha \bigg) +\half   (d^c \omega)_{\alpha \mu \bar{\nu}} (d^c \omega)^{\mu \bar{\nu}}{}_\beta~, \quad d^c = i(\partial-\bar{\partial}) ~.
\eeq
We now substitute \eqref{eq:truncatedH} and \eqref{eq:trT-improved}.
\beq\notag
\widehat \Ric_{\alpha \beta} \=\! - \nabla_\alpha^{\Ch} \partial_\beta \widehat\Phi - \nabla^{\Ch}_\beta \partial_\alpha \widehat\Phi + \half  H_{\alpha \mu \bar{\nu}} H^{\mu \bar{\nu}}{}_\beta~.
\eeq
Converting the Chern connection $\nabla^\Ch$ to the Levi-Civita connection $\nabla$ gives:
\beq \label{eq:einstein-ab}
\widehat \Ric_{\alpha \beta} + 2 \widehat\nabla_\alpha \widehat\nabla_\beta \widehat \Phi  -\frac{1}{4} H_{\alpha pq} H_\beta{}^{pq} = 0~.
\eeq
In \eqref{eq:einstein}, there is also the term $\frac{\ap}{4} \cE_{\alpha \beta}$, which is
\beq \notag
\frac{\ap}{4}  \tr F_{\alpha m} F_\beta{}^m - \frac{\ap}{4}  \tr R^\H{}_{\alpha m} R^\H{}_\beta{}^m \ \overset{\ \ap^2}{=}  - \frac{\ap}{4} g^{\mu \bar{\nu}} \tr \widehat{R}^\H{}_{\alpha \mu} \widehat{R}^\H{}_{\beta \bar{\nu}}  - \frac{\ap}{4}  g^{\mu \bar{\nu}} \tr \widehat{R}^\H{}_{\alpha \bar{\nu}} \widehat{R}^\H{}_{\beta \mu}~.
\eeq
This term is order $\cO(\ap^3)$. To see this, we can compute the Hull curvature using the explicit formulas in \eqref{defn:RH}. The only non-zero component of $\widehat{R}^{\H}{}_{\alpha \beta}{}^m{}_n$ is $\widehat{R}^{\H}{}_{\alpha \beta}{}^\mu{}_{\bar{\nu}}$. Hence
\beq \label{eq:RHjunk}
- \frac{\ap}{4}  \tr \widehat{R}^\H{}_{\alpha m} \widehat{R}^\H{}_\beta{}^m \overset{\ \ap^2}{=}\!  - \frac{\ap}{4} g^{\mu \bar{\nu}} \tr \widehat{R}^\H{}_{\alpha \mu}{}^\sigma{}_{\bar{\delta}} \widehat{R}^\H{}_{\beta \bar{\nu}}{}^{\bar{\delta}}{}_\sigma  - \frac{\ap}{4}  g^{\mu \bar{\nu}} \tr \widehat{R}^\H{}_{\alpha \bar{\nu}}{}^{\bar{\delta}}{}_\sigma \widehat{R}^\H{}_{\beta \mu}{}^\sigma{}_{\bar{\delta}}~.
\eeq
From here we can use
\beq\notag
\widehat{R}^{\H}{}_{\mu \nu}{}^\alpha{}_{\bar{\beta}} \=\! - (\ii \partial \bar{\partial} \widehat{\omega})_{\mu \nu}{}^\alpha{}_{\bar{\beta}}~, \quad \widehat{R}^{\H}{}_{\beta \bar{\nu}}{}^{\bar{\delta}}{}_\sigma \= \nabla^{\Ch}_\beta H_{\bar{\nu}}{}^{\bar{\delta}}{}_\sigma~,
\eeq
to reduce \eqref{eq:RHjunk} to $\cO(\ap^3)$. Thus $\ap \cE_{\alpha \beta} = \cO(\ap^3)$. Finally, we note that
\beq \notag
\ap^2 R^\H{}_{\alpha p}{}^{\bar{\beta}}{}_q \cE^{pq} \overset{\ \ap^2}{=} 0 ~,
 \eeq
from the identities on $R^\H$ listed above and $\ap \cE_{\alpha \beta} = \cO(\ap^3)$. This shows \eqref{eq:einstein} when $m,n$ are both holomorphic indices.

\subsubsection{Mixed directions}

Next, we consider \eqref{eq:einstein} when $m,n$ are mixed barred/unbarred indices. As before, the starting point is the general formula for the Ricci curvature in complex geometry, which expresses the Levi-Civita Ricci curvature of a hermitian metric $\omega$ in terms of the first Chern class representative:
\beq \label{eq:einstein1}
  \begin{split}
  \Ric_{\alpha \bar{\beta}} &\= R^{\rm Ch}{}_{\alpha \bar{\beta}}{}^\mu{}_\mu -(\ii \partial \bar{\partial} \omega)_\mu{}^\mu{}_{\bar{\beta} \alpha} + \frac{1}{2} \nabla_\alpha^{\rm Ch} (d^c \omega)_\mu{}^\mu{}_{\bar{\beta}} - \frac{1}{2} \nabla_{\bar{\beta}}^{\rm Ch} (d^c \omega)_\mu{}^\mu{}_\alpha  \\
                            &+ \frac{1}{2} (d^c \omega)_{\alpha \mu \bar{\nu}} (d^c \omega)^{\mu \bar{\nu}}{}_{\bar{\beta}} + \frac{1}{4} (d^c \omega)_{\alpha \bar{\mu} \bar{\nu}} (d^c \omega)^{\bar{\mu} \bar{\nu}}{}_{\bar{\beta}} \\
  &- \frac{1}{2} (d^c \omega)_\alpha{}^\mu{}_{\bar{\beta}} (d^c \omega)_\nu{}^\nu{}_\mu - \frac{1}{2} (d^c \omega)_\alpha{}^{\bar{\mu}}{}_{\bar{\beta}} (d^c \omega)_\nu{}^\nu{}_{\bar{\mu}} ~.
\end{split}
  \eeq
Here $d^c \omega = i (\partial - \bar{\partial})\omega$. This formula holds on any hermitian manifold $(X,\omega)$ and can be calculated directly from the conventions stated in the appendix. We can then rewrite \eqref{eq:einstein1} in our particular setup, where we substitute \eqref{eq:trChern}, $H=d^c \widehat{\omega}$ and \eqref{eq:trT-improved}.
\beq \notag
\begin{split}
   \widehat{\Ric}_{\alpha \bar{\beta}} &\=\! -2 \partial_\alpha \partial_{\bar{\beta}} \widehat{\Phi} - (\ii \partial \bar{\partial} \widehat{\omega} )_\mu{}^\mu{}_{\bar{\beta} \alpha}  \\
  &+ \frac{1}{2} H_{\alpha \mu \bar{\nu}} H^{\mu \bar{\nu}}{}_{\bar{\beta}} + \frac{1}{4} H_{\alpha \bar{\mu} \bar{\nu}} H^{\bar{\mu} \bar{\nu}}{}_{\bar{\beta}} + H_\alpha{}^\mu{}_{\bar{\beta}} \widehat\Phi_\mu - H_\alpha{}^{\bar{\mu}}{}_{\bar{\beta}} \widehat\Phi_{\bar{\mu}}~ .
\end{split}
\eeq
The Hessian of the Levi-Civita connection in holomorphic coordinates is
\beq\notag
\widehat\nabla_\alpha \widehat\nabla_{\bar{\beta}} \widehat\Phi \= \partial_\alpha \partial_{\bar{\beta}} \widehat\Phi - \frac{1}{2} H_\alpha{}^\mu{}_{\bar{\beta}} \widehat\Phi_\mu + \frac{1}{2} H_\alpha{}^{\bar{\mu}}{}_{\bar{\beta}} \widehat\Phi_{\bar{\mu}}~,
\eeq
and so using equation \eqref{eq:iddbo} for $i \partial \bar{\partial} \omega$ gives
\beq \label{eq:road2einstein}
  \widehat\Ric_{\alpha \bar{\beta}} + 2 \widehat\nabla_\alpha \widehat\nabla_{\bar{\beta}} \widehat\Phi -\frac{1}{4} H_{\alpha mn} H_{\bar{\beta}}{}^{mn} - \frac{\ap}{8} (\tr F \wedge F - \tr \widehat{R}^\Ch \wedge \widehat{R}^\Ch)_\mu{}^\mu{}_{\bar{\beta} \alpha}  \= 0~.
  \eeq
Substituting \eqref{eq:trFFRR} into \eqref{eq:road2einstein} gives
\beq
   \widehat\Ric_{\alpha \bar{\beta}} + 2 \widehat\nabla_\alpha \widehat\nabla_{\bar{\beta}} \widehat\Phi -\frac{1}{4} H_{\alpha mn} H_{\bar{\beta}}{}^{mn} + \frac{\ap}{4} \cE_{\alpha \bar{\beta}} + \frac{\ap}{4}  (\ii \Lambda_\omega \ii \partial \bar{\partial} \widehat\omega)^m{}_n R^\Ch{}_{\alpha \bar{\beta}}{}^n{}_m \overset{\ \ap^2}{=} 0 ~ . \label{eq:troublesome}
   \eeq
The last term is not in the form given in \eqref{eq:einstein}. We manipulate the corresponding term and obtain
   \beq \label{quadBS0}
\frac{\ap^2}{8} R^\H{}_{\alpha}{}^p{}_{\bar{\beta}}{}^q \cE_{pq}  \overset{\ \ap^2}{=}
\frac{\ap^2}{8} R^\H{}_{\alpha}{}^\nu {}_{\bar{\beta}}{}^{\bar{\mu}} \cE_{\nu \bar{\mu}} \overset{\ \ap^2}{=} \frac{\ap^2}{8} R^\Ch{}_{\alpha \bar{\beta}}{}^\nu{}_\mu \cE^\mu{}_{\nu}
   \eeq
   since $\cE_{\mu \nu} = \cO(\ap^2)$ and $ R^\H{}_{\alpha}{}^{\bar{\nu}} {}_{\bar{\beta}}{}^\mu = \cO(\ap)$, $\cE$ is symmetric, and K\"ahler symmetries for $R^\H$ hold up to errors of order $\ap$. We note that multiplying \eqref{eq:trFFRR} by $\ap$ gives
   \beq \label{quadBS1}
\frac{\ap^2}{2} \ii \Lambda_\omega (\tr F \wedge F - \tr R^\Ch \wedge R^\Ch ){}^\mu{}_{\nu} \overset{\ \ap^2}{=} - \ap^2 \cE^\mu{}_\nu ~.
   \eeq
Substitute \eqref{quadBS1} into \eqref{quadBS0} and apply \eqref{eq:iddbo} to bring out $\ii \partial \bar{\partial} \omega$.
\beq \label{eq:quadBS}
\frac{\ap^2}{8} R^\H{}_{\alpha}{}^p{}_{\bar{\beta}}{}^q \cE_{pq}  \overset{\ \ap^2}{=}  \frac{\ap}{2} R^\Ch{}_{\alpha \bar{\beta}}{}^\nu{}_\mu \ii \Lambda_\omega ( \ii \partial \bar{\partial} \widehat{\omega})^\mu{}_\nu ~.
\eeq
Substituting \eqref{eq:quadBS} into \eqref{eq:troublesome}, while noting the double count in trace over $m,n$ versus $\mu,\nu$, gives \eqref{eq:einstein}.

\subsubsection{Ricci hat}
\label{s:riccihat}

It remains to show why \eqref{eq:einstein} implies
\beq \label{eq:einstein2}
\Ric_{mn}+2\nabla_m\nabla_n\Phi-\frac14H_{mpq}H_n{}^{pq}  + \frac{\ap}{4} \cE_{mn} + \Delta^{\rm conn}_{mn}  \overset{\ \ap^2}{=} 0 ~,
\eeq
where
\beq
\Delta^{\rm conn}_{mn}  \= -\frac{\ap}{4}e^{2\Phi} \nabla^p \Bigl( e^{-2\Phi}\left(
  \nabla_m\cG_{np}  +\nabla_n\cG_{mp}  -2\nabla_p\cG_{mn}  \right) \Bigr) ~,
\eeq
with the notation $\cG_{mn} = \Ric_{mn} + 2 \nabla_m \nabla_n \Phi  - \frac{1}{4} H_{mab} H_n{}^{ab}$. First, the formula for the linearization of the Ricci curvature gives the following Taylor expansion: expand the Ricci curvature of $\widehat{g} = g+k$ to obtain
 \beq \notag
\widehat{R}_{mn} \= R_{mn} + \frac{1}{2} \bigg(  \nabla^p \nabla_m k_{np} + \nabla^p \nabla_n k_{mp} - \nabla^p \nabla_p k_{mn} - \nabla_m \nabla_n \tr k\bigg) + \mathcal{Q}(k,k) ~.
\eeq
In our case, $\widehat{g} = g + \frac{\ap^2}{4} \cE$, and so
 \beq \label{eq:hatric}
\widehat{R}_{mn} \overset{\ \ap^2}{=} R_{mn} + \frac{\ap^2}{8} \bigg(  \nabla^p \nabla_m \cE_{np} + \nabla^p \nabla_n \cE_{mp} - \nabla^p \nabla_p \cE_{mn} - \nabla_m \nabla_n \tr \cE \bigg) ~.
\eeq
Second, \eqref{eq:einstein} implies $\cG_{mn} = - \frac{\ap}{4} \cE_{mn} + \cO(\ap^2)$. Therefore, using $d \Phi = \cO(\ap)$, we obtain
\beq \label{eq:newdeltaconn}
\Delta^{\rm conn}_{mn}  \overset{\ \ap^2}{=} \frac{\ap^2}{16} \left(
\nabla^p \nabla_m\cE_{np}  + \nabla^p \nabla_n\cE_{mp}  -2 \nabla^p\nabla_p\cE_{mn}  \right) ~.
\eeq
We now manipulate \eqref{eq:einstein} by substituting \eqref{eq:hatric} and $\widehat{\Phi} = \Phi + \frac{3 \ap^2}{64} \tr \cE$, and then adding \eqref{eq:newdeltaconn}. The result is
\beq \notag
\begin{split}
  &\ \Ric_{mn}+2\nabla_m\nabla_n\Phi-\frac14H_{mpq}H_n{}^{pq}  + \frac{\ap}{4} \cE_{mn} + \Delta^{\rm conn}_{mn}  \\
  &\overset{\ \ap^2}{=} - \frac{\ap^2}{16} (\nabla_m \nabla^p  \cE_{np}  +  \nabla_n \nabla^p \cE_{mp}) + \frac{\ap^2}{32} \nabla_m \nabla_n \tr \cE ~.
  \end{split}
  \eeq
  The term $ R^\H{}_{m}{}^p{}_n{}^q \cE_{pq}$ was absorbed by commuting covariant derivatives $[\nabla^p,\nabla_m] \cE_{np}$ and using zeroth order Ricci-flatness. The graviton equation \eqref{eq:einstein2} now follows from substituting
\beq \label{eq:divE}
\nabla^p \cE_{pn} = \frac{1}{4} \nabla_n \tr \cE + \cO(\ap),
\eeq
which can be verified by the instanton conditions $F_p{}^p=0$ and $\widehat{R}^\H{}_p{}^p=\cO(\ap)$ and the Bianchi identity.

\subsection{Dilaton equation}
\label{s:dilatoneom}
Finally, we must derive the dilaton equation
\beq \label{eq:dilaton}
 R - 4(\nabla \Phi)^2 + 4 \nabla^2 \Phi - \frac{1}{2} |H|^2 + \frac{\ap}{4} \tr |F|^2  -  \frac{\ap}{4} \tr |R^\H|^2 \overset{\ \ap^2}{\=} 0~.
 \eeq
Taking the trace of the graviton equation \eqref{eq:einstein} does not immediately produce the dilaton equation. So instead we start by deriving an identity for $\nabla^2 \Phi$ by using special identities from conformally balanced metrics on complex manifolds. 

We start with the definition of the Laplacian in holomorphic coordinates:
\beq\notag
\widehat\nabla^2 \log \norm{\Omega}_{\widehat{g}} \= \widehat{g}^{\alpha \bar{\beta}} \partial_\alpha \partial_{\bar{\beta}} \log \norm{\Omega}_{\widehat{g}} - (d^c \widehat\omega)_\alpha{}^{\mu \alpha} \partial_\mu \log \norm{\Omega}_{\widehat{g}} + (d^c \widehat\omega)_\alpha{}^{\bar{\mu} \alpha} \partial_{\bar{\mu}} \log \norm{\Omega}_{\widehat{g}}~.
\eeq
On a general conformally balanced geometry $(X,\widehat\omega,\Omega)$ satisfying $d (\norm{\Omega}_{\widehat{g}} \widehat{\omega}^2)=0$, we can apply \eqref{eq:RCH-hol} and \eqref{eq:R-inst}, hence there holds
\beq\notag
\begin{split}
   \widehat\nabla^2 \log \norm{\Omega}_{\widehat{g}} & \=\! - (\ii \partial \bar{\partial} \widehat\omega)_\alpha{}^{\alpha \mu}{}_\mu + (d^c \widehat\omega)_{\alpha \mu \bar{\nu}} (d^c \widehat{\omega})^{\mu \bar{\nu} \alpha} \\
 &\quad - (d^c \widehat{\omega})_\alpha{}^{\mu \alpha} \partial_\mu \log \norm{\Omega}_{\widehat{g}} + (d^c \widehat{\omega})_\alpha{}^{\bar{\mu} \alpha} \partial_{\bar{\mu}} \log \norm{\Omega}_{\widehat{g}}~.
  \end{split}
\eeq
In our setup, the function $\log \norm{\Omega}_{\widehat{g}}$ is related to the dilaton via $\partial_i \log \norm{\Omega}_{\widehat{g}} = - 2 \partial_i \widehat\Phi$, $H=d^c \widehat{\omega}$, and our conventions are $|H|^2=\frac{1}{6} H_{mnp} H^{mnp}$. Thus
\beq\notag
 \widehat\nabla^2 \widehat\Phi \= \frac{1}{2} (i \partial \bar{\partial} \widehat\omega)_\alpha{}^{\alpha \mu}{}_\mu - \half  |H|^2 - H_\alpha{}^{\mu \alpha} \partial_\mu \widehat\Phi + H_\alpha{}^{\bar{\mu} \alpha} \partial_{\bar{\mu}} \widehat\Phi~.
 \eeq
 Since $H_\mu{}^\mu{}_\lambda = - 2 \partial_\lambda \widehat\Phi$, this becomes
 \beq\notag
 \widehat\nabla^2 \widehat\Phi - \frac{1}{2} (\ii \partial \bar{\partial} \widehat\omega)_\alpha{}^{\alpha \mu}{}_\mu + \half  |H|^2 - 2 |\widehat\nabla \widehat\Phi|^2\=0~.
 \eeq
We now substitute the equation \eqref{eq:iddbo} for $i \partial \bar{\partial} \widehat\omega$ and \eqref{eq:trFFRR0}.
 \beq \notag
 \widehat\nabla^2 \widehat\Phi - 2 |\widehat\nabla \widehat\Phi|^2 - \frac{\ap}{8} \tr |F|^2 + \frac{\ap}{8} \tr |\widehat{R}^\Ch|^2+ \half  |H|^2  \overset{\ \ap^2}{=} 0~.
\eeq
Here we used 
\beq\notag
\ap \tr (g^{\mu \bar{\nu}} \widehat{R}^{\Ch}{}_{\mu \bar{\nu}})^2 \= \cO(\ap^3)~,
\eeq
which follows from \eqref{eq:badguy}. We now remove hats by substituting $\widehat\Phi = \Phi + \frac{3 \ap^2}{64} \tr \cE$, and exchange $R^\Ch$ for $R^\H$ up to lower order terms.
\beq \label{eq:dilaton1}
 2 \nabla^2 \Phi - 4 |\nabla \Phi|^2 + \frac{3 \ap^2}{32} \nabla^2 \tr \cE - \frac{\ap}{4} \tr |F|^2 + \frac{\ap}{4} \tr |R^\H|^2+ |H|^2  \overset{\ \ap^2}{=} 0~.
\eeq
We can now combine identity \eqref{eq:dilaton1} with the trace of the graviton equation to obtain the dilaton equation. The trace of the graviton equation \eqref{eq:einstein2} is
\beq \label{eq:dilaton2}
 R+ 2 \nabla^2 \Phi -\frac{3}{2} |H|^2 + \frac{\ap}{2}  \tr |F|^2 - \frac{\ap}{2} \tr |R^\H|^2 + g^{mn} \Delta^{\rm conn}_{mn} \overset{\ \ap^2}{=} 0~,
 \eeq
 where \eqref{eq:newdeltaconn} gives
 \beq \notag
g^{mn} \Delta^{\rm conn}_{mn} \overset{\ \ap^2}{=} \frac{\ap^2}{8} ( \nabla^p \nabla^n \cE_{np} - \nabla^p \nabla_p \tr \cE) ~. 
 \eeq
After substituting \eqref{eq:divE}, the traced graviton equation \eqref{eq:dilaton2} becomes
\beq \label{eq:dilaton3}
R - \frac{3 \ap^2}{32} \nabla^2 \tr \cE + 2 \nabla^2 \Phi -\frac{3}{2} |H|^2 + \frac{\ap}{2}  \tr |F|^2 - \frac{\ap}{2}  \tr |R^\H|^2  \overset{\ \ap^2}{=} 0~.
 \eeq
Adding equation \eqref{eq:dilaton1} to equation \eqref{eq:dilaton3} gives
 \beq\notag
  R+ 4 \nabla^2 \Phi - 4 |\nabla \Phi|^2 -\frac{1}{2} |H|^2 + \frac{\ap}{4}  \tr |F|^2 - \frac{\ap}{4} \tr |R^\H|^2 \overset{\ \ap^2}{=} 0~.
 \eeq
We conclude that the dilaton equation is satisfied up to terms of order $\cO(\ap^3)$.

 \section{Conclusions}

\subsection{Geometric flows}
There is a version of Ricci flow adapted to heterotic string compactifications. The idea is as follows: in practice, one can construct various explicit examples of complex geometries $(X,\Omega, \widehat\omega)$ and bundles $E \rightarrow X$ solving
\beq \label{eq:partialsusy}
F^{0,2} \=0~, \quad d \Omega \= 0~, \quad \dd (\norm{\Omega}_{\widehat{g}} \widehat\omega^2)\= 0~.
\eeq
These are in a sense intermediate supersymmetric configurations; the full supersymmetry equations \eqref{eq:Fo2}, \eqref{eq:iddbo} are not satisfied, as we are missing
\beq \label{eq:partialsusy2}
F \wedge \widehat\omega^2 \= 0~, \quad \ii \partial \bar{\partial} \widehat\omega \= \frac{\ap}{8} \Bigl[\tr \bigl( \widehat{R}^\Ch \w \tr \widehat{R}^\Ch\bigr) -\bigl( \tr F \w F\bigr) \Bigr]~.
\eeq
The method then is to start with concrete examples solving the partial constraints \eqref{eq:partialsusy} and deform them along a flow which seeks to reach an on-shell configuration solving the full equations of motion. The anomaly flow, introduced in \cite{Phong2018,Phong2018B}, is a type of Ricci flow which does this: it is a flow that remains in the class of complex geometries solving \eqref{eq:partialsusy} and whose fixed-points solve \eqref{eq:partialsusy2}. The flow equations are:
\begin{align}
  \partial_t ( \norm{\Omega}_{\widehat{g}(t)}  \widehat\omega(t)^2) &\= \ii \partial \bar{\partial} \widehat\omega(t) - \frac{\ap}{8} ( \tr \widehat{R}^{\Ch} \wedge \widehat{R}^\Ch - \tr F_{h(t)} \wedge F_{h(t)})~,\\
  h^{-1} \partial_t h &\=\! - \ii \Lambda_{\widehat\omega(t)} F_{h(t)}~,
\end{align}
where the pair $(\widehat\omega(t),h(t))$ is flowing, and the holomorphic volume form $\Omega$ on $X$ and complex structure on $E$ remains fixed. Here $h(t)$ is a hermitian metric on the holomorphic bundle $E$ with Chern curvature $F_h = \bar{\partial}( h^{-1} \partial h)$ and its flow is given by the Donaldson heat flow \cite{0529.53018}. The initial metric $\widehat\omega_0$ solves $d ( \norm{\Omega}_{\widehat{g}(0)} \widehat\omega(0)^2)=0$ and the equations are such that $d ( \norm{\Omega}_{\widehat{g}(t)} \widehat\omega(t)^2)=0$ at all times. The reason this is a type of Ricci flow is because the metric tensor evolves as (see e.g. \cite{Ashmore2024})
\beq \notag
\partial_t \widehat{g}_{\alpha \bar{\beta}} = \frac{e^{2 \widehat{\Phi}}}{2} \bigg[ - \widehat{\Ric}_{\alpha \bar{\beta}} - 2 \widehat\nabla_\alpha \widehat\nabla_{\bar{\beta}} \widehat\Phi +\frac{1}{4} H_{\alpha mn} H_{\bar{\beta}}{}^{mn} \bigg] + \cO(\ap)~.
\eeq
As the flow is nonlinear, there is a requirement on the scale: $|\ap \widehat{R}_{\widehat{g}(t)}| \ll 1 $ must be satisfied for the flow to be mathematically well-defined \cite{Phong2018, 2408.15514}. Explicit examples of this flow can be tested on $T^4$ fibrations over Riemann surfaces \cite{Fei2021} and $T^2$ fibrations over $K3$ surfaces \cite{Phong2018-sc}. We note that there is also another version of Ricci flow \cite{mariomolinastreets2024} adapted to heterotic string theory without the gravitational term $\tr R \wedge R$.

If the flow converges as $t \rightarrow \infty$, then limiting fixed point configurations solve \eqref{eq:partialsusy} and \eqref{eq:partialsusy2}. Our current work has demonstrated that these equations of complex geometry are consistent with supersymmetry at $\ap^2$ and solve the equations of motion after transforming the hatted fields to their unhatted counterparts via \eqref{eq:unhatting}. For the string theoretic study of the anomaly flow at leading order in $\ap$, see \cite{Ashmore2024}. 

From the mathematical perspective, the major question on the anomaly flow is to understand the class of initial data which leads to long-time existence and convergence; specific explicit examples have been constructed with a wide range of behaviour including finite-time divergence \cite{Fei2021} as well as infinite-time convergence \cite{Phong2018-sc}. Regardless of the behavior of this particular flow, an outstanding question is to determine when an intermediate configuration solving \eqref{eq:partialsusy} can be deformed to solve the full system including \eqref{eq:partialsusy2}; the significance of this problem in pure mathematics and differential geometry has been underscored by S.-T. Yau \cite{liyau2005, Fu:2006vj}. 

\subsection{Torsional \texorpdfstring{$K3$}{K3}  solutions}
Let us comment on the assumption of a smooth $\ap \to 0$ limiting metric, which is used throughout our analysis. There are in fact examples in string theory (and not  supergravity) that do not satisfy this assumption. But if this $\ap \to 0$ limit does not produce genuine limiting fields on the threefold, such solutions do not have sigma model descriptions and must be described by other methods such as duality. 

Nonetheless, the example of a $T^2$ fibration over a $K3$ receives interest and so we comment very briefly on the set-up. The Calabi-Yau threefold given by a $T^2$ fibered over a $K3$ base was suggested in \cite{Dasgupta:1999ss} to have significance in string theory, and it has been widely studied in both string theory and pure mathematics \cite{Fu:2006vj, Becker:2006et, Becker:2009df,Melnikov:2014ywa,Garcia-Fernandez:2018qcl}. In particular, Fu-Yau \cite{Fu:2006vj} constructed solutions to the nonlinear constraint \eqref{eq:iddbo} for $i \partial \bar{\partial} \omega$ over these geometries.

Let $X$ be the total space of a $U(1) \times U(1)$ principal bundle over a $K3$ surface $S$. Let $(\o_S,\O_S)$ be a Calabi-Yau structure on $S$. The hermitian metric and holomorphic volume form on the threefold $X$ are:
\beq\label{eq:K3T2herm}
\o \= e^{2\phi} \o_S + \frac{\ii a}{2}\, \ITh \w  \IThb~, \quad \O_X \= \sqrt{a}\, \O_S \w \ITh~,
\eeq
where $a$ is a constant denoting the volume of the torus. Here $\phi: S \rightarrow \mathbb{R}$ is a function on the base and $\ITheta = \ITheta^1 + \ii \ITheta^2$ and $\ITheta^i$ is a connection 1-form on each $U(1)$-bundle factor. Flux quantization has the implication that $a$ is an integer multiple of $2\pi \ap$ \cite{Melnikov:2014ywa}. This setup is such that the metric and Hermitian form on the threefold are not well-defined in the large-radius (i.e., $\ap \to 0$) limit as the geometry collapses down to the base $(S, e^{2 \phi} \o_S)$.

If the supergravity and supersymmetry equations studied here are to approximate the string theory coming from the sigma model, one must have a smooth $\ap \to 0$ limit. As a result, these backgrounds do not correspond to weakly coupled sigma models—linear or non-linear—that flow to conformal field theories. Instead, they are believed to be defined only indirectly, through dualities with type IIB string theory or M-theory. 
Consequently, without a sigma model to rely on, the $\ap^2$-analysis discussed here is not obviously appropriate to study  these geometries. We likely need new computational tools, such as duality to M-theory, and  leave this open for future investigation.

\subsection{Conclusions}

We have analysed the Bergshoeff--de Roo supersymmetry equations through order $\ap^2$ assuming a smooth $\ap\to 0$ limit  of heterotic compactifications on $SU(3)$ manifolds. The internal space remains a complex threefold with trivial canonical bundle, and the Bismut connection has holonomy contained in $SU(3)$. The flux--torsion, holomorphic-volume and conformally balanced equations take the compact form \eqref{eq:introconf} in terms of the redefined fields $(\widehat g,\widehat\omega,\widehat\Phi)$, while the Green--Schwarz Bianchi identity admits the resolved form in \eqref{surprising-bianchi}.

Written in the hatted variables, these tensorial supersymmetry equations have the same explicit form through order $\ap^2$ as the first-order Strominger system. This does not mean that second-order effects are absent. The field redefinition itself contains $\ap^2$ corrections, and the graviton equation in the hatted basis retains the explicit curvature contraction displayed in \eqref{eq:graviton_hatted}.

A central  result concerns the  Hull connection. Its chain-rule variation in the known BdR action produces the genuine order-$\ap^2$ contribution $\Delta^{\rm conn}_{mn}$ to the graviton equation. Since the Killing-spinor equation is a global differential equation, its curvature identity is automatic and cannot supply an independent condition. Evaluating that identity through order $\ap^2$ reproduces precisely $\Delta^{\rm conn}_{mn}$. Supersymmetry together with the Green--Schwarz Bianchi identity therefore implies the corrected equations of motion without imposing an auxiliary tangent-bundle instanton condition.

Indeed, supersymmetry relates the $(0,2)$ component and Hermitian trace of the physical Hull curvature to $\dd H$, so $R^\H$ is generically not an instanton on the torsional perturbative branch. If the instanton condition is nevertheless imposed, the trace relation, the Lefschetz isomorphism and compactness give $\dd H=\cO(\ap^3)$ and $H=\cO(\ap^3)$. Thus the auxiliary condition is not only unnecessary; it restricts solutions to being torsion free through $\ap^2$, which is too strong.

It is natural to ask how the $\ap$ corrections modify the moduli space. In particular, one could study the corrections to the moduli space's natural \K metric and its associated \K potential governing the moduli and matter field dynamics, as constructed in \cite{Candelas:2016usb,McOrist:2016cfl,McOrist:2019mxh}. Based on \cite{Witten:1985bz},  it is unlikely for the perturbative $\ap$-corrections to modify the dimension of the moduli space and so the modifications to the metric are not going to be topological or dimension changing in nature. In studies of mirror symmetry of Calabi-Yau manifolds, the \K potential is modified by an $\ap^3 \zeta(3)$ term. Are there additional terms for heterotic theories?  At $\ap^2$, the results here suggest there might be some protection of the  \K potential from $\ap^2$ corrections, at least when it is written in terms of the hatted basis. We hope to report on this in future work. It would be of interest to further investigate the associated universal geometric structures  \cite{Candelas:2018lib,McOrist:2024glz} and their capacity to encode the effects of $\ap$-corrections on the background geometry. 

 The conditions that derive from an action functional (labelled the superpotential see e.g. \cite{LopesCardoso:2003dvb,delaOssa:2015maa,McOrist:2016cfl,Ashmore:2018ybe})
 $$
 W \= \int \O \w (H+\ii \dd \o)~,
 $$
  include $\dd \O = 0$, $F^{0,2} = 0$, $H=\ii (\del-\delb)\o$ and $R^{0,2} = 0$~. While $W$ captures a subset of the first order $\ap$ supersymmetry equations, it remains to determine how the $\ap^2$ corrections are included in $W$; its clear the hatted basis plays a role. Investigating these issues, and how the finite deformations of $W$ described in \cite{Ashmore:2018ybe} are affected, would be interesting to pursue as future work.

It would also be of interest to investigate the structure and implications of the $\ap^2$ supersymmetry algebra in the context of compactifications on $\text{G}_2$-- and $\text{Spin}(7)$--holonomy manifolds, as well as on special geometries such as $\text{K3} \times T^2$, which preserve extended supersymmetry. For $\text{G}_2$ and $\mathrm{Spin}(7)$ backgrounds, the geometry is governed by a closed (or co-closed) defining form—either the associative 3-form in the case of $\text{G}_2$, or the Cayley 4-form in the $\text{Spin}(7)$ case—leading to nontrivial torsion classes when fluxes are introduced. In such settings, the preservation of supersymmetry often requires gauge fields to satisfy an instanton condition, i.e., that their curvature lies in a specific subbundle of two-forms determined by the holonomy. Understanding how this condition is modified in the presence of $\ap$-corrections, and how it is reflected in the structure of the BdR algebra, is an important  question for future work.

 \subsection*{Acknowledgements}
 We would like to thank X. de la Ossa, J. Knapp, I. Melnikov, G.  Tartaglino-Mazzucchelli, and E. Svanes for enlightening conversations. We thank the referees for many comments and suggestions which greatly improved the paper.
 
 {\bf Funding}
We would like to thank MATRIX Institute Research Program `The Geometry of Moduli Spaces in String Theory' for hospitality where this project was conceived. SP is supported by a NSERC Discovery Grant. JM is supported by an ARC Discovery Project Grant DP240101409. The authors have no competing, financial or non-financial interests to declare that are relevant to the content of this article.

{\bf Availability of data and materials}: Not applicable.
 
{\bf Code Availability}: Not applicable.

{\bf Declarations}:

{\bf Conflict of interest}: Not applicable.
 
{\bf Ethics approval}: Not applicable. 

{\bf Consent to participate}: Not applicable. 

{\bf Consent for publication}: Not applicable.

\appendix

\section{Gamma matrix results}
We list here some useful gamma matrix identities. Our conventions for gamma matrices are $\{ \gamma_m, \gamma_n \} = 2 g_{mn} I$. 

We have the following gamma matrix results:
\begin{align}\label{eq:gamma_rel1}
& [\g^m, \g_r] \= 2\g^m{}_r~,&\qquad  &\{ \g^m, \g_r\} \= 2\d^m{}_r~,\nn\\
& [\g^{mn}, \g_r] \= -4 \d^{[m}{}_r \g^{n]}~,&  &\{ \g^{mn}, \g_r \} \= 2\g^{mn}{}_r~,\nn\\
& [\g^{mnp}, \g_r] \=  2\g^{mnp}{}_r ~,&  &\{ \g^{mnp}, \g_r \} \= 6 \d^{[m}{}_r \g^{np]}~, 
\end{align}
and
\begin{align}\label{eq:gamma_rel2}
& [\g^{mn}, \g_{rs}] \= -8 \d^{[m}{}_{[r} \g^{n]}{}_{s]}~,&  &\{ \g^{mn}, \g_{rs} \} \= 2\g^{mn}{}_{rs} - 4\d^{[mn]}{}_{rs}~,\nn\\
& [\g^{mnp}, \g_{rs}] \= 12 \d^{[m}{}_{[r} \g^{np]}{}_{s]}~,&  &\{ \g^{mnp}, \g_{rs} \} \= 2\g^{mnp}{}_{rs} - 12\d^{[mn}{}_{rs}\g^{p]}~.
\end{align}
These results are independent of dimension. We now restrict to an $SU(3)$ manifold, so that we work over a 6-dimensional manifold $X$. In this case, we take gamma matrices representing ${\rm Cliff}(6)$ to be pure imaginary: $\gamma_m^\dagger = \gamma_m$. Let $\eta$ be a pure Weyl spinor, so that with respect to the almost complex structure $J^k{}_\ell = i \eta^\dagger \gamma^k{}_\ell \eta$ we have the decomposition $T_{\mathbb{C}} X \= \ccT_X^{(1,0)} \oplus \overline{\ccT_X^{(1,0)}}$ with $\ccT_X^{(1,0)} = \{ v \in T_{\mathbb{C}} X : \gamma(v) \eta = 0 \}$. If $\mu, \nu$ denotes indices along $\ccT_X^{(1,0)}$, then
$$
 \gamma^\mb \eta \= \gamma_\m \eta \= 0~, \quad \{ \g^\m, \g^\nb\} \= 2g^{\m\nb}~.
$$
Then we have some useful results
\begin{align}\label{eq:spinor_rel}
& \g^\mb \eta \= 0~,&\quad  & \g^{\m\nb} \eta \= -g^{\m\nb}\eta ~,&\quad &\g^{\mb\nb} \eta \= 0~, \nn\\[3pt]
& \g^{\m\n\rb} \eta \= g^{\m\rb} \,\g^\n \eta -  g^{\n\rb} \,\g^\m\eta ~,&\quad  &  \g^{\m\nb\rb} \eta \=0  ~, &\quad & \g^{\mb\nb\rb} \eta \= 0~,\nn\\
&  \g^{\m\n} \g^\r \eta \= \g^{\m\n\r}\eta ~,  &\quad  & \g^{\m\nb}\g^{\r} \eta \= 2g^{\nb\r} \,\g^\m \eta - g^{\m\nb} \,\g^\r\eta ~,&\quad & \g^{\mb\nb} \g^\r \eta \= 0~.
\end{align}
In addition
\beq\label{eq:spinor_rel2}
\begin{split}
   \g^{\m\n}\g^{\r\s\tb} \eta &\= \left(  g^{\r\tb} \g^{\m\n\s} - g^{\s\tb} \g^{\m\n\r} \right)\eta~,    \\
    \g^{\m\nb}\g^{\r\s\tb} \eta &\= 2(g^{\r\tb}g^{\s\nb} - g^{\s\tb} g^{\r\nb} )\g^\m \eta + g^{\m\nb} \left(g^{\s\tb} \g^\r - g^{\r\tb} \g^\s \right)\eta~,    \\
    \g^{\mb\nb}\g^{\r\s\tb} \eta &\= 0~.
\end{split}
\eeq

\section{Dictionary of notation to Bergshoeff-de Roo}
\label{s:BdRNotation}
In Appendix A of BdR \cite{Bergshoeff:1989de} the action and supersymmetry variations correct to $\ap^2$ are written out. Here we provide a simple dictionary of notation. The left-hand side is always BdR; the right hand side is our notation, largely the same as AQS \cite{Anguelova:2010ed}. 
\beq\label{eq:BdRdict}
\begin{split}
& \phi \= e^{2\Phi / 3}~, \qquad \phi^{-1} \del_m \phi \= \frac{2}{3} \del_m \Phi~, \quad \phi^{-3} \= e^{-2\Phi}~,\\
 &\omega_{mab} \= - \Th^\LC_{mab}~, \quad \wt R(\omega) \= - R(\Th^\LC)~, \\
 &H^{BdR}_{mnp} \= \frac{1}{3\sqrt{2}} H_{mnp}~, \quad B^{BdR}_{mn} \= \frac{1}{\sqrt{2}} B_{mn}~,\\
& \Omega^\pm_{mab} \=\!- \Th^\mp_{mab}~, \qquad \alpha \= \beta \= -\frac{\ap}{4}~,\\
& T_{mnpq} \= - \frac{1}{3!} (\dd H)_{mnpq}~,\\
& e \= \det(e_m{}^a) \= \sqrt{g}~,
\end{split}
\eeq
Here $e$ is the vielbein determinant; we set $\k_{10}^2=1$. The Riemann curvature is defined in BdR (see Appendix A of BdR) as 
\beq
\wt R(\o)_{mn}{}^{ab} \= 2 \del_{[m} \o_{n]}{}^{ab}  - \o_{[m}{}^{ac} \o_{n]}{}_c{}^b~, 
\eeq
and so the map is $\wt R(\o) = - R(\Th^\LC)$, where the RHS is defined in a conventional fashion $R(\Th^\LC) = \dd \Th^\LC + (\Th^\LC)^2$.

The Bergshoeff--de Roo Lagrangian density, once transformed to our notation, is
\beq
\begin{split}
 \cL &\= e\phi^{-3} \left\{ -\half R(\o) - \frac{3}{4} H_{mnp} H^{mnp} + \frac{9}{2} (\phi^{-1} \del_m \phi)^2 + \cdots\right\}\\
&\= \frac{\sqrt{g}\,e^{-2\Phi}}{2} \left\{ R(\Th^\LC) - \half |H|^2 + 4(\del_m\Phi)^2 + \frac{\ap}{4}\tr |F|^2 - \frac{\ap}{4}\tr |R^\H|^2 + \cdots\right\}~.
\end{split}
\eeq
The field strength $F,R^\H$ are anti-hermitian with trace negative definite $\tr T^A T^B = - \delta^{AB}$. It is useful to note that $\O^{-\,ab} \dd \O^{-\,ab} = - \tr \O^- \dd \O^-$ as $\O^-$ is antisymmetric in its endomorphism indices. We discuss this in subsection to come.

The three-form becomes
\beq
\begin{split}
 H_{mnp} \= 3 \del_{[m} B_{np]}  + \frac{\ap}{4} \CS(A)_{mnp} -  \frac{\ap}{4} \CS(\Th^+)_{mnp}~.
\end{split}
\eeq
These conventions match Green-Schwarz's anomaly cancellation paper \cite{Green:1984sg}, see below.
 The gravitino and dilatino variations become (correct to $\ap^2$):
\beq
\begin{split}
 \d \Psi_m &\= \left(\del_m + \frac{1}{4} \G^{ab}\left( \Th^-_{mab} + \ap P_{mab} \right)\right)\ve~,\quad P_{mab}\= \frac{1}{4} e^{2\Phi} \nabla^{-\,p} (e^{-2\Phi} (\dd H)_{pmab})~,\\
 \d \l &\= -\frac{1}{2\sqrt{2}} \left\{ \delslash \Phi - \half \Hslash + \frac{3\ap}{2}  \Pslash \right\}\ve~, \\
 \d \chi &\=\! -\frac{1}{4} \G^{mn} F_{mn} \ve~.
 \end{split}
\eeq
As written here there is one important difference as compared with AQS:   in $P_{mab}$ BdR write  the adjoint of $\nabla^-$  contracted on the first index, not the second index as written in AQS. The normalisation of this term differs slightly, $1/4$ here vs $6$ in AQS.

\subsection{The trace, the action and null energy condition}
There is ambiguity in parts of the physics literature surrounding the sign of the trace in the action and Bianchi identity. To that end, we head direct to the source: the Green-Schwarz anomaly cancellation paper \cite{Green:1984sg}.  The action is originally presented in the Einstein frame, which we have converted to the string frame and adapted to our notation. \footnote{The action written in Green-Schwarz is in Einstein frame. A dictionary of conventions to our notation is $\phi^{GS} = e^{\Phi/2}$, $\k^2 / g^2 = \ap/4$ and we set Newton's constant $\k=1$.  Also $\frac{\ap}{4} H^{GS}_{mnp} = H_{mnp}$ which similar rescaling for B-field.  The string frame metric is a conformal rescaling:$g^S_{mn} = g^E_{mn} e^{\Phi/2}$. As for BdR \cite{Bergshoeff:1989de}, the curvature scalar has opposite sign. } 
\beq\label{eq:GSaction}
\begin{split}
 S \= \frac{1}{2} \int \vol e^{-2\Phi} \left\{ R - \frac{1}{12} H_{mnp} H^{mnp} + 4 (\del_m\Phi)^2 - \frac{\ap}{4} F^A_{mn} F^{A\,mn}  + \cdots\right\} ~,
\end{split}
\eeq
where $F = F^A T^A$ and the generators $T^A$ are antihermitian, so that $F = \dd A + A^2$ with a similar definition for the spin connection. Anomaly cancellation then requires
\beq\label{eq:GSBianchi}
H \= \dd B + \frac{\ap}{4} CS(A) - \frac{\ap}{4} CS(\Th)~, \qquad \dd H \= \frac{\ap}{4} \tr F\w F -\frac{\ap}{4} \tr R\w R~.
\eeq
As the gauge group is compact, one has the possibility of working with a hermitian connection and field strength $\hat F = d\hat A - i \hat A^2$ whose generators can be normalised so that $\tr T^A T^B$ is positive definite. This would be at the expense of modifying the sign of $\tr F\w F$ in the Bianchi identity \eqref{eq:GSBianchi}, so that anomaly cancellation continues to hold and we continue to get the topological relation $c_2(X)=c_2(E)$.\footnote{Occasionally in the physics literature, $F$ is written as antihermitian in terms of $A$ but the trace is normalised so that $\Tr T^A T^B = \d^{AB}$ is positive definite; in this case the simplest interpretation is that $\Tr$ is taken to mean the negative of the matrix trace.}

We will always work with antihermitian $F$ and $R$, and we utilise the principle of BdR  \cite{Bergshoeff:1989de} that $F$ and $R$ appear on the same footing. Generators are normalised to $\tr T^A T^B = - \d^{AB}$. After including the curvature term, we get the following action
\beq\label{eq:MPaction}
\begin{split}	
 S \= \frac{1}{2} \int \vol e^{-2\Phi} \left\{ R - \frac{1}{12} H_{mnp} H^{mnp} + 4 (\del_m\Phi)^2 + \frac{\ap}{4} \tr |F|^2 - \frac{\ap}{4} \tr |R^\H|^2   + \cdots\right\} ~.
\end{split}
\eeq
This agrees with the first order action in $\ap$ written down by Bergshoeff--de Roo \cite{Bergshoeff:1989de}. Integrating the Bianchi identity we get $c_2(V) = c_2(X)$.  For this subsection only, as we are determining the overall sign, we work to first order in $\ap$. 

There are some physics calculations we can do to verify the relative sign of $\tr |F|^2$ in the supergravity action. First,  we compute the energy-momentum tensor $T_{mn}$ and check a positive energy theorem from general relativity. The sign of the gauge field contribution must ensure positive energy density. The simplest test is the null energy condition (NEC), requiring:
$$
k^m k^n T_{mn} \ge 0 \quad \text{for all null } k^m \text{ with } k^m k^n g_{mn} \= 0.
$$
From Einstein's equations we can deduce $T$:
\beq
R_{mn} - \half g_{mn} R \= T_{mn}~.
\eeq
 If we start from the action \eqref{eq:GSaction}, then we get $R_{mn}$ and $R$ follow from varying with respect to $g_{mn}$ and the dilaton:
\begin{equation}\begin{split}
&  R - 4(\nabla \Phi)^2 + 4 \nabla^2 \Phi - \half |H|^2 + \frac{\ap}{4} \tr |F|^2  -  \frac{\ap}{4} \tr |R^\H|^2 + \cO(\ap^3) \= 0~,\\[3pt]
&\cG_{mn}+\frac{\ap}{4}\cE^\H_{mn}+\Delta^{\rm conn}_{mn}\=\cO(\ap^3)~,\\[6pt]
\end{split}
\label{EOM}\end{equation}
\vskip5pt
where recall \eqref{eq:Delta-connection_b}
\beq\notag
\begin{split}
\cG_{mn}&\=\Ric_{mn}+2\nabla_m\nabla_n\Phi-\frac14H_{mpq}H_n{}^{pq}~,\\
\D_{mn}^{\text conn} &\= \frac{\ap}{4}e^{2\Phi} \nabla^p \Bigl( e^{-2\Phi}\left(
 \nabla_m\cG_{np}  +\nabla_n\cG_{mp}  -2\nabla_p\cG_{mn}  \right) \Bigr) +\cO(\ap^3)~.\end{split}
\eeq

 We find
\beq
\begin{split}
 T_{mn} &\= - 2\left( \nabla_m \nabla_n\Phi + g_{mn} ((\nabla \Phi)^2 -  \nabla^2 \Phi )\right) + \frac{1}{4} \left(H_{mpq} H_n{}^{pq} -  g_{mn}  |H|^2\right) \\
 &\quad -\frac{\ap}{4} \left( \tr F_{mp} F_n{}^p - \half g_{mn} \tr |F|^2\right) +\frac{\ap}{4} \left( \tr R^\H_{mp} R^\H_n{}^p - \half g_{mn} \tr |R^\H|^2\right)~.
\end{split}
\eeq
For simplicity, we now set the dilaton to a constant, in which case, there is no difference between Einstein and string frame.  The NEC $k^m k^n T_{mn} \ge 0$ becomes
\beq
k^m k^n T_{mn} \= \qrt \left( H_{mpq} k^m \right)^2 - \frac{\ap}{4} \tr (F_{mp} k^m) (F_{n}{}^p k^n) + \frac{\ap}{4} \tr (R^\H_{mp} k^m) (R^\H_{n}{}^p k^n)~.
\eeq
Rewrite in terms of generators of the Lie algebra so that terms are manifestly squares:
\beq
k^m k^n T_{mn} \= \qrt \left( H_{mpq} k^m \right)^2 +  \frac{\ap}{4} (F^A_{mp} k^m)^2 -   \frac{\ap}{4} (R^{\H\,A'}_{mp} k^m)^2~,
\eeq
where $A$ denotes a basis of the gauge group and $A'$ runs over the adjoint of $SO(9,1)$. 
We see that the first two terms are manifestly positive definite. Note that the sign changed as we evaluated the trace; this confirms we have the correct relative sign in the supergravity action. Furthermore, the $ \tr R^\H \w R^\H$ term contributes a negative energy density. This is consistent with heterotic compactifications that are dual type IIB compactifications (or F-theory), in which the $ \tr R^\H \w R^\H$ term is believed to be mapped to the gravitational contribution of O7-planes and these are believed to have a negative contribution to matter energy density \cite{Dasgupta:1999ss}. In turn this maps to an 8-derivative term in M-theory known as $X_8$. Such negative density contributions via higher derivative (equivalently $\ap$) corrections are important also in type IIA and massive type IIA string theory  \cite{McOrist:2012yc,Maxfield:2013wka}, facilitating the existence of flux solutions similar to the $K3$ torsional solutions in heterotic.

It is also worth pointing out the moduli space metric for a compactification to $d=4$, written up \citeM,\citeSG is evaluated with the convention that the operator $\Tr$ has action that is positive definite on antihermitian generators, as is done so in for example \cite{Anguelova:2010ed, Martelli:2010jx,Becker:2009df}\footnote{For  example $\Tr$ could be taken to be proportional to minus the matrix trace. An advantage of this notation is that quantities that appear in say the energy momentum tensor, or the moduli space metric, that are positive come with a plus sign. The disadvantage is its evaluation is not simply taking the matrix trace.}. We write down how the metric looks in the usual convention that $\tr$ is the matrix trace. This metric is derivable just from the action \eqref{eq:MPaction} and is given by 
\beq\begin{split}
 g\#_{\a\bb} &\= \frac{1}{V}\int_X \Big( \D_\a{}^\m\star\D_{\bb}{}^{\nb}\,g_{\m\nb} + \frac{1}{4} \ccZ_\a^{(1,1)}\star\ccZb_{\bb}^{(1,1)} - \frac{\ap}{4}\tr{ ( \fD_\a A \star \fD_{\bb}A ) } \Big) + \cdots\\[0.1cm]
\end{split}\raisetag{2cm}\eeq
This metric governs the kinetic terms of $d=4$ massless scalar fields. Since moduli can mix under diffeomorphisms, the gauge bundle's contribution should match the form and sign of metric and complex structure moduli terms.

\section{Hermitian geometry}

\subsection{Conformally balanced metrics}
Let $X$ be a complex manifold with ${\rm dim}_{\mathbb{C}} X = 3$. The tangent bundle splits as
\[
T_{\mathbb{C}} X \= \ccT_X^{(1,0)} \oplus \ccT_X^{(0,1)}~,
\]
and we will use Greek indices (e.g. $\mu,\nu$) to denote directions along $\ccT_X^{(1,0)}$. We will sometimes use the notation
\[
  \dd^c \=\! \ii (\partial - \bar{\partial})~.
\]
Let $\omega = \ii g_{\mu \bar{\nu}} \, dx^{\mu \bar{\nu}} \in \Omega^{1,1}(X,\mathbb{R})$ be a hermitian metric, so that the components satisfy $\overline{g_{\mu \bar{\nu}}} = g_{\nu \bar{\mu}}$ and $g_{\mu \bar{\nu}}$ is a positive-definite local matrix. The components of a differential form in holomorphic coordinates are given by
\[
\eta \= \frac{1}{p! q!}  \eta_{\mu_1 \cdots \mu_p \bar{\nu}_1 \cdots \bar{\nu}_q} dx^{\mu_1 \cdots \mu_p \bar{\nu}_1 \cdots \bar{\nu}_q}, \quad \eta \in \Omega^{p,q}(X)~.
\]
Roman indices (e.g. $i,j$) are used full indices spanning all of $T_{\mathbb{C}} X$. Our conventions for norms of forms are:
\[
|\eta|^2 \= \frac{1}{k!} \eta_{i_1 \cdots i_k} \eta^{i_1 \cdots i_k}, \quad \eta \in \Omega^k(X)~.
\]
We use the notation $i \Lambda_\omega: \Omega^{p+1,q+1}(X) \rightarrow \Omega^{p,q}(X)$ for metric contraction:
\[
(i \Lambda_\omega \eta)_K \= g^{\mu \bar{\nu}} \eta_{\mu \bar{\nu} K}, \quad K = \mu_1 \cdots \mu_{p} \bar{\nu}_1 \cdots \bar{\nu}_{q}~.
\]
In complex dimension 3, we note the following wedge product identity
\beq \label{eq:wedge-id}
\eta \wedge \omega \= \half  (\Lambda_\omega \eta) \wedge \omega^2, \quad \eta \in \Omega^3(X)~.
\eeq
We now discuss the conformally balanced condition, which generalizes the K\"ahler condition on $\omega$. Let $f$ be an arbitrary function and consider
\[
\partial (e^{-2 f} \omega^2) \= 2 e^{-2 f} \bigg[  \partial \omega \wedge \omega - \partial f \wedge \omega^2 \bigg]~.
\]
Substitute \eqref{eq:wedge-id} with $\eta = \partial \omega$.
\beq  \label{eq:dconfbal}
\partial (e^{-2 f} \omega^2) \= 2 e^{-2 f} \bigg[ -\half  (i \Lambda_\omega) (i \partial \omega) -  \partial f \bigg] \wedge \omega^2~.
\eeq
We see that
\beq \label{def:confbal}
\bigg[ d ( e^{-2f} \omega^2) = 0 \bigg]  \ \Leftrightarrow \ \bigg[ (\partial - \bar{\partial}) f \=\! - \frac{\ii}{2}  \Lambda_\omega T \bigg] \ \Leftrightarrow  \ \bigg[ T_\mu{}^\mu{}_\lambda \=\! -2 \partial_\lambda f \bigg]~,
\eeq
where we use the notation $T \in \Omega^3(X)$ for 
\[
T \= \ii (\partial - \bar{\partial})\omega~, \quad T_{\mu \nu \bar{\rho}} \=\! - \partial_\mu g_{\nu \bar{\rho}} + \partial_\nu g_{\mu \bar{\rho}}~.
\]
A metric $\omega$ is conformally balanced if it solves any of the equivalent equations in \eqref{def:confbal}.

\subsection{Connection Symbols} \label{app:ConnectionSymbols}
Given a hermitian manifold $(X,g)$, we can ask whether there is an optimal natural connection $\nabla$ on $T_{\mathbb{C}} X$. Unless $g$ is a K\"ahler metric, there is no agreed-upon optimal connection. We present here several connections which all play various roles in string theory and in the main text. Our conventions are as follows: a vector field $V \in \Gamma(T_{\mathbb{C}} X)$ is written as
\[
V \= V^k \partial_k \= V^\mu \partial_\mu + V^{\bar{\mu}} \partial_{\bar{\mu}}~,
\]
where $k$ denotes $T_{\mathbb{C}} X$ indices and $\mu$ denotes $\ccT_X^{(1,0)}$ indices. Our notation for a connection $\nabla$ acting on $V \in \Gamma(T_{\mathbb{C}} X)$ is
\[
\nabla_i (V^k \partial_k) \= (\nabla_i V^k) \partial_k~, \quad \nabla_i V^k \= \partial_i V^k + \Gamma_i{}^k{}_j V^j~,
\]
where $\Gamma_i{}^k{}_j$ will be specified by the connection under consideration. It what follows, we only specify $\Gamma_\mu{}^k{}_j$ with the understanding that $\Gamma_{\bar{\mu}}{}^{\bar{k}}{}_{\bar{j}} = \overline{\Gamma_\mu{}^k{}_j}$.

\subsubsection*{Levi-Civita}
The Levi--Civita connection $\nabla^{\LC}$ is the unique metric compatible connection with no torsion. Unfortunately, $\nabla^{\LC}$ does not preserve $\ccT_X^{(1,0)}$ (unless $g$ is K\"ahler) and is not recommended for calculations in holomorphic coordinates.

\beq\begin{split}\label{eq:LCsymbols}
 &\G^{\LC}{}_{\m}{}^{\n}{}_\r   \=  g^{\n\sb} \, \del_\m g_{\r\sb} - \frac{1}{2} T_{\m}{}^{\n}{}_{\r}  ~,\\
 &\G^{\LC}{}_{\m}{}^{\nb}{}_{\r}  \=  0 ~,\\
 &\G^{\LC}{}_{\m}{}^{\n}{}_{\rb}   \=  \frac{1}{2} T_{\m}{}^\n{}_{\rb} ~,\\
 &\G^{\LC}{}_{\m}{}^{\nb}{}_{\rb}  \=\!  -\frac{1}{2} T_{\m}{}^{\nb}{}_{\rb} ~.
\end{split}\eeq

\subsubsection*{Chern}
The Chern connection is the unique connection which satisfies both $\nabla_k g_{\mu \bar{\nu}} = 0$ and $\nabla_{\bar{\mu}} V^\alpha = \partial_{\bar{\mu}} V^\alpha$, making it well-suited for calculations in holomorphic coordinates.
\beq\label{eq:Chsymbols}
\begin{split}
 &\G^{\Ch}{}_{\m}{}^{\n}{}_\r   \=  g^{\n\sb} \, \del_\m g_{\r \sb} ~,\\
 &\G^{\Ch}{}_{\m}{}^{\nb}{}_{\r}  \=  0 ~,\\
 &\G^{\Ch}{}_{\m}{}^{\n}{}_{\rb}  \=  0 ~,\\
 &\G^{\Ch}{}_{\m}{}^{\nb}{}_{\rb}  \=  0 ~.
\end{split}\eeq

\subsubsection*{Hull}

A connection $\G^\Hu$ appears in the heterotic action and Bianchi identity which we will refer to as the Hull connection. It is defined by $\G_m^\Hu = \G_m^{\LC} + \half H_m$. Let us assume $H = d^c \omega$. Then this connection $\G^{\Hu}$ does not preserve $\ccT_X^{(1,0)}$ and its connection symbols are:

\beq\begin{split} \label{defn:RH}
 &\G^{\Hu}{}_{\m}{}^{\n}{}_\r  \=  \G^{\Ch}{}_{\m}{}^{\n}{}_\r ~,\\
 &\G^{\Hu}{}_{\m}{}^{\nb}{}_{\r}  \=  0 ~,\\
 &\G^{\Hu}{}_{\m}{}^{\n}{}_{\rb}    \=  (d^c \omega)_\m{}^\n{}_{\rb} ~,\\
 &\G^{\Hu}{}_{\m}{}^{\nb}{}_{\rb}  \=  0 ~.
\end{split}\eeq

We remark that in the context of the present paper, the formula for the 3-form is $H = d^c \omega + 2 \ap P$. We will identify $H = d^c \widehat{\omega}$ for a corrected hermitian metric $\widehat{\omega}$, and these formulas will be applied to $\widehat{\G}^\Hu$.

\subsubsection*{Bismut}
It was noticed by Yano \cite{YanoBook} that specifying the ansatz $\G_m^\Bi = \G_m^{\LC} - \half T_m$ where $T \in \Omega^3(X)$ is a priori an arbitrary 3-form, together with the requirement that $\nabla^\Bi$ preserves $\ccT_X^{(1,0)}$, uniquely determines the 3-form in the ansatz as $T = i (\partial-\bar{\partial})\omega$. See Proposition \ref{uniqueness-bismut} for details. This connection in non-K\"ahler complex geometry is sometimes called the Bismut connection.
\medskip
\beq\begin{split}
 &\G^{\Bi}{}_{\m}{}^{\n}{}_\r   \=  \G^{\Ch}{}_\m{}^\n{}_\r - T_{\m}{}^{\n}{}_{\r} ~,\\
 &\G^{\Bi}{}_{\m}{}^{\nb}{}_{\r}  \=  0 ~,\\
 &\G^{\Bi}{}_{\m}{}^{\n}{}_{\rb}  \=  0 ~,\\
 &\G^{\Bi}{}_{\m}{}^{\nb}{}_{\rb}  \=\!  - T_{\m}{}^{\nb}{}_{\rb} ~.
\end{split}\eeq

\subsection{Curvature}
From a connection $\nabla$ on $T_{\mathbb{C}} X$, the curvature tensor $R \in \Omega^2( {\rm End} \, T_{\mathbb{C}} X)$ will be denoted
\[
R_{pq}{}^i{}_j = \partial_p \Gamma_q{}^i{}_j + \Gamma_p{}^i{}_\ell \Gamma_q{}^\ell{}_j - (p \leftrightarrow q)~.
\]
From here, one can compute the curvatures of all the connections listed above in holomorphic coordinates. The Ricci curvature and scalar curvature of the Levi-Civita connection are denoted
\[
\Ric_{mn} \=\! - R^{\LC}{}_{m i}{}^i{}_n~, \quad R \= R_m{}^m~.
\]
The 4-form $\tr R \wedge R$ for various connections will be used in the main text, which is
\[
\tr R \wedge R \= \frac{1}{4} R_{mn}{}^i{}_j R_{pq}{}^j{}_i \, \dd x^{mnpq}~.
\]
The coordinate expression for the Chern curvature is the simplest.
\[
R^{\Ch}{}_{\mu \bar{\nu}}{}^\alpha{}_\beta \=\! - \partial_{\bar{\nu}} ( g^{\alpha \bar{\sigma}} \partial_\mu g_{\beta \bar{\sigma}})~.
\]
One can take four different traces of $R^{\Ch}$ along the holomorphic directions to get four analogs of Ricci curvature. Two of these traces will play a role in the main text.
\begin{enumerate}
\item The first trace is
\beq \label{eq:trR1}
R^{\Ch}{}_{\mu \bar{\nu}}{}^\alpha{}_\alpha \=\! - \partial_\mu \partial_{\bar{\nu}} \log \det g_{\rho \bar{\sigma}}~.
\eeq
Its significance is that it defines the first Chern class. If $X$ admits a holomorphic volume form $\Omega$, then
\beq \label{eq:RCH-hol}
R^{\Ch}{}_{\mu \bar{\nu}}{}^\alpha{}_\alpha \= \partial_\mu \partial_{\bar{\nu}} \log \norm{\Omega}^2_\omega~.
\eeq
Here $\Omega \overset{\rm loc}{=} f(x) dx^{123}$ in local holomorphic coordinates $x^\mu$ with $f(x)$ a local non-vanishing holomorphic function, and $\norm{\Omega}^2_\omega \overset{\rm loc}{=} |f|^2 (\det g_{\mu \bar{\nu}})^{-1}$.

\item The second  trace that will appear in our analysis is
  \[
(i \Lambda_\omega R^{\Ch})^\alpha{}_\beta \= R^{\Ch}{}_{\mu}{}^\mu{}^\alpha{}_\beta~.
  \]
  This quantity vanishes if the Chern connection is an instanton. We can compare both traces via the exchange relation
\beq \label{exchange-relation}
R^{\Ch}{}_\mu{}^\mu{}_{\bar{\beta} \alpha} \= R^{\Ch}{}_{\alpha \bar{\beta}}{}^\mu{}_\mu - (i \partial \bar{\partial} \omega)_\mu{}^\mu{}_{\bar{\beta} \alpha} + \nabla_\alpha^{\Ch} T_\mu{}^\mu{}_{\bar{\beta}} - \nabla_{\bar{\beta}}^{\Ch} T_\mu{}^\mu{}_\alpha + T_{\alpha \mu \bar{\nu}} T^{\mu \bar{\nu}}{}_{\bar{\beta}}~.
\eeq
  
\end{enumerate}

We end with a lemma containing an identity which will be used in the main text.

\begin{prop}
  Let $X$ be a complex threefold with hermitian metric $\omega$ and holomorphic volume form $\Omega$. Suppose
  \[
{\rm d} (\norm{\Omega}_g \omega^2) \=0~.
  \]
  Then
   \begin{align} 
R^{\Ch}{}_\mu{}^\mu{}_{\bar{\beta} \alpha} &\=\! - (\ii \partial \bar{\partial} \omega)_\mu{}^\mu{}_{\bar{\beta} \alpha} + T_{\alpha \mu \bar{\nu}} T^{\mu \bar{\nu}}{}_{\bar{\beta}}~, \label{eq:R-inst}\\
g^{\mu \bar{\nu}} R^{\Hu}{}_{\mu \bar{\nu} ij} &\=\! - (\ii \partial \bar{\partial} \omega)_\mu{}^\mu{}_{ij}~, \quad R^{\Hu}{}_{\mu \nu ij} = - (\ii \partial \bar{\partial} \omega)_{\mu \nu ij}~. \label{eq:RH-inst}
\end{align}
\end{prop}

Identity \eqref{eq:R-inst} follows from \eqref{exchange-relation} upon substituting \eqref{def:confbal} and \eqref{eq:trR1}. Identity \eqref{eq:RH-inst} then follows from converting Hull to Chern via \eqref{defn:RH}. The interpretation of \eqref{eq:RH-inst} is the deviation of the Hull connection of a conformally balanced metric from being an instanton. We refer to \eqref{eq:RhdH} for this identity on $R^{\Hu}$ derived from the BdR supersymmetry algebra with identification $dH = -2 \ii \partial \bar{\partial} \omega + \cO(\ap^2)$.


\providecommand{\href}[2]{#2}\begingroup\raggedright\endgroup

\end{document}